\documentclass[usenatbib]{mn2e}
\usepackage{apjfonts,amsfonts,amsmath,amssymb,bm,ctable,verbatim,xcolor}

\newcommand{\fireone}{{FIRE-1}}
\newcommand{\firetwo}{{FIRE-2}}
\newcommand{\firex}{{FIRE-3}}
\newcommand{\msun}{M_{\sun}}

\newcommand{\Alf}{{Alfv\'en}}
\newcommand{\Hmol}{{\rm H}_{2}}
\newcommand{\HI}{{\rm H{\small I}}}

\newcommand{\GIZMO}{{\small GIZMO}}
\newcommand{\gizmourl}{\href{http://www.tapir.caltech.edu/~phopkins/Site/GIZMO.html}{\url{http://www.tapir.caltech.edu/~phopkins/Site/GIZMO.html}}}
\newcommand{\FIREurl}{\href{http://fire.northwestern.edu}{\url{http://fire.northwestern.edu}}}
\newcommand{\FIREpublicurl}{\href{http://flathub.flatironinstitute.org/fire}{\url{http://flathub.flatironinstitute.org/fire}}}
\newcommand{\acknowledgments}[1]{\begin{small}\section*{Acknowledgments}\end{small}{\noindent #1}\vspace{5pt}}
\newcommand{\datastatement}[1]{\begin{small}\section*{Data Availability Statement}\end{small}{\noindent #1}\vspace{5pt}}

\title[\firex\ Updates]{\firex: Updated Stellar Evolution Models, Yields, \&\ Microphysics and Fitting Functions for Applications in Galaxy Simulations}

\author[Hopkins et al.]{
\vspace{0.2cm}
\parbox[t]{\textwidth}{Philip F.~Hopkins$^1$, 
Andrew Wetzel$^2$, 
Coral Wheeler$^3$, 
Robyn Sanderson$^4$,
Michael Y. Grudi{\'c}$^{5}$,
Omid Sameie$^6$, 
Michael Boylan-Kolchin$^6$, 
Matthew Orr$^7$,
Xiangcheng Ma$^{8}$, 
Claude-Andr{\'e} Faucher-Gigu{\`e}re$^{9}$, 
Du\v{s}an Kere\v{s}$^{10}$, 
Eliot Quataert$^{11}$, 
Kung-Yi Su$^7$, 
Jorge Moreno$^{12}$, 
Robert Feldmann$^{13}$, 
James S. Bullock$^{14}$, 
Sarah R. Loebman$^{15}$,
Daniel Angl{\'e}s-Alc{\'a}zar$^{16}$, 
Jonathan Stern$^{17}$, 
Lina Necib$^{18}$, 
\&\ Christopher C. Hayward$^{7}$
} \vspace*{4pt} \\
$^1$ TAPIR, Mailcode 350-17, California Institute of Technology, Pasadena, CA 91125, USA. E-mail:phopkins@caltech.edu \\
$^2$ {Department of Physics and Astronomy, University of California, Davis, CA 95616, USA} \\
$^3$ {Department of Physics and Astronomy, California State Polytechnic University, Pomona, Pomona, CA 91768, USA} \\
$^4$ {Department of Physics and Astronomy, University of Pennsylvania, 209 South 33rd Street, Philadelphia, PA 19104, USA} \\ 
$^5$ {Carnegie Observatories, 813 Santa Barbara St, Pasadena, CA 91101, USA; NASA Hubble Fellow}\\
$^6$ {Department of Astronomy, The University of Texas at Austin, 2515 Speedway, Stop C1400, Austin, Texas 78712-1205, USA} \\
$^7$ {Center for Computational Astrophysics, Flatiron Institute, 162 5th Ave., New York, NY 10010 USA} \\ 
$^8$ {Department of Astronomy and Theoretical Astrophysics Center, University of California Berkeley, Berkeley, CA 94720} \\
$^{9}$ {Department of Physics and Astronomy and CIERA, Northwestern University, 2145 Sheridan Road, Evanston, IL 60208, USA} \\ 
$^{10}$ {Department of Physics, Center for Astrophysics and Space Science, University of California at San Diego, 9500 Gilman Drive, La Jolla, CA 92093, USA} \\ 
$^{11}$ {Department of Astrophysical Sciences, Princeton University, Peyton Hall, Princeton, NJ 08544, USA} \\
$^{12}$ {Department of Physics and Astronomy, Pomona College, Claremont, CA 91711, USA} \\ 
$^{13}$ {Institute for Computational Science, University of Zurich, Winterthurerstrasse 190, Zurich CH-8057, Switzerland} \\ 
$^{14}$ {Department of Physics and Astronomy, 4129 Reines Hall, University of California, Irvine, CA 92697, USA} \\
$^{15}$ {Department of Physics, University of California, Merced, 5200 N. Lake Road, Merced, CA 95343, USA} \\
$^{16}$ {Department of Physics, University of Connecticut, 196 Auditorium Road, U-3046, Storrs, CT 06269, USA} \\
$^{17}$ {School of Physics and Astronomy, Tel Aviv University, Tel Aviv 69978, Israel} \\
$^{18}$ {Department of Physics \&\ Kavli Institute for Astrophysics and Space Research, Massachusetts Institute of Technology, Cambridge, MA 02139, USA}
}

\date{}
\begin{document}
\maketitle

\begin{abstract}
Increasingly, uncertainties in predictions from galaxy formation simulations (at sub-Milky Way masses) are dominated by uncertainties in stellar evolution inputs. In this paper, we present the full set of updates from the \firetwo\ version of the Feedback In Realistic Environments (FIRE) project code, to the next version, \firex. While the transition from \fireone\ to \firetwo\ focused on improving numerical methods, here we update the stellar evolution tracks used to determine stellar feedback inputs, e.g.\ stellar mass-loss (O/B and AGB), spectra (luminosities and ionization rates), and supernova rates (core-collapse and Ia), as well as detailed mass-dependent yields. We also update the low-temperature cooling and chemistry, to enable improved accuracy at $T \lesssim 10^{4}\,$K and densities $n\gg 1\,{\rm cm^{-3}}$, and the meta-galactic ionizing background. All of these synthesize newer empirical constraints on these quantities and updated stellar evolution and yield models from a number of groups, addressing different aspects of stellar evolution. To make the updated models as accessible as possible, we provide fitting functions for all of the relevant updated tracks, yields, etc, in a form specifically designed so they can be directly ``plugged in'' to existing galaxy formation simulations. We also summarize the default \firex\ implementations of ``optional'' physics, including spectrally-resolved cosmic rays and supermassive black hole growth and feedback.
\end{abstract}

\vspace{-0.5cm}
\begin{keywords}
galaxies: formation \&\ evolution --- stars: formation --- methods: numerical --- ISM: structure 
\end{keywords}

\vspace{-0.3cm}
\section{Introduction}

It is now well-established that ``feedback'' from stars -- e.g.\ coupling of stellar radiation, outflows/mass-loss, supernovae (SNe), etc., to ambient interstellar medium (ISM) gas -- plays an essential role in galaxy formation. In the absence of stellar feedback, most of the gas in the cosmic web would rapidly accrete onto galaxies, cool on a timescale short compared to the dynamical time, collapse and fragment and turn into stars or brown dwarfs \citep{bournaud:2010.grav.turbulence.lmc,tasker:2011.photoion.heating.gmc.evol,hopkins:rad.pressure.sf.fb,dobbs:2011.why.gmcs.unbound,krumholz:2011.rhd.starcluster.sim,harper-clark:2011.gmc.sims}, producing galaxies with properties grossly discrepant from observations \citep{katz:treesph,somerville99:sam,cole:durham.sam.initial,springel:lcdm.sfh,keres:fb.constraints.from.cosmo.sims} almost independent of the ``details'' of star formation \citep{white:1991.galform,keres:fb.constraints.from.cosmo.sims}. Meanwhile observed galaxies are seen to turn their gas into stars at a rate of just a few percent per dynamical time \citep{kennicutt98}, with molecular clouds disrupting owing to feedback after just a few percent of their mass becomes stars \citep{williams:1997.gmc.prop,evans:1999.sf.gmc.review,evans:2009.sf.efficiencies.lifetimes}, and then galaxies appear to expel a large fraction of their mass into the circum-galactic medium (CGM; \citealt{aguirre:2001.igm.metal.evol.sims,pettini:2003.igm.metal.evol,songaila:2005.igm.metal.evol,martin:2010.metal.enriched.regions,oppenheimer:outflow.enrichment,werk:2014.cos.halos.cgm,tumlinson:2017.cgm.review}) in observed galactic winds \citep{martin99:outflow.vs.m,martin06:outflow.extend.origin,heckman:superwind.abs.kinematics,newman:z2.clump.winds,sato:2009.ulirg.outflows,chen:2010.local.outflow.properties,steidel:2010.outflow.kinematics,coil:2011.postsb.winds}. 

In the past decade there has been remarkable progress capturing these feedback processes in simulations which attempt to capture the multi-phase complexity of the ISM and CGM \citep{hopkins:fb.ism.prop,kim:tigress.ism.model.sims,grudic:sfe.gmcs.vs.obs,benincasa:2020.gmc.lifetimes.fire,keating:co.h2.conversion.mw.sims}. These simulations have begun to resolve the self-consistent generation of galactic outflows and fountains alongside accretion onto galaxies \citep{narayanan:co.outflows,angles.alcazar:particle.tracking.fire.baryon.cycle.intergalactic.transfer,hayward.2015:stellar.feedback.analytic.model.winds,muratov:2016.fire.metal.outflow.loading,hafen:2018.cgm.fire.origins,2019arXiv191001123H,hopkins:2020.cr.outflows.to.mpc.scales,ji:fire.cr.cgm} and the turbulent processes within the ISM \citep{hopkins:frag.theory,hopkins:2012.intermittent.turb.density.pdfs,guszejnov:imf.var.mw,escala:turbulent.metal.diffusion.fire,guszejnov:universal.scalings,rennehan:turb.diff.implementation.fancy,gurvich:2020.fire.vertical.support.balance}. 

Specifically, there have been major advances in both numerical methods and understanding the key physics of how supernovae \citep{martizzi:sne.momentum.sims,gentry:clustered.sne.momentum.enhancement,rosdahl:2016.sne.method.isolated.gal.sims,hopkins:sne.methods,2018MNRAS.478..302S,kawakatu:2020.obscuration.torus.from.stellar.fb.in.torus}, stellar mass-loss \citep{wiersma:2009.enrichment,conroy:2014.agb.heating.quenching,2018A&ARv..26....1H}, and stellar radiation \citep{hopkins:rad.pressure.sf.fb,hopkins:2019.grudic.photon.momentum.rad.pressure.coupling,hopkins:radiation.methods,wise:2012.rad.pressure.effects,rosdahl:m1.method.ramses,2018ApJ...859...68K,emerick:rad.fb.important.stromgren.ok} couple to the ISM. One such example is the Feedback In Realistic Environments (FIRE) project \citep{hopkins:2013.fire},\footnote{\FIREurl} which represents an attempt to synthesize all of the major known stellar feedback channels directly from stellar evolution models, to combine them with the known ISM thermo-chemical cooling processes and cosmological initial conditions into predictive simulations of galaxy formation. 

However, most of the galactic models above, including FIRE, utilize stellar evolution ``inputs'' -- e.g.\ stellar population synthesis models that predict key inputs for feedback such as supernova rates, mass-loss rates and yields, stellar population spectra -- which themselves rely on stellar evolution libraries whose isochrones and mass-loss assumptions are, in turn, calibrated to observations that are often several decades old. Common basic assumptions (that all massive stars are single and non-rotating) are almost certainly incorrect, and some key empirical ingredients (e.g.\ calibration of giant-star mass-loss rates) have been revised by more than an order-of-magnitude in the past decade. Indeed, stellar astrophysics has seen a revolution in the last decade, with truly transformative, qualitatively new data (and orders-of-magnitude increase in data volume) coming from time-domain surveys such as \textit{Kepler} \citep{2014PASP..126..398H,2015MNRAS.452.2127S}, astrometric distances from \textit{Gaia} \citep{2017ApJ...844..102H,2019A&A...623A.110G}, massive spectroscopic surveys such as SDSS and APOGEE \citep{2014ApJS..211...17A,2017AJ....154...94M,2019ARA&A..57..571J}, alongside an enormous number of time-domain studies focused on binaries and explosions such as PTF/ZTF and ASAS-SN \citep{2018MNRAS.477.3145J,2019PASP..131a8002B}, and now accompanying gravitational wave constraints from LIGO.
These radical improvements in observations have been accompanied by explosive growth in theory and modeling, utilizing new codes and techniques and data to fundamentally revise our understanding of e.g.\ basic stellar evolution and interior dynamics, rotation, binarity, mass-loss, and the pre-explosion physics that is crucial for nucleosynthetic yields \citep{2015ApJS..220...15P,2016ApJ...823..114N,2017ApJ...835..173S,2017ApJ...835...77M,2019ARA&A..57...35A,2019arXiv191212300A}. 

Meanwhile there have been a number of other important advances for galactic ``inputs,'' e.g.\ better constraints on the redshift evolution and shape of the meta-galactic UV background \citep[e.g.,][]{khaire19_lowz,worseck19_HeII,gaikwad20_consistent} and timing of reionization from CMB and Gunn-Peterson measurements \citep[e.g.][and references therein]{2019ApJ...878...12H,2020A&A...641A...1P}, and an explosion of data on cold atomic and molecular gas from facilities such as ALMA that can probe the detailed thermo-chemical state of ISM metals and gas at densities $n \gg 1\,{\rm cm^{-3}}$ and $T \ll 10^{4}\,$K \citep[e.g.][]{2018A&ARv..26....5C}.

The collection of these advances warrants updating the basic input assumptions of our previous galaxy formation simulations. In this paper, we therefore synthesize and present the updated set of stellar evolution libraries, yields, cooling functions, and other assumptions that underpin the FIRE simulations, constituting the ``\firex'' version of the FIRE simulation code. In \S~\ref{sec:overview}, we provide a brief overview of previous FIRE versions and motivations for this study, and describe the range of applicability of the fitting functions provided here. \S~\ref{sec:updates} describes the updates to treatment of fluid dynamics (\S~\ref{sec:fluids}), the UV background (\S~\ref{sec:uvb}), stellar evolution tracks (\S~\ref{sec:stellar.evol.tracks}), Solar abundances (\S~\ref{sec:solar.z}), cooling physics (\S~\ref{sec:cooling}), treatment of HII regions (\S~\ref{sec:hii}) and other radiative feedback channels (\S~\ref{sec:radiation}), SNe (\S~\ref{sec:sne}), stellar mass-loss (\S~\ref{sec:mass.loss}), star formation criteria (\S~\ref{sec:sf.criteria}), and nucleosynthetic yields (\S~\ref{sec:yields}). We then summarize the \firex\ implementations of ``optional'' physics which will be used in some (but not all) \firex\ runs, specifically explicit evolution of cosmic rays (\S~\ref{sec:crs}) and black hole accretion/feedback (\S~\ref{sec:bhs}). In \S~\ref{sec:galaxy.compare} we briefly compare the effects of the updates to the default \firex\ model on galaxy formation simulations, and conclude in \S~\ref{sec:conclusions}. Some additional tests are presented in Appendix~\ref{sec:alternatives}, and additional details of the mechanical feedback implementation are in  Appendix~\ref{sec:appendix:sne}.

\section{Overview \&\ Background}
\label{sec:overview}

\subsection{\fireone\ and \firetwo}
\label{sec:overview.fire12}

The first version of the FIRE code -- \fireone\ -- attempted to synthesize the core physics of stellar feedback from SNe (Ia \&\ II), stellar mass-loss (O/B \&\ AGB mass-loss), and radiation (photo-heating, ionization, and pressure) together with detailed cooling physics from $\sim 10-10^{10}\,$K, into fully-cosmological simulations of galaxy formation \citep{hopkins:2013.fire}. Subsequent work used this code to study a wide variety of topics ranging from detailed properties of dwarf galaxies \citep{onorbe:2015.fire.cores,chan:fire.dwarf.cusps,elbadry.2015:core.transformation.stellar.kinematics.gradients.in.dwarfs}, elemental abundance patterns in galaxies \citep{ma:2015.fire.mass.metallicity}, galactic outflows and the circum-galactic medium \citep{faucher-giguere:2014.fire.neutral.hydrogen.absorption,muratov:2015.fire.winds,hafen:2016.lyman.limit.absorbers,vandevoort:sz.fx.hot.halos.fire}, the origins of star formation scaling relations \citep{orr:stacked.vs.bursty.sf.fire,sparre.2015:bursty.star.formation.main.sequence.fire}, and high-redshift and massive galaxy populations \citep{feldmann.2016:quiescent.massive.highz.galaxies.fire,feldmann:colors.highz.quiescent.massivefire.gals,oklopcic:clumpy.highz.gals.fire.case.study.clumps.not.long.lived}. 

These and related papers which developed the numerical methods \citep{hopkins:rad.pressure.sf.fb,hopkins:fb.ism.prop} represented an initial attempt to directly take the outputs of stellar evolution models and apply them in galaxy-scale simulations to model salient stellar feedback rates. However, the input stellar evolution models from {\small STARBURST99} \citep{starburst99}, while widely-used at the time, generally relied on stellar evolution tracks (e.g.\ \citealt{BC03}), which assumed non-rotating, non-binary stellar populations, with mass-loss rates which were calibrated to older observations that have since been revised strongly (towards lower mass-loss rates; see e.g.\ \citealt{2014ARA&A..52..487S}). Many of the key isochrone inputs and outputs were calibrated to stellar observations from before 1990 \citep[e.g.][]{1983A&A...121...77W,1987A&A...188...74W,1984ApJ...278L..41O,1985ApJ...292..640K,1995A&A...297..727B}. As a result, even some of the first \fireone\ studies noted that some quantities: e.g.\ detailed elemental abundance patterns within galaxies, or the escape fraction of ionization photons, were quite different if one considered more ``state-of-the-art'' stellar evolution and/or yield models instead \citep[see][]{ma:2015.fire.escape.fractions,ma:2016.disk.structure}. Moreover, some of the approximations used in \fireone\ (and \firetwo) -- for example, treating core-collapse SNe yields as IMF-averaged, as opposed to tracking different yields from different stellar mass progenitors, or treating low-temperature molecular gas with a simple sub-grid molecular fraction estimator -- significantly limit the predictive power for modeling e.g.\ different detailed internal abundance spreads, or cold gas observables like CO \citep{bonaca:gaia.structure.vs.fire,escala:turbulent.metal.diffusion.fire,muley:2020.fire.nugrid.yield.tests,keating:co.h2.conversion.mw.sims}

Nonetheless, the next version of FIRE, \firetwo\ \citep{hopkins:fire2.methods},\footnote{The \firetwo\ simulations are publicly available via \citet{wetzel:fire2.public.release} at \\FIREpublicurl} attempted for the sake of consistency to keep the physical inputs (e.g.\ stellar evolution assumptions, feedback rates, etc.) fixed to \fireone\ values as much as possible, while updating instead the numerical methods. This represented a major update to numerical accuracy, utilizing a new hydrodynamic solver with a flexible arbitrary Lagrangian-Eulerian method \citep{hopkins:gizmo} as opposed to the Pressure-SPH method used for \fireone\ \citep{hopkins:lagrangian.pressure.sph}, enabling the accurate numerical addition of new physics such as magnetic fields \citep{hopkins:mhd.gizmo}, anisotropic diffusion \citep{hopkins:gizmo.diffusion}, and cosmic rays \citep{chan:2018.cosmicray.fire.gammaray,su:turb.crs.quench,hopkins:cr.mhd.fire2}. It also made major improvements to the accuracy of the gravitational force integration, and numerical treatment/coupling of both mechanical (SNe and mass-loss; \citealt{hopkins:sne.methods}) and radiative \citep{hopkins:radiation.methods} feedback ``injection'' which improved convergence. Together these enabled order-of-magnitude higher-resolution simulations reaching $\sim 30\,M_{\odot}$ resolution in small dwarfs and $\sim 3000\,M_{\odot}$ resolution in Local-Group (MW+Andromeda) halos \citep{wetzel.2016:latte,garrisonkimmel:local.group.fire.tbtf.missing.satellites,wheeler:ultra.highres.dwarfs}, and produced a large number of detailed results studying a wide variety of galaxy properties, including behavior of different ISM phases \citep[e.g.][]{elbadry:HI.obs.gal.kinematics,moreno:2019.fire.merger.suite,moreno:2021.galaxy.merger.sims,gurvich:2020.fire.vertical.support.balance,benincasa:2020.gmc.lifetimes.fire}, detailed galactic structure and scaling relations \citep[e.g.][]{elbadry:fire.morph.momentum,Sanderson2020,orr:ks.law,wellons20_rotation,yu:2021.fire.bursty.sf.thick.disk.origin,Bellardini2021}, satellite and dark matter properties \citep[e.g.][]{garrison.kimmel:2019.sfh.local.group.fire.dwarfs,garrisonkimmel:local.group.fire.tbtf.missing.satellites,necib:2019.dm.velocity.distribution.fire.vs.obs,samuel:2020.plane.of.satellites.fire, Santistevan2020}, and properties of the circum-galactic medium \citep[e.g.][]{hafen:2018.cgm.fire.origins, 2019arXiv191001123H,ji:fire.cr.cgm,li:2021.low.z.fire.cgm.probes,stern21_ICV}.

\subsection{\firex}
\label{sec:overview:fire3}

In this new version of FIRE, \firex, our goal is not to change our core numerical methods, nor to change the fundamental physics being simulated. Instead, it is to update the ``known'' microphysics, particularly the treatment of stellar evolution, yields, ISM cooling and chemistry, to more accurate and complete inputs that enable more detailed observational predictions. We stress that this is not a numerical methods paper. All of the relevant numerical methods for the default version of \firex\ are described and extensively tested in the lengthy \firetwo\ numerical methods papers, specifically the series \citet{hopkins:fire2.methods,hopkins:sne.methods,hopkins:radiation.methods}, but also including the updates for  various ``non-default'' physics (e.g.\ black holes, cosmic rays, etc.) where relevant in \citet{daa:BHs.on.FIRE,ma:fire.reionization.epoch.galaxies.clumpiness.luminosities,su:discrete.imf.fx.fire,garrisonkimmel:local.group.fire.tbtf.missing.satellites,chan:2018.cosmicray.fire.gammaray,hopkins:cr.mhd.fire2}. We do make some minor numerical modifications to the ``default'' treatment in \firex\ for improved accuracy, all of which we describe below, but even these are all modifications proposed and tested specifically in the \firetwo\ numerical methods papers above. We also stress that all of these updates are driven by more accurate theoretical and empirical inputs to the ``microphysics,'' rather than any ``desired'' result on galactic scales.

As with previous versions of FIRE, we implement all physics here in the {\small GIZMO} code \citep{hopkins:gizmo}.

\begin{figure}
	\includegraphics[width=0.98\columnwidth]{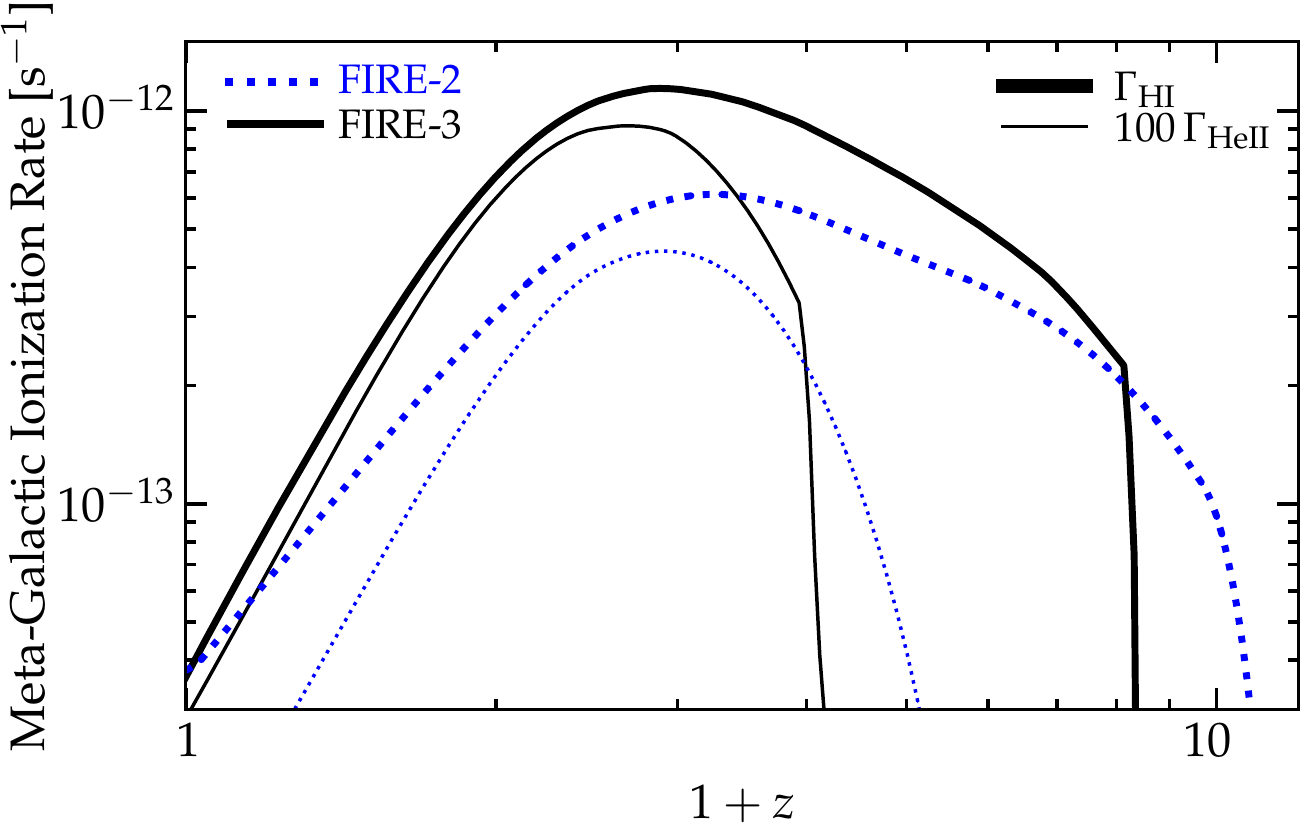}
	\vspace{-0.1cm}
	\caption{Comparison of the meta-galactic UV background (UVB) photoionization rates versus redshift $z$ for HI ($\Gamma_{\rm HI}$; {\em thick}) and HeII ($\Gamma_{\rm HeII}$, multiplied by 100 for ease of comparison; {\em thin}), in \firetwo\ ({\em blue dotted}) and \firex\ ({\em black solid}), per \S~\ref{sec:uvb}. Updated observational constraints modestly increase the photoionization rates at intermediate redshifts. The sharp drops in the HI and HeII ionization rates in \firex\ at $z>7$ and $z>3$, respectively, come from more accurately accounting for the large optical depth of the Universe to these photons when HI or HeII ionization is not yet complete \citep[see][]{cafg:2020.uv.background}. 
	\label{fig:uvb}}
\end{figure}

\subsection{Utility \&\ Range of Applicability of Inputs}
\label{sec:overview:applicability}

For the sake of making the updated models public in the most useful form for other galaxy formation simulations, we have endeavored to reduce as much as possible the new stellar evolution and cooling physics to simple, easily-implemented fitting functions, which can be immediately inserted into different numerical simulation codes and/or semi-analytic models for galaxy and star formation.

The types of models for which these are applicable are those with resolution broadly in the range $\sim 10-10^{6}\,{\rm M_{\odot}}$ (mass) or $\sim 0.1 - 100\,{\rm pc}$ (force/spatial), which attempt to explicitly treat/resolve some of the multi-phase structure of the ISM and/or CGM (e.g.\ the existence of giant molecular cloud complexes), and spatially-and-time resolved galactic star formation and SNe/stellar feedback events (e.g.\ the time between individual SNe in a single star particle is generally much larger than its numerical timestep), but with insufficient resolution to actually model/predict {\em individual} (proto)stellar collapse and masses and accretion/evolution tracks (aka forward-modeling stellar accretion and the IMF). For significantly higher-resolution simulations, different algorithms such as those in e.g.\ STARFORGE \citep{guszejnov:2020.starforge.jets,grudic:starforge.methods} are required which can correctly treat every star as an evolving-mass sink particle and deal with accretion and feedback from individual proto and main-sequence stars along independent stellar evolution tracks. For significantly lower-resolution simulations, different algorithms like those in MUFASA or SIMBA \citep{dave.2016:mufasa.fire.inspired.cosmo.boxes,dave:mufasa.followup.gas.metal.sfr.props.vs.time,2019MNRAS.486.2827D,thomas:bhs.in.simba.sims} which treat stellar feedback as a continuous, collective processes integrating implicitly over (rather than trying to directly numerically resolve) different star-forming regions and ISM phases, are more appropriate.

\section{Updates from \firetwo\ to \firex}
\label{sec:updates}

We now describe all updates from \firetwo\ to \firex. ``Default'' \firetwo\ should be understood to be the version presented and studied extensively in \citet{hopkins:fire2.methods}. Any details that we do {\em not} explicitly describe as modified in this section remain identical in \firex\ and \firetwo.

\subsection{Fluid Dynamics, Magnetic Fields, Conduction, and Viscosity}
\label{sec:fluids}

{\bf Improved Face Error Detection:} The fluid dynamics solver is largely unchanged, using the same meshless finite-mass (MFM) method as \firetwo. The only change is a slightly improved treatment of special cases, discussed in \citet{hopkins:gizmo}, where one simultaneously has (1) strong fluxes, (2) elements with extremely different densities (hence kernel/cell sizes) interacting directly, and (3) a pathological spatial configuration of neighboring elements (e.g.\ all $\sim 32$ nearest cell centers nearly aligned in a plane). In this case, the matrix inversion procedure needed to determine the effective faces for hydrodynamic fluxes becomes ill-conditioned, and floating point errors can lead to artificially large (or small) fluxes. We have improved the procedure from \citet{hopkins:gizmo} for dealing with such cases, by simultaneously (1) pre-conditioning the matrices to reduce floating-point errors, (2) adaptively expanding the neighbor search to ensure dimensionless condition numbers (as defined in Appendix~C of \citealt{hopkins:gizmo}) $\lesssim 10^{3}$, and (3) limiting the effective face area to the maximum possible geometric area between neighbor cells. This newer treatment is the default behavior in the public {\small GIZMO} code \citep[see][]{hopkins:gizmo.public.release},\footnote{The public version of {\small GIZMO} is available at \gizmourl} but was not implemented in \firetwo\ to ensure code-consistency. But in any case, this occurs rarely and has small effects. 

{\bf Additional ``Default'' Physics:} In \firex, certain physics that were ``optional'' in \firetwo\ are now ``default.''  This includes turbulent transport of passive scalars (e.g. metals), following \citet{hopkins:fire2.methods} Appendix~F3 \citep[see][for validation tests]{colbrook:passive.scalar.scalings,su:2016.weak.mhd.cond.visc.turbdiff.fx,escala:turbulent.metal.diffusion.fire}, and kinetic magnetohydrodynamics (MHD), i.e.\ magnetic fields \citep[see][for the \GIZMO\ MHD methods]{hopkins:mhd.gizmo,hopkins:cg.mhd.gizmo} with fully-anisotropic Spitzer-Braginskii conduction and viscosity as in \citet{hopkins:gizmo.diffusion} (with the coefficients given in \citealt{hopkins:cr.mhd.fire2}, scaling appropriately with the plasma state, accounting for saturation and limitation by plasma instabilities; \citealt{cowie:1977.evaporation,komarov:whistler.instability.limiting.transport,squire:2017.max.braginskii.scalings}). As shown in \citet{su:2016.weak.mhd.cond.visc.turbdiff.fx,su:fire.feedback.alters.magnetic.amplification.morphology,hopkins:cr.mhd.fire2,chan:2018.cosmicray.fire.gammaray,ji:fire.cr.cgm}, while important for predicting certain properties and regulating e.g.\ cosmic ray transport, these physics generally have small effects on bulk galaxy properties. \firex\ also includes, by default, the abundance ``tracers'' (following an additional set of passive scalars to model trace metal species) model of Wetzel et al., in prep.

\subsection{UV Background}
\label{sec:uvb}

\firetwo\ adopted the meta-galactic ultraviolet background (UVB) spectrum, as a function of redshift, from \citet{faucher-giguere:2009.ion.background}. Since this time observational constraints on the UVB have greatly improved.
For example, the tabulation used in \firetwo\ produced a redshift of HI reionization $z\sim10$, consistent with WMAP-7 constraints \citep[][]{WMAP7_cosmology} but too high compared to more recent Planck data and other measurements which imply a lower reionization midpoint $z\lesssim8$ \citep[e.g.][]{planck18_cosmology}. 
For \firex, we update the assumed UVB to the more recent \citet{cafg:2020.uv.background} model which synthesizes and better reproduces a number of different empirical constraints including updated luminosity functions, stellar spectra including binary stars, obscured and non-obscured AGN \citep[following][]{shen:bolometric.qlf.update}, intergalactic HI and He II photoionization rates measured at $z\sim0-6$, and the local X-ray background. 
As demonstrated in \citet{onorbe:reion.correction} and \citet{puchwein19_uvb}, UVB models that assume the intergalactic medium (IGM) is optically thin will produce earlier reionization than intended if applied without correction in simulations like ours to the pre-reionization Universe (which is opaque to ionizing photons and not yet in ionization equilibrium). The UVB tabulation we use for \firex\ therefore adopts the ``effective'' photoionization and photoheating rates calibrated to match correctly the Planck 2018 reionization optical depth and recent constraints from quasar absorption spectra on HeII reionization \citep[][]{khaire17_emissivity, worseck19_HeII}. These effective rates produce more accurate redshifts of HI and HeII reionization in the simulations; these correspond to the sharp drops in the HI and HeII photoionization rates at $z>7$ and $z>3$ for the \firex\ model in Figure \ref{fig:uvb}.\footnote{The updated UVB model is available in the \texttt{TREECOOL} file with the public {\small GIZMO} code (\gizmourl) or alternatively at \href{https://galaxies.northwestern.edu/uvb-fg20}{\url{https://galaxies.northwestern.edu/uvb-fg20/}}} 

The effects of the changed background intensity and shape are generally minor except for e.g.\ detailed CGM or Ly$\alpha$ forest studies. The most important effect of the updated UVB is later reionization, which can significantly influence dwarf galaxy star formation histories in the ultra-faint regime especially. These effects (and the role of remaining uncertainties in the UVB) will be studied in detail in future work.

\begin{figure}
	\includegraphics[width=0.98\columnwidth]{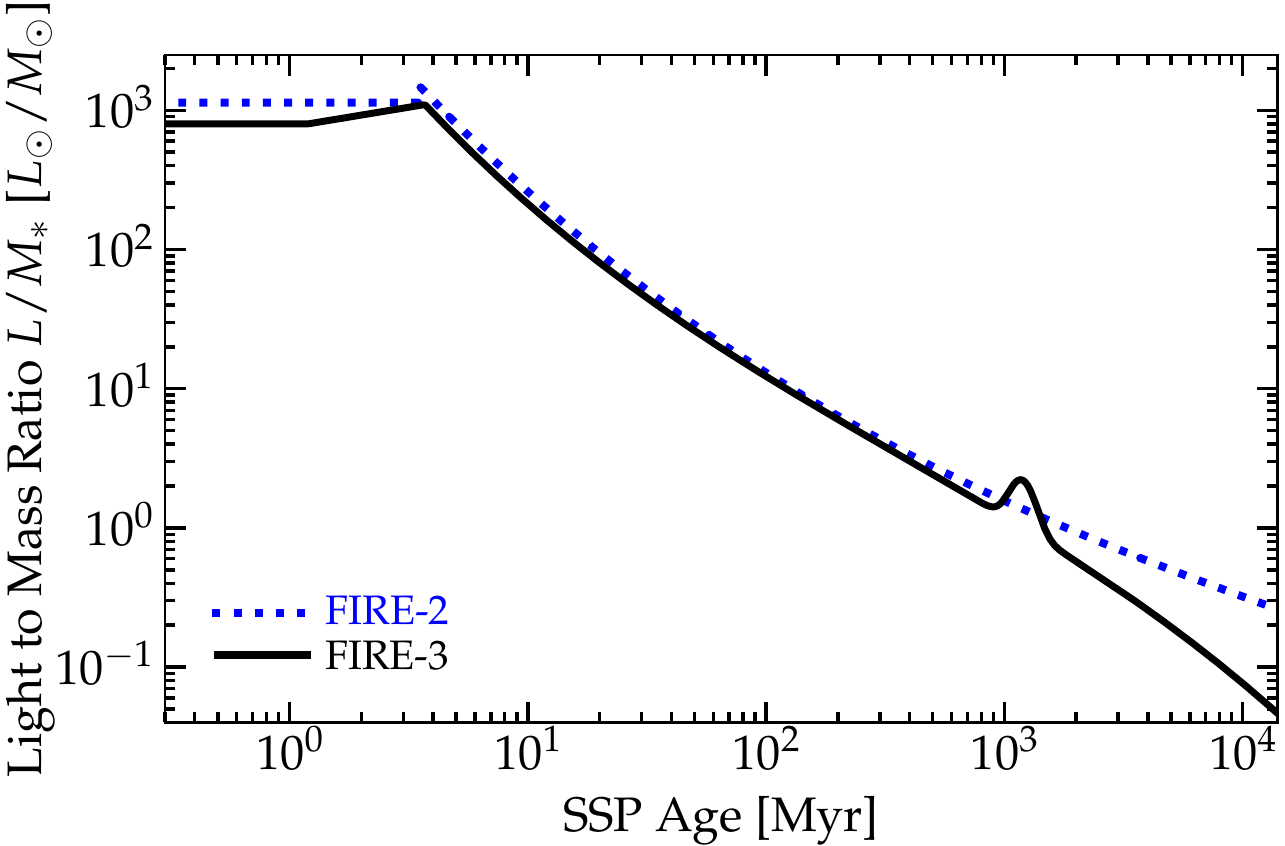}
	\includegraphics[width=0.98\columnwidth]{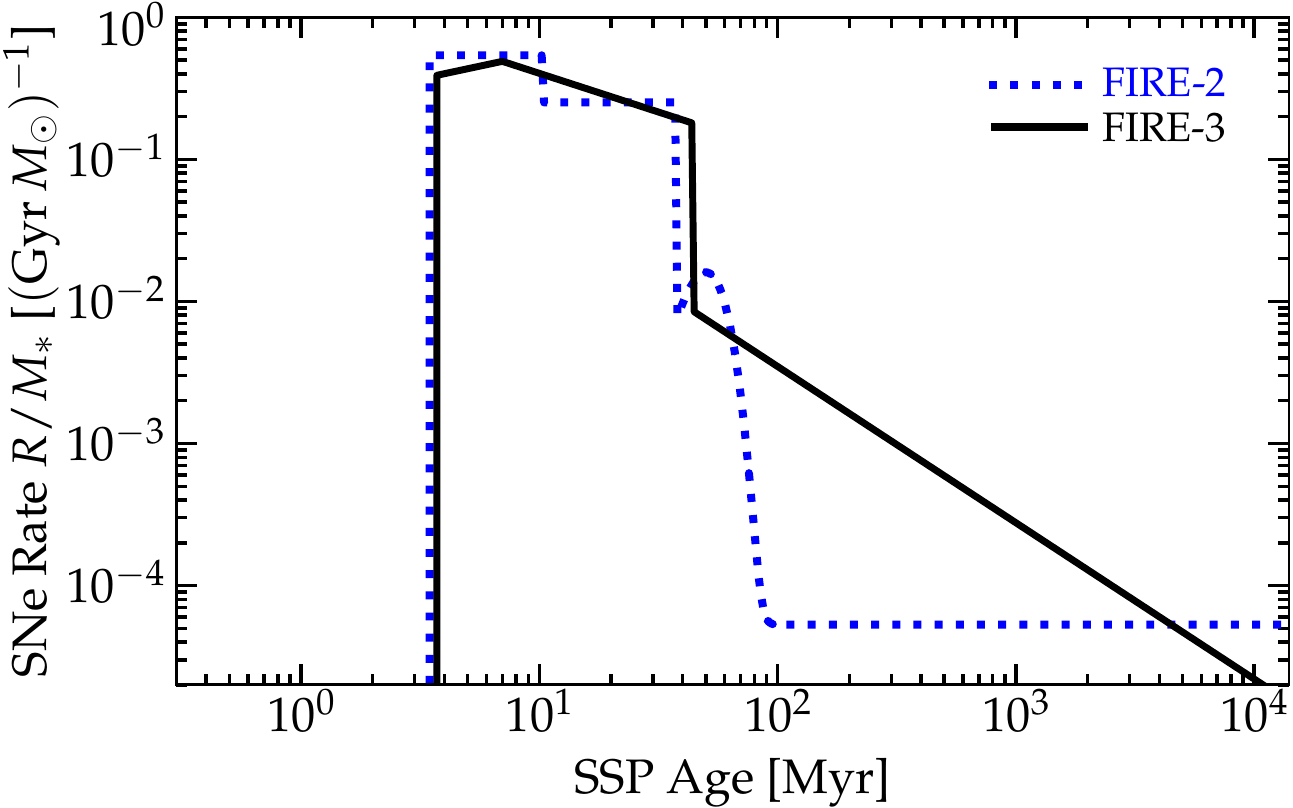}
	\includegraphics[width=0.98\columnwidth]{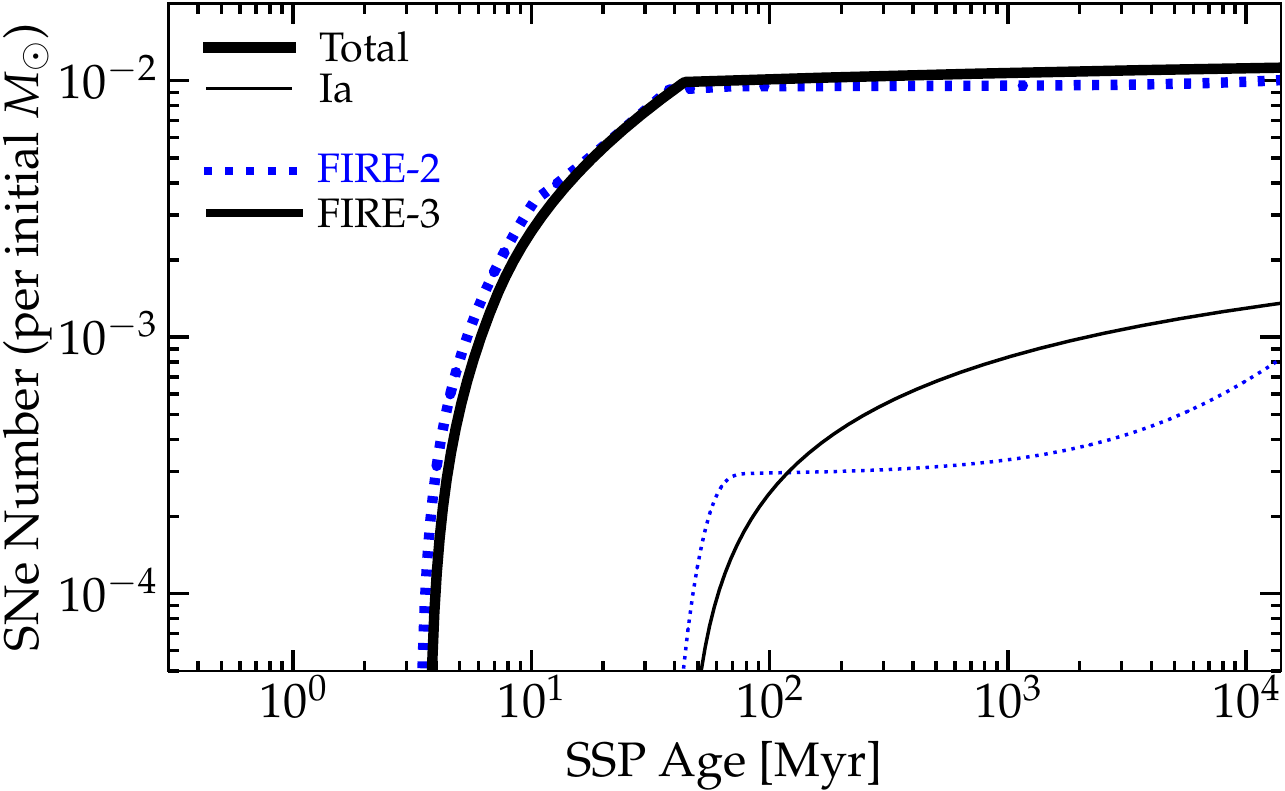}
	\vspace{-0.1cm}
	\caption{Comparison of the single-stellar-population (SSP) bolometric light-to-mass-ratio $L/M_{\ast}$ ({\em top}), SNe rate ({\em middle}), and cumulative SNe number ({\em bottom}), vs.\ SSP age, at solar abundances, in \firetwo\ ({\em blue dotted}) and \firex\ ({\em black solid}), per \S~\ref{sec:stellar.evol.tracks}. These depend only weakly on metallicity. Rotation and updated calibrations for stellar isochrones lead to relatively small changes in either $L/M_{\ast}$ or the core-collapse SNe rate (SNe before $\lesssim 44\,$Myr). The Ia rate ({\em middle}) is appreciably different: the more realistic and well-motivated delay-time-distribution (DTD) scaling as $t^{-1.1}$ in \firex\ leads to a significant enhancement in the Ia rate between $\sim 100-3000\,$Myr, although the Hubble-time-integrated Ia number is only a factor $\sim 1.6$ larger ({\em bottom}). 
	\label{fig:tracks}}
\end{figure}

\subsection{Stellar Evolution Tables}
\label{sec:stellar.evol.tracks}

{\bf Updated Isochrones \&\ Stellar Models:} Our stellar feedback models take their inputs in the form of SNe and stellar mass-loss rates, stellar luminosities, and spectra, directly from standard stellar evolution models as a function of stellar population age, mass, and metallicity. The tabulations adopted in \firetwo\ are all described in detail in the Appendices of  \citet{hopkins:fire2.methods}. We have re-fit all of the salient stellar evolution tables for improved (1) physical accuracy (using more recent and detailed stellar evolution models), (2) numerical accuracy (using more accurate fitting functions), and (3) consistency (using more recent models that make simultaneous predictions for more diverse quantities). Wherever possible, we use the results from the January 2021 version of STARBURST99 \citep{2014ApJS..212...14L}, adopting a 3-part \citet{kroupa:2001.imf.var} IMF (with slopes $(0.3,\,1.3,\,2.3)$ from $(0.01-0.08,\,0.08-0.5,\,0.5-100)\,M_{\odot}$), an $8\,M_{\odot}$ SNe cutoff ($120\,M_{\odot}$ BH formation cutoff), the preferred ``evolution'' wind model, using the updated Geneva 2013 rotating stellar model isochrones (which are designed to reproduce many of the effects attributed to binarity as well, in massive stellar populations, and therefore show much smaller differences compared to models like BPASS from \citealt{BPASS}, as compared to the older models), sampled as densely as possible at all metallicities available. We have also compared the results using all available isochrone sets published in the last decade in either STARBURST99 or BPASS, to ensure we do not fit any spurious features. All quantities below are IMF-integrated, with $M_{\ast} = M_{\ast}(t)$ the star-particle mass at time $t$ ($t$ its age at a given timestep) -- i.e.\ all quantities are corrected to be multiplied by the instantaneous stellar mass, without needing to correct ``back to'' the initial mass -- and $\tilde{z}$ the appropriate metal abundance of the star particle (defined precisely below).

{\bf Stellar Luminosities:} The updated bolometric luminosities per unit stellar mass $L/M_{\ast}$ are well fit by:
\begin{align}
\frac{L/M_{\ast}}{L_{\odot}/M_{\odot}} =& 
\begin{cases}
a_{L,1} \hfill & \ (t \le t_{L,1}) \\ 
a_{L,1}\,(t/t_{L,1})^{\psi_{L,1}} \hfill & \ (t_{L,1} < t \le t_{L,2}) \\ 
a_{L,2}\,(t/t_{L,2})^{\psi_{L,2}}\,f_{L,2} \hfill & \ (t_{L,2} < t) 
\end{cases}
\end{align}
where 
\begin{align}
\nonumber \psi_{L,1} &\equiv \frac{\ln{(a_{L,2}/a_{L,1})}}{\ln{{(t_{L,2}/t_{L,1})}}} , \\ 
\nonumber \psi_{L,2} &\equiv -1.82\,\left[ 1 - 0.1\,\left\{ 1 - 0.073\,\ln{\left(\frac{t}{t_{L,2}} \right)} \right\}\,\ln{\left( \frac{t}{t_{L,2}} \right)} \right] , \\  
\nonumber f_{L,2} &\equiv 1+1.2\,\exp{\left[- \frac{(\ln{( t/t_{L,3} )})^{2}}{a^{2}_{L,3}}  \right]} ,   
\end{align}
with $(a_{L,1},\,a_{L,2},\,a_{L,3})=(800,\,1100\,\tilde{z}^{-0.1},\,0.163)$ and $(t_{L,1}, \,t_{L,2},\,t_{L,3})=(1.2,\,3.7,\,1200)\,$Myr.

The bolometric ionizing photon flux is given by $L_{\rm ion} \equiv f_{\rm ion}\,L$ with a fraction 
$f_{\rm ion} = f_{{\rm ion},1}$ at $t<t_{{\rm ion},1}$, $f_{\rm ion}=f_{{\rm ion},1}\,(t/t_{{\rm ion},1})^{-2.9}$ for $t_{{\rm ion},1} \le t \le t_{{\rm ion},2}$, and $f_{\rm ion}=0$ at $t>t_{{\rm ion},2}$ 
with $f_{{\rm ion},1}=0.5$ and $(t_{{\rm ion},1},\, t_{{\rm ion},2}) = (3.5,\,150)$\,Myr. The ratio of FUV/NUV/optical flux to bolometric, needed for our five-band radiation treatment, remains identical to \firetwo. 

These fits are similar to \firetwo, as shown in Fig.~\ref{fig:tracks}, with slightly reduced (by $\sim 40\%$) zero-age main sequence (ZAMS) bolometric luminosities (but only $\sim20\%$ reduced in FUV+ionizing bands, with harder spectra favored by many observations; see e.g.\ \citealt{steidel:2016.hard.stellar.spectra}) owing to rotation, and a slightly stronger optical feature (the ``bump'' at $t\sim1\,$Gyr, owing to giants, though this has little effect on the integrated output) followed by slightly more rapid decline (with the integrated photon energies integrated to $10\,$Gyr reduced by just $\sim 20\%$).

{\bf Core-Collapse Rates:} The updated core-collapse supernova (CCSNe) rate is well-fit by:
\begin{align}
\frac{R_{\rm CC}/M_{\ast}}{{\rm Gyr}^{-1}\,M_{\odot}^{-1}} =& 
\begin{cases}
0 \hfill & \ (t < t_{S,1}\ {\rm or}\ t > t_{S,3}) \\ 
a_{S,1}\,(t/t_{S,1})^{\psi_{S,1}} \hfill & \ (t_{S,1} \le t \le t_{S,2}) \\ 
a_{S,2}\,(t/t_{S,2})^{\psi_{S,2}} \hfill & \ (t_{S,2} \le t \le t_{S,3}) 
\end{cases}
\end{align}
where $\psi_{S,1}\equiv \ln{(a_{S,2}/a_{S,1})}/\ln{{(t_{S,2}/t_{S,1})}}$, $\psi_{S,2}\equiv \ln{(a_{S,3}/a_{S,2})}/\ln{{(t_{S,3}/t_{S,2})}}$, $(a_{S,1},\, a_{S,2},\, a_{S,3})=(0.39,\,0.51,\,0.18)$, $(t_{S,1},\, t_{S,2},\, t_{S,3})=(3.7,\,7.0,\,44)$\,Myr. 
In \firetwo\ each CC SNe carried an identical IMF-averaged ejecta mass $8.72\,M_{\odot}$, metal yield, and ejecta energy $E=10^{51}$\,erg. In \firex\ the ejecta masses, yields, and energies vary as described below (\S~\ref{sec:yields}). These are treated numerically (accounting for thermal and kinetic energy in detail) per \S~\ref{sec:sne}.

While this interpolates the declining CC SNe rates more accurately between $t\sim10-40\,$Myr compared to our \firetwo\ fit (both are compared in Fig.~\ref{fig:tracks}), the duration of the CC phase, total number, mass, and energetics of CC SNe are all identical to within $< 10\%$ (the total CC energy is $\sim 7\%$ larger at Solar metallicity).

{\bf Ia Rates:} The Ia rate is not given by STARBURST99: we update the \firetwo\ model taken from \citet{mannucci:2006.snIa.rates} (which has a small prompt component, and under-predicts the cosmological Ia rate by a factor $\sim 2$ compared to more recent observational estimates) to the fits to various constraints on the Ia rate from \citet{maoz:Ia.rate}, with an integrated number $1.6 \times10^{-3}$ of Ia's per solar mass over a Hubble time, and a $t^{-1.1}$ dependence including both prompt and delayed contributions. The ``initial'' delay time (time of first Ia) remains unconstrained so long as it is $\lesssim 100\,$Myr; for consistency, we set this to be equal to the time of the {\em last} CCSNe. The Ia rate is then:
\begin{align}
\frac{R_{\rm Ia}/M_{\ast}}{{\rm Gyr}^{-1}\,M_{\odot}^{-1}} =& 
\begin{cases}
0 \hfill & \ (t < t_{{\rm Ia},1}) \\ 
a_{{\rm Ia},1}\,(t/t_{{\rm Ia},1})^{\psi_{{\rm Ia},1}} \hfill & \ (t_{{\rm Ia},1} \le t ) 
\end{cases}
\end{align}
where $t_{{\rm Ia},1}=t_{S,3}=44\,$Myr, $\psi_{{\rm Ia},1}=-1.1$, $a_{{\rm Ia},1}=0.0083$. All Ia's carry $10^{51}$\,erg of initial ejecta kinetic energy (coupled numerically in the same method as CCSNe), a total ejecta mass $=1.4\,M_{\odot}$, and yields given below.

This Ia delay time distribution differs appreciably from the \firetwo\ fit, as seen in Fig.~\ref{fig:tracks} (for additional tests of this model in FIRE, see Gandhi et al. in prep.): the onset time for Ia and rates at $\sim 50\,$Myr and $\sim5\,$Gyr are similar, but the smooth decline means the Ia rate in the new fits between $\sim 0.08 - 2\,$Gyr is much higher, while the rate at $\sim10\,$Gyr is a factor $2-3$ lower. Integrating these rates from age 0 to age 1, 10, $13.7$\,Gyr, the new fit gives a factor of 2.4, 1.8, and 1.5 (respectively) more Ia's per unit mass.

\begin{figure}
	\includegraphics[width=0.98\columnwidth]{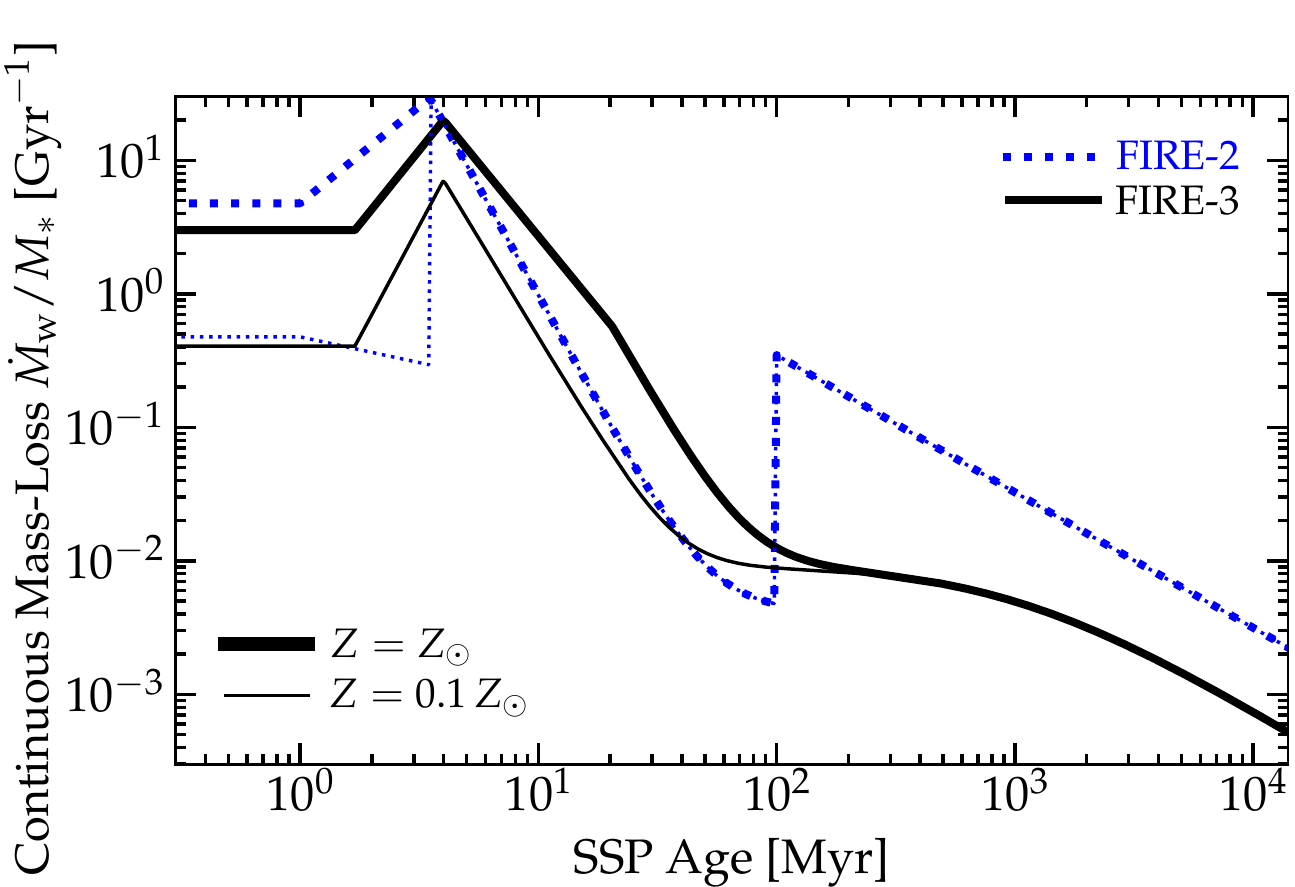}
	\includegraphics[width=0.98\columnwidth]{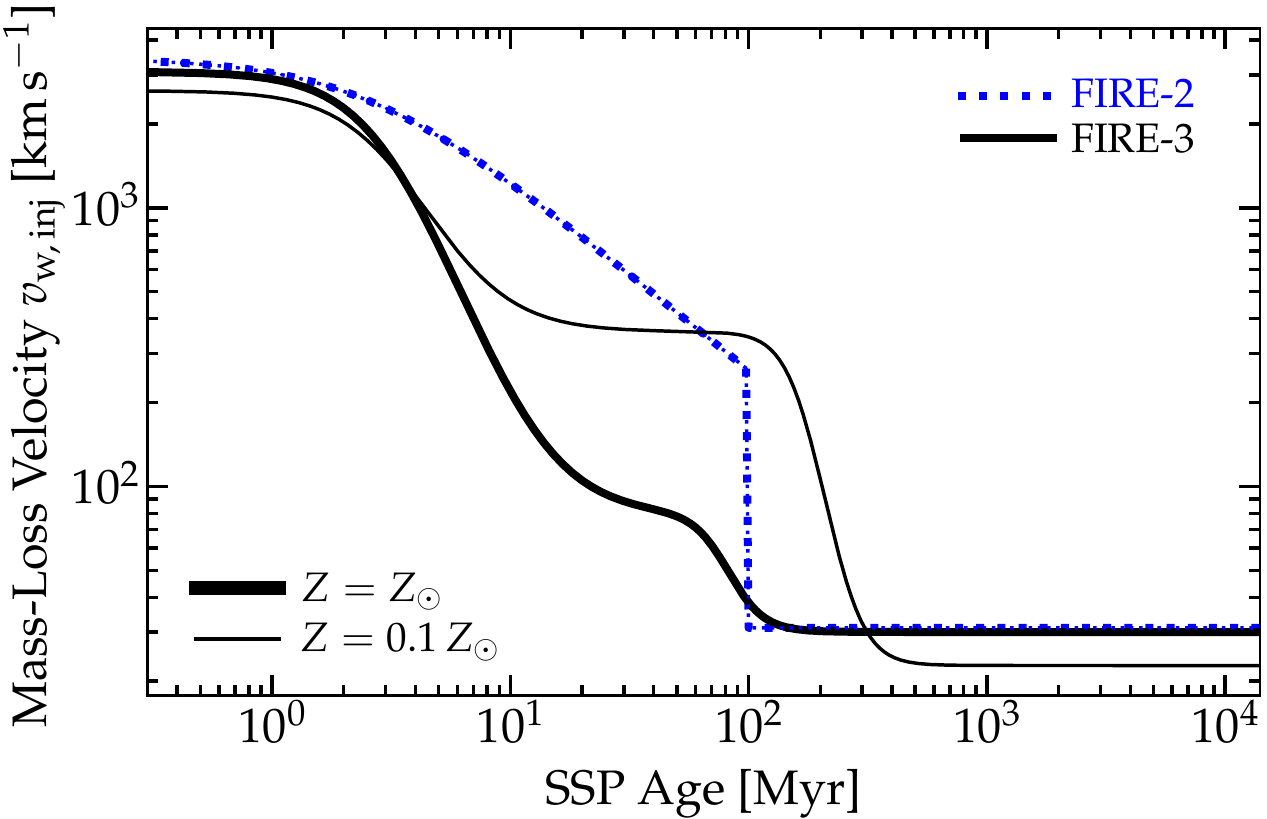}
	\vspace{-0.1cm}
	\caption{Continuous mass-loss rate $\dot{M}_{\rm w}$ (e.g.\ including O/B and AGB outflows, but not SNe; {\em top}) and mean initial ejecta mass-loss velocity $v_{\rm w}$ (e.g.\ kinetic luminosity $(1/2)\,\dot{M}_{\rm w}\,v_{\rm w}^{2}$; {\em bottom}) from \S~\ref{sec:stellar.evol.tracks}, vs.\ SSP age in \firetwo\ ({\em blue dotted}) and \firex\ ({\em black solid}), as Fig.~\ref{fig:tracks}. We compare two different initial metallicities (assuming solar abundance ratios). The O/B winds ($\lesssim 40\,$Myr) are similar, with the \firex\ models interpolating more smoothly in $Z$ or [Fe/H] dependence owing to better model isochrone sampling, and rotation slightly delaying/weakening some outflows. The major change is to the AGB winds, which in the models used for \firetwo\ turned on very strongly at $\gtrsim 100\,$Myr. The actual input AGB-star observations used to calibrate these stellar tracks were almost all from before 1990; the tracks used for the \firex\ model have been updated with observations from the past two decades which have significantly reduced the inferred mass-loss rates, and all models we survey using observations from the last decade show similar reductions.
	\label{fig:winds}}
\end{figure}

\begin{figure}
	\includegraphics[width=0.98\columnwidth]{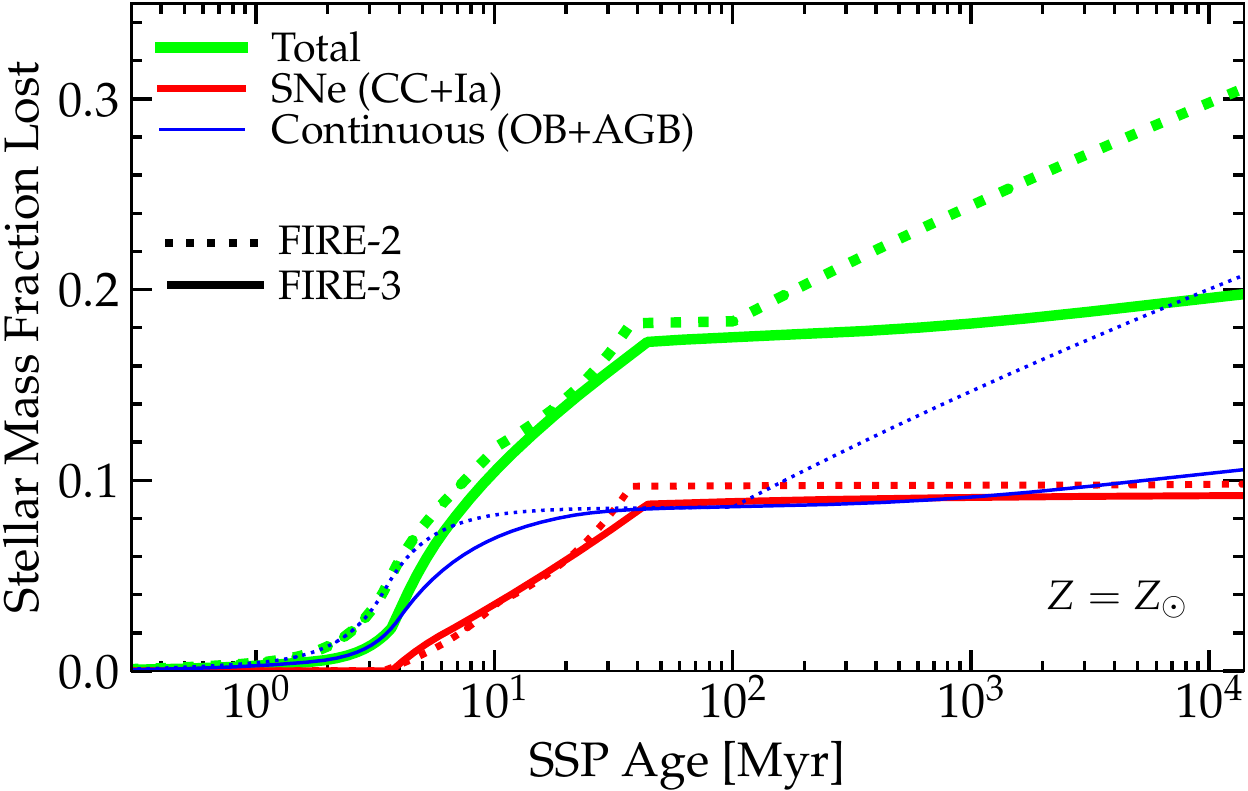}
	\vspace{-0.1cm}
	\caption{Cumulative fractional mass-loss (fraction of the initial SSP mass returned to the ISM as a function of time) vs.\ SSP age in \firetwo\ ({\em dotted}) and \firex\ ({\em solid}), at Solar metallicity, as Fig.~\ref{fig:winds}. O/B outflows (dominant at $t \ll 20\,$Myr), then core-collapse SNe are similar to each other and in \firetwo\ and \firex, returning $\sim 20\%$ of the SSP mass. The Hubble-time-integrated AGB mass-loss (dominant at $\gtrsim 100\,$Myr) is reduced by a factor of a few, from $\sim 10\%$ to $\sim 2\%$.
	\label{fig:winds.cumulative}}
\end{figure}

{\bf OB \&\ AGB Mass-Loss Rates:} The updated total stellar mass-loss rate (including all massive-star/OB and late-time, lower-mass AGB mass loss processes) is well-fit by:
\begin{align}
\frac{\dot{M}_{w}/M_{\ast}}{{\rm Gyr}^{-1}} = & 
\begin{cases}
a_{w,1} \hfill & \ (t \le t_{w,1}) \\ 
a_{w,1}\,(t/t_{w,1})^{\psi_{w,1}} \hfill & \ (t_{w,1} < t \le t_{w,2}) \\ 
a_{w,2}\,(t/t_{w,2})^{\psi_{w,2}} \hfill & \ (t_{w,2} < t \le t_{w,3}) \\
a_{w,3}\,(t/t_{w,3})^{\psi_{w,3}} \hfill & \ (t_{w,3} < t ) \\
\end{cases} \\ 
\nonumber &+ a_{A,1}\,\left\{ \left[ 1 + (t/t_{\rm A})^{1.1} \right]\,\left[ 1 + a_{A,2}\,(t/t_{\rm A})^{-1} \right] \right\}^{-1}
\end{align}
where $\psi_{w,1}\equiv \ln{(a_{w,2}/a_{w,1})}/\ln{{(t_{w,2}/t_{w,1})}}$, $\psi_{w,2}\equiv \ln{(a_{w,3}/a_{w,2})}/\ln{{(t_{w,3}/t_{w,2})}}$, $\psi_{w,3}=-3.1$, 
$(t_{w,1},\, t_{w,2},\, t_{w,3},\, t_{\rm A}) = (1.7,\, 4.0,\, 20,\, 1000)\,$Myr, and $(a_{w,1},\, a_{w,2},\, a_{w,3},\, a_{A,\,1},\, a_{A,\,2}) = (3\,\tilde{z}^{0.87}, \, 20\,\tilde{z}^{0.45}, \, 0.6\,\tilde{z},\, 0.01, \, 0.01)$. The first (piecewise power-law) term represents the dominant contribution from line-driven OB winds, hence its strong scaling with progenitor metallicity, while the second ($a_{A}...$ term) represents the dominant contribution from AGB outflows, which in modern models depends negligibly on progenitor abundances.\footnote{Note that we ignore the detailed timing of some particular narrow pulsational-related features in the AGB mass loss rates, as these vary enormously in timing and strength between models, and in the models used here contribute negligibly to the integral. However we do ensure the best-fit total mass loss rate integrated to a Hubble time exactly matches the full stellar evolution model.} 
The updated continuous input momentum ($\dot{M}_{w}\,v_{\rm w,\,inj}$) and kinetic energy ($\dot{M}_{w}\,v_{\rm w,\,inj}^{2}/2$) fluxes associated with this mass-loss at injection are fit by:
\begin{align}
\frac{v_{\rm w,\,inj}}{{\rm km\,s^{-1}}}
= &\,
\tilde{z}^{0.12}  \left[ 
\frac{3000}{1+(t/t_{v,1})^{2.5}}
+ \frac{600}{1 + \tilde{z}^{3}\,(t/t_{v,2})^{6} + 11.2\,\tilde{z}^{1.5}}
+ 30
\right]
\end{align}
with $(t_{v,1},\,t_{v,2})=(3,\,50)\,$Myr. Yields are also updated, as described below.

Though qualitatively similar, this differs quantitatively from the \firetwo\ fits, as shown in Figs.~\ref{fig:winds} \&\ \ref{fig:winds.cumulative}: the metallicity dependence is more accurate, and following a broad range of recent direct and indirect observational constraints \citep[see e.g.][]{2010ApJ...722L..64K,2012ApJ...748...47M,2013MNRAS.428.1479Z,2014ARA&A..52..487S,2018A&ARv..26....1H}, the integrated stellar mass loss rates are significantly reduced compared to older models. We stress that this result is robust across all models we have surveyed in this paper that have been calibrated to observations from the last $\sim$\,decade. In contrast the model tracks used to generate the stellar populations for \firetwo\ were calibrated largely to decades-older observations, as described in \S~\ref{sec:overview}. At solar abundances, pre-SNe ($t\lesssim 4\,$Myr) OB wind strengths are reduced by $\sim 40\%$ in these \firex\ fits compared to \firetwo, but their duration is extended, owing to rotation (producing similar total mass-loss integrated to $100\,$Myr, but factor $\sim2-3$ lower mass-loss at low metallicities). The AGB mass-loss rates are systematically lower compared to \firetwo\ (by factors $\sim 10-20\,$ at $\sim100\,$Myr and factor $\sim3$ at $\sim 10\,$Gyr), reducing the total (time-and-IMF-integrated) OB+AGB mass lost by a factor $\sim 2$ at solar abundances (and as much as $\sim 10$ in extremely metal-poor populations). As a result, in the newer models, integrated AGB mass loss from $0.1-10\,$Gyr is just $\sim 2\%$ of the initial SSP mass.

{\bf Abundances for Stellar Evolution Scalings:} The reference stellar evolution models do include progenitor metallicity dependence, but exclusively assume solar abundance ratios. To scale with metallicity we must therefore choose a reference abundance. In \firetwo, we defined the reference abundance for stellar evolution tables as $Z/Z_{\odot}$. However, the actual stellar physics processes which scale significantly with abundance (e.g.\ resonant-line-driven mass-loss rates and ionizing photon production rates) are strongly dominated by iron, since this dominates the opacities in the relevant portions of the spectrum. In \firex\ we therefore scale the stellar evolution models specifically off the iron abundance, defining $\tilde{z} \equiv 10^{\rm [Fe/H]}$. This can be important for winds and radiative feedback in dwarf galaxies, where [$\alpha$/Fe]$\,>0$ typically (so e.g.\ $Z/Z_{\odot}$ can be much larger than $\tilde{z}$).

\subsection{Reference \&\ Initial Metallicities}
\label{sec:solar.z}

\firetwo\ adopted the older \citet{anders.grevesse:1989.solar.abundances} reference values for solar composition, with e.g.\ $Z\sim 0.02$. In \firex\ we follow the newer stellar evolution models and scale to the \citet{asplund:2009.solar.composition} proto-solar abundances (see also \citealt{lodders:updated.solar.abundances.review}), with mass fractions of (Z, He, C, N, O, Ne, Mg, Si, S, Ca, Fe) $=$ (0.0142, 0.2703, 2.53e-3, 7.41e-4, 6.13e-3, 1.34e-3, 7.57e-4, 7.12e-4, 3.31e-4, 6.87e-5, 1.38e-3).

In \firetwo, owing largely to difficulty capturing metal-free cooling, we initialized our simulations with a uniform metallicity floor $Z = 10^{-4} - 10^{-5}\,Z_{\odot}$ with solar abundance ratios. In \firex\ this is no longer necessary, so we begin from pure primordial gas (potentially relevant for abundances of ultra-faint dwarfs at early times).

\subsection{Cooling \&\ Neutral Gas Physics}
\label{sec:cooling}

In \firetwo, gas cooling at low temperatures ($\lesssim 10^{4}\,$K) and/or high densities ($\gtrsim 1\,{\rm cm^{-3}}$) was treated using a particularly simple fitting function to tabulated {\small CLOUDY} \citep{ferland:1998.cloudy} results, as a function of the local density, temperature, and metallicity; all details are given in the appendices of \citet{hopkins:fire2.methods}. In \firex, we update this to follow the thermal state of the cold ISM in more detail. Note temperatures $T$ below are in ${\rm K}$ and units are cgs.

\begin{figure}
	\includegraphics[width=0.99\columnwidth]{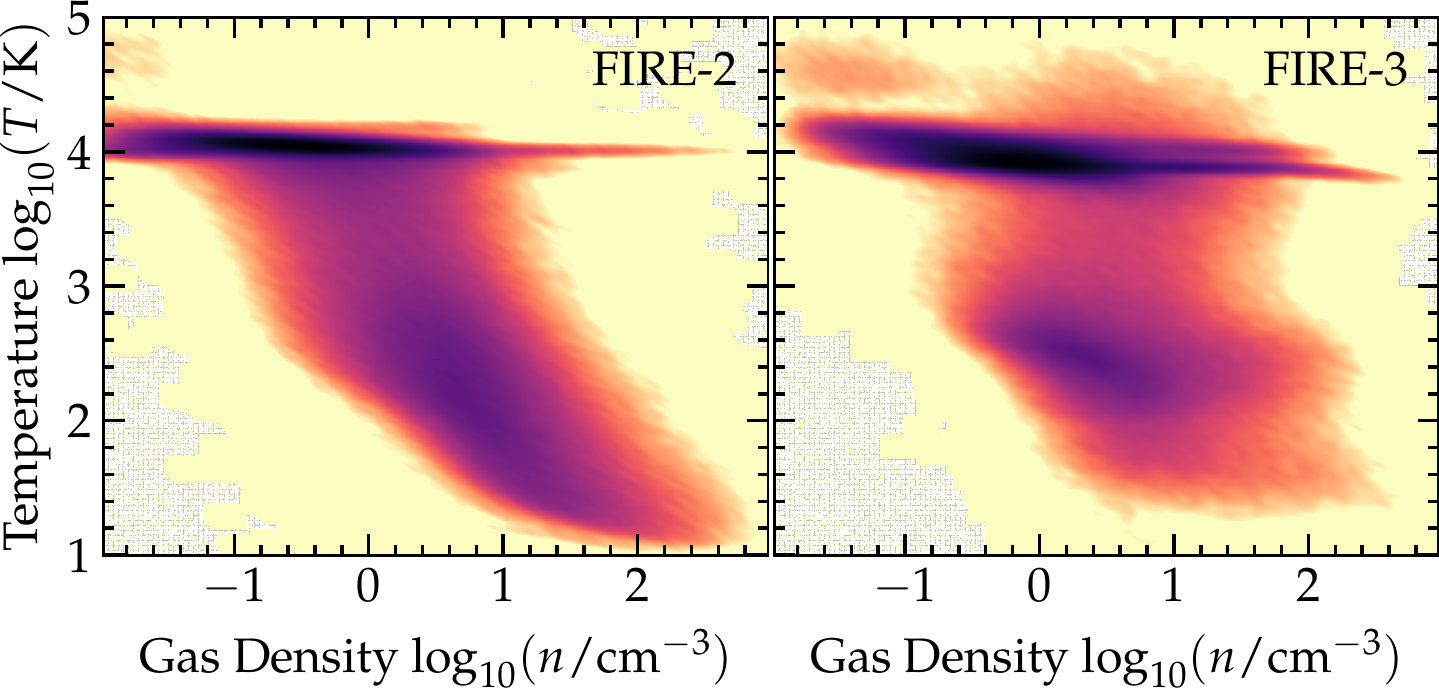}
	\vspace{-0.1cm}
	\caption{Phase diagram (density $n\equiv \rho/m_{p}$, vs.\ temperature $T$) of ISM gas (within $<10\,$kpc of the galaxy center) in simulations where we re-start a high-resolution \firetwo\ simulation of a Milky-Way like galaxy ({\bf m12i} from \citealt{hopkins:fire2.methods}) at $z=0.05$ and evolve it to $z=0$ with \firex\ (not following full cosmological evolution but allowing us to see the effect on the ISM for identical initial conditions). We focus on warm/cold gas and compare \firetwo, where low-temperature cooling was parameterized by a simple polynomial fit, to \firex, where we update the cooling to follow non-equilibrium molecular chemistry, detailed molecular and atomic and metal-line and free-electron processes, self-consistently calculate dust temperatures from ISM radiation, and more accurately calculate UVB self-shielding and HII region equilibrium temperatures (see \S~\ref{sec:cooling}). While the total mass of ``cold'' gas is similar, we see it has a quite different phase structure, with a much clearer separation between warm neutral and cold neutral  or molecular medium phases. These are not dynamically important (thermal pressure is never dominant in the cold, dense ISM) but has significant implications for observational diagnostics of neutral gas.
	\label{fig:phase}}
\end{figure}

\begin{itemize}

\item{\bf Cooling from Mostly Neutral Gas:} In \firex, we model the cooling from mostly neutral molecular gas using the combination of detailed scalings from several sources. We replace the simple fitting function for $\Lambda_{\rm Cold}$ in \citet{hopkins:fire2.methods} Appendix B with $\Lambda_{\rm neutral} = (\Lambda_{\rm Z} + \Lambda_{\rm H_{2}} + \Lambda_{\rm HD})\,{\rm erg\,s^{-1}\,cm^{3}}$, representing molecular+fine structure cooling from metals, $\Hmol$ and HD. Here $\Lambda_{\rm H_{2}} \equiv f_{\rm mol}\,\Lambda_{\rm H_{2}}^{\rm thin} / (1+\tilde{n})$ where $f_{\rm mol}$ is the H molecular-to-nucleon number ratio (so scales from $0<f_{\rm mol}<1/2$), $\tilde{n} \equiv {\Lambda_{\rm H_{2}}^{\rm thin}}/{\Lambda_{\rm H_{2}}^{\rm thick}}$, $\Lambda_{\rm H_{2}}^{\rm thin} \equiv \sum_{i}\,f_{i}\,\Lambda_{\rm H_{2},\,i}^{\rm thin}$ sums over the cooling rates from interactions between $\Hmol$ and species $i$ = ($e^{-}$, H$^{+}$, ${\rm H}_{2}$, H, He) with number-to-nucleon ratio $f_{i}$. For $\Lambda_{\rm H_{2},\,i}^{\rm thin}$ and $i$ = ($e^{-}$, H$^{+}$, ${\rm H}_{2}$, He) we adopt the fits from \citet{2008MNRAS.388.1627G} assuming a fixed 3:1 ortho-para ratio (Table~8 therein), with $\log_{10}{[\Lambda_{\rm H_{2},\,H}^{\rm thin}]} \equiv -103+97.59\,x+48.05\,x^{2}+10.8\,x^{3}-0.9032\,x^{4}$ from \citet{galla.palla:primordial.cooling} where $x\equiv \log_{10}{T}$. We take $\Lambda_{\rm H_{2}}^{\rm thick} \equiv 10^{-22}\,({\rm cm^{-3}}/n_{\rm H})\,[ 6700\,e^{-5.86/T_{3}} + 16000\,e^{-11.7/T_{3}} + 0.03\,e^{-0.51/T_{3}} + 9.5\,T_{3}^{3.76}\,e^{-0.0022/T_{3}^{3}}\,(1+0.12\,T_{3}^{2.1})^{-1} ]$ for $T_{3}=T/1000$ \citep{hollenback.mckee:co.cooling}. For HD, $\Lambda_{\rm HD} \equiv f_{\rm HD}\,\Lambda_{\rm HD}^{\rm thin} / (1 + \tilde{n}_{\rm HD})$, $\tilde{n}_{\rm HD} \approx (f_{\rm HD}/f_{\rm mol})\,\tilde{n}$, $f_{\rm HD} \approx {\rm MIN}(0.00126\,f_{\rm mol},\, 0.00004\,f_{\rm neutral})$, $f_{\rm neutral}$ the neutral H fraction, , $\Lambda_{\rm HD}^{\rm thin} \equiv 10^{-25}\,[ (1.555 + 0.1272\,T^{0.77})\,e^{-128/T} + (2.406 + 0.1232\,T^{0.92})\,e^{-255/T}]\, e^{-T_{3}^{2}/25}$, following \citet{galla.palla:primordial.cooling}. For metals, $\Lambda_{\rm Z} \equiv z_{\rm C}\,[f_{a}\,\Lambda_{\rm Z,\,{\rm atomic}} + (1-f_{a})\,\Lambda_{Z,\,{\rm mol}}]$, $\Lambda_{Z,\,{\rm atomic}} \equiv f_{a}\,10^{-27}\,[(0.47\,T^{0.15} + 4890\,x_{e}\,T^{-0.5})\,e^{-91.211/T} + 0.0208\,e^{-23.6/T}]$ (from \citealt{2002MNRAS.337.1027W,2005ApJ...620..537B,2016MNRAS.456.2586H}, adopting a depletion factor of $0.5$ for C onto dust), $x_{e}$ is the free electron fraction per H nucleon. Here $f_{a} = 1 / [1 + (n_{\rm H}/G_{0}\,340\,{\rm cm^{-3}})^{2}\,T^{-0.5}]$ \citep{tielens:2005.book} where $G_{0}$ is the local FUV radiation flux (as propagated for photo-electric heating) in Habing units. $\Lambda_{Z,\,{\rm mol}} \equiv 2.73\times10^{-31}\,T^{1.5}/(1+n_{\rm H}/n_{\rm crit})$ (from \citealt{hollenback.mckee:co.cooling})\footnote{Specifically $\Lambda_{Z,\,{\rm mol}}$ comes from \citet{hollenback.mckee:co.cooling} for  for CO cooling, re-calibrated by a factor of $1.4$ to better fit the results from the full chemical network of \citet{glover:2011.molecules.not.needed.for.sf,2014MNRAS.437....9G} including CH, OH, $\Hmol$O, and other species. Note \citet{glover:2011.molecules.not.needed.for.sf} show the resulting cooling rate for the same C abundance is nearly identical in the range of interest regardless of the form in which C and O and ``locked,'' so we do not distinguish between these particular C and O molecular species.} with $n_{\rm crit} = 1.9 \times 10^{4}\,{\rm cm^{-3}}\,T^{0.5}/(1 + \Sigma_{\rm gas}/\Sigma_{\rm crit})$, $\Sigma_{\rm crit} = 3 \times10^{-5}\,{\rm g\,cm^{-2}}\,T/z_{\rm C}$, where $z_{\rm C} \equiv 10^{\rm [C/H]}$ is the carbon abundance and $\Sigma_{\rm gas}$ is the column density integrated to infinity with the same Sobolev approximation used for shielding \citep{hopkins:fire2.methods}. We have verified that these expressions give similar results for similar conditions and processes to e.g. more complicated networks like those in \citet{glover:2011.molecules.not.needed.for.sf,richings:2014.cooling.tables,gong:molecular.chem.cooling}. 

\item{\bf Molecular Fractions:} Although our neutral gas cooling functions do not include all possible processes, the dominant uncertainty is not the neglect of particular chemical channels, but the molecular fraction $f_{\rm mol}$. In \firetwo\ we estimated $f_{\rm mol}$ using the particularly simple expression from \citet{krumholz:2011.molecular.prescription}, which is a function only of metallicity $Z$ and $\Sigma_{\rm gas}$ (making assumptions about how typical radiation fields and gas density structure behaves with these quantities), and does not correctly extrapolate to low $f_{\rm mol} \ll 1$ (it gives unphysical values at low density or low $Z$), which can be important for primordial gas cooling. 

In \firex, we update this to explicitly follow and allow for non-equilibrium $\Hmol$ chemistry, with a simplified H-only chemical network depending on the local gas, dust, cosmic ray, and radiation fields. After operator-splitting advection and hydrodynamic terms, we approximate the rate equation for the evolution of the molecular hydrogen $\Hmol$ number density by:
\begin{align}
\nonumber \frac{{\rm d}\bar{n}_{\Hmol}}{{\rm d}t} =\  & \alpha_{Z}\,f_{\rm dg}\,C_{2,\,{\rm dg}}\,\bar{n}_{\rm n}\,\bar{n}_{\HI} + \alpha_{\rm GP}\,C_{2}\,\bar{n}_{\HI}\,\bar{n}_{{\rm H}^{-}}   + \alpha_{\rm 3B}\,C_{3}\,\bar{n}_{\HI}^{2}\,\left(\bar{n}_{\HI} + \frac{\bar{n}_{\Hmol}}{8} \right) \\
\label{eqn:molfrac} & - \Gamma_{\Hmol}\,\bar{n}_{\Hmol} - \xi_{\Hmol}\,\bar{n}_{\Hmol} - \sum_{i}\,\beta_{\Hmol,\,i}\,C_{2}\,\bar{n}_{\Hmol}\,\bar{n}_{i} 
\end{align}
Here, we must account for finite-resolution effects: $\bar{n}_{x} \equiv \langle n_{x} \rangle_{i} = N_{i,\,x} / V_{i}$ is the volume-averaged number density of $x$ within a simulation cell $i$ (what is actually evolved), and $C_{m} \equiv \langle n^{m} \rangle / \langle n \rangle^{m}$ is the $m$-th order clumping factor. 

In the above, several of the rate factors ($\alpha$, $\beta$) are compiled from a variety of sources \citep{hollenback.mckee:co.cooling,1986ApJ...311L..93D,1998ApJ...499..793M,glover:2007.low.metallicity.cooling.h2.tables,2008MNRAS.388.1627G,2010ApJ...709..308M,2012MNRAS.425.3058C,2013ApJ...773L..25F,2014ApJ...790...10S,indriolo:2015.cr.ionization.rate.vs.galactic.radius,2017MolAs...9....1W,nickerson:molec.frac.ramses.rhd}, and we collect and update them here. Above, 
$\alpha_{Z} \equiv 3\times10^{-18}\,T^{0.5}\,\{[1+0.04\,(T+T_{\rm dust})^{0.5} + 0.002\,T + 8\times10^{-6}\,T^{2}]\,[1 + 10^{4}\,\exp{(-600/T_{\rm dust})}]\}^{-1}$ 
is the dust reaction rate coefficient \citep{glover:2007.low.metallicity.cooling.h2.tables} with $f_{\rm dg} \equiv \rho_{\rm dust} / 0.01\,\rho_{\rm gas}$ the solar-scaled dust-to-gas ratio which we take to be $=Z/Z_{\odot}$, $T_{\rm dust}$ is the dust temperature (defined below), $C_{2,\,{\rm dg}}$ is the micro-physical dust-gas clumping factor (which we take $=C_{2}$ for simplicity, although micro-physical simulations of dust-gas interactions suggest it could deviate by a large factor in either direction; see \citealt{moseley:2018.acoustic.rdi.sims,hopkins:2019.mhd.rdi.periodic.box.sims}), $n_{\rm n}$ is the number density of nucleons, and $n_{\HI}$ is the number density of atomic $\HI$. 
Next the $\alpha_{\rm GP}$ term represents the gas-phase formation rate which depends on the abundance of H$^{-}$, for which we follow \citet{glover:2007.low.metallicity.cooling.h2.tables} and assume local equilibrium with $\bar{n}_{{\rm H}^{-}} \approx k_{1}\,\bar{n}_{\HI}\,\bar{n}_{e} / [(k_{2}+k_{16})\,\bar{n}_{\HI} + (k_{5}+k_{17})\,\bar{n}_{{\rm H}^{+}} + k_{15}\,\bar{n}_{e} + R_{51}]$ with $\alpha_{\rm GP}=k_{2}$, where $\bar{n}_{e}$ is the number density of free thermal electrons and $n_{{\rm H}^{+}}$ is the ionized H fraction, and the coefficients $k_{n}(T)$ and $R_{51}$ (which represents photo-dissociation scaling with $\Gamma$ below) are taken from   \citet{glover:2007.low.metallicity.cooling.h2.tables}. The term $\alpha_{\rm 3B} = 6\times 10^{-32}\,T^{-0.25} +  2 \times10^{-31}\,T^{-0.5}$ represents three-body formation. 
The $\sum_{i}$ term sums over collisional dissociations between $\Hmol$ and species $i=$\,($e^{-}$, H$^{+}$, $\Hmol$, HI, HeI, HeII), with coefficients $\beta_{\Hmol,\,i}$ from \citet{2008MNRAS.388.1627G} (Eqs.~7-11, 24-25, 37-39 in Table~A1\footnote{Note \citet{2008MNRAS.388.1627G} Eq.~A1-7 should read $\ln$, not $\log$, per \citet{savin:2004.dissociation.fits}}). 
Meanwhile $\xi_{\Hmol}$ represents ionization \&\ dissociation by CRs: this is calculated directly from the resolved CR spectra following \citet{1972Phy....60..145G,Mann94} in our simulations which explicitly include cosmic rays, otherwise we assume a uniform $\xi_{\Hmol}=7.5\times10^{-16}\,{\rm s^{-1}}$ as inferred from observations of Milky Way molecular clouds \citep{indriolo:2012.cr.ionization.rate.vs.cloud.column}. $\Gamma_{\Hmol}$ represents dissociation+ionization by Lyman-Werner (LW) and ionizing photons (expressions from \citealt{glover:2007.low.metallicity.cooling.h2.tables}), for which we use the explicitly-evolved local LW and ionizing photon intensities in the simulations. 

We model the un-resolved clumping factors as in \citet{lupi:2018.h2.sfr.rhd.gizmo.methods}\footnote{See Appendix~\ref{sec:alternatives:mol} for an example of the effects of the $C_{m}$ terms on the gas phase diagram, and \citet{lupi:2018.h2.sfr.rhd.gizmo.methods} for explicit demonstration of the importance of accounting for these terms in simultaneously reproducing molecular and atomic scalings with star formation rate in simulations at these resolution scales.} by making the standard assumption that turbulence generates a log-normal local density PDF giving $C_{m} = \exp{[m\,(m-1)\,S/2]}$ where $S = \ln{[1 + (b\,\mathcal{M})^{2}]}$  \citep[e.g.][]{vazquez-semadeni:1994.turb.density.pdf,scalo:1998.turb.density.pdf,federrath:2010.obs.vs.sim.turb.compare,hopkins:2012.intermittent.turb.density.pdfs} where $S$ is the log-variance, $b$ a geometric constant reflecting the compressive to solenoidal ratio (here $b=1/2$ as expected for a ``natural mix'' of modes; \citealt{schmidt:2008.turb.structure.fns,federrath:2008.density.pdf.vs.forcingtype}), and $\mathcal{M}$ is the sonic Mach number of a cell $i$ estimated in the same manner as our usual local turbulent velocity estimation ($\mathcal{M})_{i} = \| \nabla \otimes {\bf v} \|_{i}\,\Delta x_{i}/ c_{s,\,i}$). For $\Gamma$, we separate the ionizing radiation (handled identically to HI ionizing radiation for convenience here; see \citealt{hopkins:fire2.methods} for details) and LW radiation. For the LW band we must calculate the mean incident radiation field on each molecule accounting both for shielding by dust and local self-shielding by $\Hmol$. The former is self-consistently accounted for in our radiation-transport approximations in-code, giving the dust-attenuated incident intensity $\tilde{I}_{\rm LW}$. For the latter we attempt to account for turbulence as well as thermal broadening and their effects on line overlap following \citet{2014ApJ...795...37G}: we take $\Gamma_{\rm LW} = 3.3\times10^{-11}\,S_{\Hmol}\,G_{0}$ with $G_{0}$ given by the dust-attenuated $\tilde{I}_{\rm LW}$ in Habing units and $S_{\Hmol} = (1-\omega_{2})\,(1+\tilde{\ell}^{1/2}\,x/b_{5})^{-2} + \omega_{2}\,(1+x)^{-0.5}\,\exp{[-(1+x)^{0.5}/1180]}$ is the $\Hmol$ self-shielding factor 
with $\omega_{2}=0.035$, $x \equiv N_{\Hmol} / 5\times10^{14}\,{\rm cm^{-2}}$, $N_{\Hmol} \equiv \bar{n}_{\Hmol}\,L$ the $\Hmol$ column, $L$ is given by our Sobolev-type density gradient length calculation $=\Delta x_{i} + \rho_{i} / |\nabla \rho|_{i}$, $b_{5} \equiv \sqrt{2}\,\sigma_{\rm v,\,therm}$ (where $\sigma_{\rm v,\,therm}$ is the thermal Maxwellian velocity dispersion of molecular $\Hmol$ at the cell temperature $T$), and $\tilde{\ell} \equiv {\rm MIN}[\ell_{s} / L,\,1] \approx [1 + \mathcal{M}^{2}]^{-1}$ accounts for the difference between the (unresolved) sonic length $\ell_{s}$ ($\sim 0.1\,$pc, typically) and $L$ to model the effects of a supersonic (\citealt{burgers1973turbulence}-type) cascade on the velocity coherence of line overlap.  

We update Eq.~\ref{eqn:molfrac} fully-implicitly every timestep alongside our cooling and chemistry calculation: we can assume local equilibrium if desired (setting ${\rm d}\bar{n}_{\Hmol}/{\rm d}t=0$ and solving for $\bar{n}_{\Hmol}/\bar{n}_{\HI}$), but in our default simulations we allow for non-equilibrium chemistry by evolving $n_{\Hmol}$ directly. Note that our Riemann solver allows for arbitrary equations-of-state, and we account explicitly for a variable gas adiabatic index depending on the molecular fraction \citep{grudic:starforge.methods}.

\item{\bf Free Electron Fractions:} In \firetwo, we evolved the free electron fraction $x_{e}$ explicitly in detail, including collisional and photo-ionization for H and He, but neglecting free electrons from heavier species. This can significantly under-estimate $x_{e}$ in dense, cold gas, which is generally not important for the cooling rates (except for $\lesssim 100\,$K gas at $n\sim 10^{2}-10^{4}\,{\rm cm^{-3}}$), but can be important for accurately capturing effects like ambipolar diffusion and Hall MHD. In \firex, we calculate the contribution from these electrons and CR ionization following \citet{wardle.2007:mhd.protoplanetary.disk.review,keith:2014.planetary.disk.ionization.model} and add them to the free electron budget. This involves solving for the equilibria for the heavy ion number density $n_{i}$, free electron density $n_{e}$, and mean charge per dust grain $Z_{g}$: $\dot{n}_{i}=\zeta\,n_{n}-k_{ei}\,n_{i}\,n_{e}-k_{ig}\,n_{g}\,n_{i}$, $\dot{n}_{e} = \zeta\,n_{n}-k_{ei}\,n_{i}\,n_{e} - k_{eg}\,n_{g}\,n_{e}$, $\dot{Z}_{g} = k_{ig}\,n_{i}-k_{eg}\,n_{e}$, subject to charge neutrality $n_{i}-n_{e}+Z_{g}\,n_{g}=0$. Here $k_{xg} = \pi\,\langle a_{g}\rangle^{2}\,\sqrt{8\,k_{b}\,T/(\pi\,m_{x})}$, $m_{i}$ the mean molecular weight of the relevant ions (dominated by Mg under conditions where these terms are significant, with total Mg abundance scaled from our explicitly-evolved abundances), the grain number $n_{g} \approx 0.01\,(Z/Z_{\odot})\,\rho / (4\pi/3\,\bar{\rho}_{g}\,\langle a_{g}\rangle^{3})$ scales with the dust-to-gas ratio (hence $Z$), $\bar{\rho}_{g}=2.4\,{\rm g\,cm^{-3}}$ is the material grain density and $\langle a_{g} \rangle$ is an effective weighted grain size taken to be $=0.1\,\mu{\rm m}$ \citep{draine:2007.pah.model.update}, $\zeta$ is the CR ionization rate defined below, $k_{ei}\approx 9.77\times10^{-8}\,{\rm cm^{3}\,s^{-1}}$ in the regimes of relevance \citep[e.g.][]{1989ApJ...345..782M}, and $n_{n}$ is the neutral number density. Comparing to the results of more detailed chemical network calculations from e.g.\ \citet{glover:2011.molecules.not.needed.for.sf,2019ApJ...872..107X} over their appropriate density ranges, our simple model agrees reasonably well in $x_{e}$ at densities from $n \ll 10^{-8}\,{\rm cm^{-3}}$ to $n\sim 10^{18}\,{\rm cm^{-3}}$.

\item{\bf Ionized Atomic Gas Cooling/Heating:} In \firetwo, we followed ionized metal-line cooling at high temperatures (in e.g.\ mostly-ionized, un-shielded gas) using the \citet{wiersma:2009.coolingtables} cooling tables from {\small CLOUDY} at $T>10^{4}\,$K (with a strict cutoff below that temperature). In \firex\ we extend these following \citet{richings:2014.cooling.tables} to include ionized gas metal-line cooling and/or heating (with rates scaling with the ionized gas fraction and free electron abundance) down to arbitrarily low temperatures. Like in \firetwo\ these are tabulated and calculated species-by-species for the separately-tracked species C, N, O, Ne, Mg, Si, S, Ca, Fe, Note that this generally has appreciable effects only on the lowest-density cool gas, because almost all the cold gas at $T\ll 10^{4}\,$K is self-shielded. The tables are included in the public {\small GIZMO} code.

\item{\bf Dust Temperatures:} Dust-gas collisions and cooling/heating are included as in \firetwo, with $\Lambda_{\rm dust} = 1.12\times10^{-32}\,{\rm erg\,s^{-1}\,cm^{3}}\,\left(T-T_{\rm dust}\right)\,T^{1/2}\,(1-0.8\,\exp{(-75/T)})\,(Z/Z_{\odot})$ (\citealt{meijerink.spaans:xray.cooling.models}; assuming a dust-to-gas ratio scaling linearly with metallicity). In \firetwo, we assumed a constant dust temperature $=30\,$K. In \firex, we calculate the dust temperature self-consistently assuming local equilibrium between absorbed and emitted radiation, including the sum of incident radiation from all tracked bands in our RHD approximation (treating the opacity in each as band-centered), plus the CMB, assuming a $\beta=1$ opacity law,\footnote{Specifically, we assume the grain absorption efficiency $Q$ scales $\propto \lambda^{-1}$, giving an opacity law $\kappa \propto T_{\rm eff}^{\beta}$ with $\beta \approx 1$ for an effective temperature $T_{\rm eff}$ of each band, which is a reasonable approximation for many grain compositions assuming peak absorption/emission wavelengths from $\sim 0.1-100\,\mu{\rm m}$ \citep{1984ApJ...285...89D,2011A&A...532A..43O} and empirically fits observations of molecular clouds in the density and temperature range of interest \citep{schnee:2014.mm.sized.grains.in.star.forming.regions,2015A&A...578A.131L,2017ApJ...849...13W}. In simulations where we instead evolve the radiation using explicit methods (e.g.\ M1, see \citealt{hopkins:radiation.methods}) we calculate dust temperatures using the detailed evolved opacities at each wavelength for the resolved spectrum as in \citet{grudic:starforge.methods}.} and assuming the incident IR radiation field has the same radiation temperature as the local dust temperature. This gives $T_{\rm dust}$ as the solution to the quintic equation: $(T_{\rm dust}/2.92\,{\rm K})^{5} \approx \sum_{i}\,(T_{{\rm rad},\,i}/{\rm K})\,(e_{{\rm rad},\,i}/{\rm eV\,cm^{-3}})$, where $T_{{\rm rad},\,i}$ is the effective radiation temperature for each band (roughly the temperature corresponding via Wien's law to the absorption-weighted mean band wavelength, for narrow bins) and $e_{{\rm rad},\,i}$ is the local radiation energy density in the band from our radiation transport, including the 5 bands we explicitly evolve (infrared, near-IR and optical, near-UV, photo-electric, and photo-ionizing), plus the CMB (at the cosmologically-evolved temperature $T_{\rm cmb}\approx2.73\,(1+z)$\,K and energy density $e_{\rm cmb} \approx 0.262\,(1+z)^{4}\,{\rm eV\,cm^{-3}}$, i.e.\ ignoring CMB attenuation), where $T_{\rm rad}$ in the IR is equal to $T_{\rm dust}$.\footnote{For reference, if one wished to adopt a constant Galactic background to estimate dust temperatures, the \citet{draine:ism.book} interstellar radiation field can be reasonably approximated for these purposes as a two-component model with $e_{{\rm rad},\,1}\approx 0.31\,{\rm eV\,cm^{-3}}$, $T_{{\rm rad},\,1} \approx 30\,$K, and $e_{{\rm rad},\,2}\approx 0.66\,{\rm eV\,cm^{-3}}$, $T_{{\rm rad},\,2}\approx 5800\,$K.} This generally has small dynamical effects in FIRE simulations, as dust-gas coupling only dominates cooling at $n \gg 10^{6}\,{\rm cm^{-3}}$. 

\item{\bf Cosmic Ray (CR) Heating:} In simulations with explicitly-evolved CR dynamics we calculate the CR heating (including the thermalized components of e.g.\ Coulomb, pionic/spallation, and ionization interactions) from the evolved CR spectra following \citet{1972Phy....60..145G,Mann94}. In our simulations without explicit CR dynamics, we calculate this from an assumed CR background spectrum updated from the older \citet{goldsmith:molecular.dust.cooling.gmcs} estimate used in \firetwo\ to the newer estimates based on the interstellar Voyager I \&\ II data, following \citet{cummings:2016.voyager.1.cr.spectra}. The net effect is that the CR thermal heating rate in neutral gas is systematically lower by a factor $\approx 0.34$ using the newer data. This has small effects, because CR heating is almost always sub-dominant at resolved scales.

\item{\bf Photo-Electric Heating:} We follow photo-electric heating as in \firetwo, tracing the UV radiation from stars to each resolution element. In \firex, we add the (appropriately self-shielded) contribution from the UVB in the relevant wavelength range (integrating over the assumed UVB spectrum). This is generally small compared to the contribution from stars within a galaxy, but can have a minor effect in rare situations with metal-enriched (dusty) warm neutral medium gas at very large galactic-centric radii (e.g. cooling of galactic outflows at galacto-centric radii $\gg 30\,$kpc). We also update the opacities as described below.

\item{\bf Temperature Floor \&\ CMB Effects:} In \firetwo\ we enforced a minimum temperature of $10\,$K, owing to cutoffs in our cooling tables. In \firex\ we allow cooling, in principle, to arbitrarily small temperatures, though a negligible amount of gas reaches $\ll 10\,$K within galaxies. Note that gas can in principle cool to temperatures below the CMB via e.g.\ adiabatic processes, but it is coupled to the CMB both directly through Compton scattering of CMB photons which we include as in \firetwo\ (see \citealt{hopkins:fire2.methods} Eq.~B20 or \citealt{rybicki.lightman:1986.radiative.processes.book}), and through dust heating/cooling where the dust temperatures are sensitive to the CMB as described above. To account for the effects of the CMB on radiative cooling as $T\rightarrow T_{\rm cmb}$ (and heating/excitation by the CMB when $T<T_{\rm cmb}$), motivated by \citet{bromm:2002.cooling.first.stars} we take $\Lambda_{\rm rad} \rightarrow \Lambda_{\rm rad}\,(T-T_{\rm cmb})/(T+T_{\rm cmb})$ for the purely-radiative molecular and atomic cooling rates $\Lambda_{\rm rad}$.

\item{\bf UVB Self-Shielding:} Because our simulations are ``zoom-ins'' we cannot self-consistently follow the UVB and so assume a uniform meta-galactic background as noted above, with a self-shielding correction. \firetwo\ used a simple local-Jeans-length based criterion for self-shielding from the UV background \citep[see][]{hopkins:fire2.methods}. In \firex, we update this to the more accurate expressions (calibrated to radiation transport experiments) from \citet{rahmati:2013.shielding.prescription.calibration} with some updates. The UVB is reduced by a factor $f_{\rm shield} = 0.98\,(1+y^{1.64})^{-2.28} + 0.02\,[1+y\,(1+[n_{\rm H}/10]^{4})]^{-0.84}$ with $y \equiv 400\, (n_{\rm H}/{\rm cm^{-3}})\,T^{-0.173}\,(\Gamma_{\rm HI}/10^{-12})^{-0.66}$ where $\Gamma_{\rm HI}$ is the usual optically-thin photo-ionization rate including both the UVB and local sources, and the $n_{\rm H}^{4}$ term is added based on our own RHD simulations of dense GMCs \citep{hopkins:radiation.methods,hopkins:2019.grudic.photon.momentum.rad.pressure.coupling,grudic:starforge.methods} to ensure sufficient shielding in dense gas $n_{\rm H} \gtrsim 100\,{\rm cm^{-3}}$ (several dex larger densities than the original \citealt{rahmati:2013.shielding.prescription.calibration} calibration).

\end{itemize}

Fig.~\ref{fig:phase} shows the effects of these changes on the phase (density-temperature) diagram of a Milky Way-like galaxy at $z=0$. Although extensive, these changes do not overall produce a large change in the dense/cold gas dynamics (as thermal pressure is essentially always sub-dominant to gravity, turbulence, magnetic and other pressures in cold ISM gas, see \citealt{hopkins:fire2.methods,gurvich:2020.fire.vertical.support.balance}), nor in the total mass of cold/dense gas, but they do lead to the cold/dense gas having a more clearly-defined phase structure (as more detailed features in the cooling functions and transitions from atomic-to-molecular can be followed). This produces, for example, a much more obvious warm-vs-cold neutral medium delineation. This could have important implications for accurate predictions of observational diagnostics of cold and/or neutral ISM gas.

\subsection{Treatment of HII Regions}
\label{sec:hii}

\firetwo\ and \firex\ implement photo-ionization heating in compact HII regions similarly. In simulations where each star particle represents an IMF-averaged population, and star and gas elements have approximately equal mass, then any compact HII regions (sourced by a single star particle) with density $\gtrsim 100\,{\rm cm^{-3}}$ are necessarily un-resolved. As a result we adopt a probabilistic approach to treat HII regions (stochastically determining if a particle is fully-ionized if the HII region mass is less then one gas cell)  to statistically capture the proper expansion dynamics of ionized bubbles. However \firex\ makes two improvements to the \firetwo\ model. 

\begin{itemize}

\item{In \firetwo, compact HII regions identified as above were simply forced to be fully-ionized and to maintain a constant minimum temperature $T=10^{4}$\,K for the duration of the timestep. In \firex, upon initial ionization, the cell has its temperature set to the minimum of either the current temperature plus the heat added from the energy of initial ionization, or the equilibrium HII region temperature from standard collisional cooling $T_{\rm eqm}/10^{4}\,{\rm K} \approx  {\rm MIN}\{6.62,\ 0.86/(1 + 0.22\,\ln{[Z/Z_{\odot}])}\}$ \citep{draine:ism.book}. It is then immediately passed to the cooling \&\ chemistry routine, where the incident ionizing radiation and FUV flux is set to the ``Stromgren value'' (value needed to maintain full ionization) and self-shielding is set to zero, where the exact ionization states and temperatures can be self-consistently solved.}

\item{In \firetwo, the gas element first selected for this stochastic ionization was selected uniformly among the nearest gas elements to the star. In \citet{hopkins:radiation.methods}, comparing this approximate method to explicit radiation-hydrodynamics calculations with the M1 solver and other tests, we showed that this tended to slightly {\em under}-estimate the effects of radiative feedback, as (given our Lagrangian numerical method with approximately equal-mass gas elements) this is biased towards ``expending'' the photons on denser, less volume-filling structures near the star (which require more photons to ionize), while more detailed calculations show ionization primarily in the lower-density cavities/bubbles/channels. In \firex, we therefore select the lowest-density neighbor of the overlapping neighbor gas cells for this procedure.}

\end{itemize}

\subsection{Treatment of Other Radiative Feedback Channels}
\label{sec:radiation}

Other radiation feedback terms (single and multiple-scattering radiation pressure, photo-electric and photo-ionization by stars outside of compact HII regions) are treated as in \firetwo. We have made a number of algorithmic improvements in \firex, to allow for more accurate integration of the short range radiation-pressure terms, allow for more accurate estimation of the absorption in the kernel around the star (independently evaluating each direction, following \citealt{hopkins:2019.grudic.photon.momentum.rad.pressure.coupling}), and to more accurately follow the metallicity dependence of opacities in each step of the long-range photon transport.\footnote{In \firetwo, the metallicity used for the short-range kernel opacity was the stellar metallicity (assuming the kernel was uniform); in \firex\ each cell uses its own metallicity and opacity.} \firex\ also adds standard Thompson and Kramers opacities from H and He for the evolved bands (FUV/NUV/optical-NIR/IR) which were in \firetwo\ assumed to have pure dust opacity (proportional to $Z$) -- these essentially set minimum opacities below $Z \lesssim 10^{-3}\,Z_{\odot}$. We have also re-calculated the effective weighted dust opacities for all bands using the new stellar spectra/SEDs from our updated stellar evolution models with the \citet{draine:2007.pah.model.update} dust models (so the dust extinction opacities at e.g.\ Lyman-Werner, photoelectric,  NUV, optical-NIR, mid-far IR are $(900,\,720,\,480,\,180,\,6.5)\,{\rm cm^{2}\,g^{-1}}$, multiplied by $Z/Z_{\odot}$). The primary effect here results from \firex\ adopting a MW-like dust model as opposed to the \firetwo\ SMC-like dust model, which lowers the UV dust opacities by a factor of $\sim 2-3$. These have small effects, generally tending to make single-scattering radiation pressure slightly weaker in low-metallicity young stellar environments, and slightly stronger in old high-metallicity environments \citep{hopkins:radiation.methods}.

\subsection{Numerical Coupling of Supernovae}
\label{sec:sne}

\firetwo\ and \firex\ both implement SNe numerically as described in \citet{hopkins:sne.methods}, expanding on the method first proposed in \citet{hopkins:2013.fire} to accurately account for SNe mass, metals, energy, and momentum fluxes, accounting for cooling and unresolved expansion, and manifestly conserving mass and momentum and ensuring numerical isotropy (avoiding imprinting preferred directions) even in arbitrary mesh configurations. In \firex, we use by default the extended method described in Appendix~E of \citet{hopkins:sne.methods}, which includes an extra neighbor loop in assigning fluxes from a SNe to neighbor cells to manifestly conserve total coupled energy (and mass and momentum) in the limit where the relative motion of the star and inflow/outflow motion of mesh cells around the star is large compared to the ejecta velocity (so non-linear corrections to the total kinetic energy of the gas from the coupled momentum of a single SNe must be included). We make some minor improvements to this including an additional predictor-corrector loop to maintain conservation even when multiple star particles attempt to update the same gas cell simultaneously, as detailed in Appendix~\ref{sec:appendix:sne}; the implementation of this algorithm is provided in the public version of \GIZMO. \firetwo\ by default used the simpler default method in the main text of \citealt{hopkins:sne.methods}, which is manifestly energy-conserving in the limit of uniform or negligible or divergence-free motion of gas towards/away from the star particle. As shown in \citet{hopkins:sne.methods}, the two methods converge to identical behavior at sufficiently high resolution, but the new method is more accurate at intermediate resolution (albeit with added computational expense).

As discussed and tested in detail in \citet{hopkins:sne.methods}, to accurately deal with unresolved ``PdV'' work during expansion of the SNe to the numerically-resolved coupling radius, we require an analytic expression for expected SN cooling radius or terminal momentum $p_{t}$ as a function of SN energy, ambient gas density and metallicity. In \fireone\ \&\ \firetwo\ we adopted that from \citet{cioffi:1988.sne.remnant.evolution}, $p_{t} \propto (E_{\rm SN}/10^{51}\,{\rm erg})^{13/14}\,(n/1\,{\rm cm^{-3}})^{-0.14}\,\tilde{Z}^{-0.21}$ (with $\tilde{Z} = {\rm max}[0.01,\,Z/Z_{\odot}]$), testing the normalization directly for $Z=Z_{\odot}$, $n=1\,{\rm cm^{-3}}$, $E_{\rm SN}=10^{51}\,{\rm erg}$. In \firex\ we update this to better match the calibration from more recent studies of resolved simulations of SNe in inhomogeneous media \citep{martizzi:sne.momentum.sims,walch.naab:sne.momentum,haid:snr.in.clumpy.ism,gentry:sne.momentum.boost}, giving $p_{t} \propto (E_{\rm SN}/10^{51}\,{\rm erg})\,(n/1\,{\rm cm^{-3}})^{-0.14}\,\tilde{Z}^{-0.12}$ (normalized to the same value for $E_{\rm SN}=10^{51}\,{\rm erg}$, $n=1\,{\rm cm^{-3}}$, $\tilde{Z}=1$). This slightly reduces the momentum of the lowest-metallicity SNe, but the effect is weak.

In \firex, we also update the ionization state of shocked gas in SNe (and stellar mass-loss) immediately, rather than waiting for the cooling step later in the timestep; this produces negligible differences.

\subsection{Numerical Coupling of Stellar Mass-Loss}
\label{sec:mass.loss}

As with SNe, we update the numerical treatment of stellar mass-loss (continuous OB/AGB-winds) using the additional  \citet{hopkins:sne.methods} Appendix~E correction terms for kinetic energy, and update the ionization states immediately in shocks; this generally has a small effect for mass-loss.

In \firetwo\ we applied the same scaling for ``PdV work'' done in an un-resolved momentum-conserving stage for SNe to stellar mass-loss: this amounts to converting some coupled thermal energy into kinetic, representing the work done by an energy-conserving shell below the resolution limit and cooling radius, when that cooling radius is unresolved. This importance of this term for SNe is reviewed in \citet{hopkins:sne.methods}. However, the scalings used \citep[see also][]{martizzi:sne.momentum.sims,rosdahl:2016.sne.method.isolated.gal.sims,haid:snr.in.clumpy.ism} assume {\em discrete} energy injection events, so quantities like the cooling radius depend on $E_{\rm injected} \sim 10^{51}\,{\rm erg}$. As such these scalings become ill-defined in the true continuum limit for continuous stellar mass-loss, if applied every timestep (where $\Delta E_{\rm injected} \sim \dot{E}\Delta t$ is distributed continuously over a resolved extended period of time). In that continuous limit the correct solution can be quite different \citep[see e.g.][]{weaver:1977.wind.bubble.expansion,mckee:bubble.expansion,murray:momentum.winds,cafg:2012.egy.cons.bal.winds}. And it has long been recognized empirically that observations of massive wind bubbles do not appear to be consistent with a large energy-conserving phase which could do significant ``PdV'' work \citep{harper.clark:stellar.bubbles.energy.missing,lopez:2010.stellar.fb.30.dor,2011ApJS..194...16T,rosen:2014.xray.energy.wind.clusters,olivier:2020.hii.region.feedback.mechanism.breakdown}, and recent high-resolution simulations of individual winds appear to support this \citep[see][and references therein]{lancaster:2021.fractal.wind.mixing,lancaster:2021.fractal.wind.bubble.sims}. Since our simulations reach resolution sufficient to resolve the stellar mass-loss timescales and outflow expansion times, in \firex\ we simply couple the injected energy and momentum for continuous mass-loss directly without assuming a sub-grid conversion/``PdV work'' model. This can have a significant effect in some of our galaxies (not all), reducing stellar masses by as much as a factor $\sim2$, in a highly non-linear manner: by effectively making early stellar mass-loss feedback weaker (because less momentum is coupled), self-gravitating gas clumps collapse slightly further and form more stars before disrupting, leading to more bursty star formation and more strongly clustered SNe explosions, which produces {\em more} effective galactic outflows \citep{fielding:sne.vs.galaxy.winds}.

\begin{figure}
	\includegraphics[width=0.98\columnwidth]{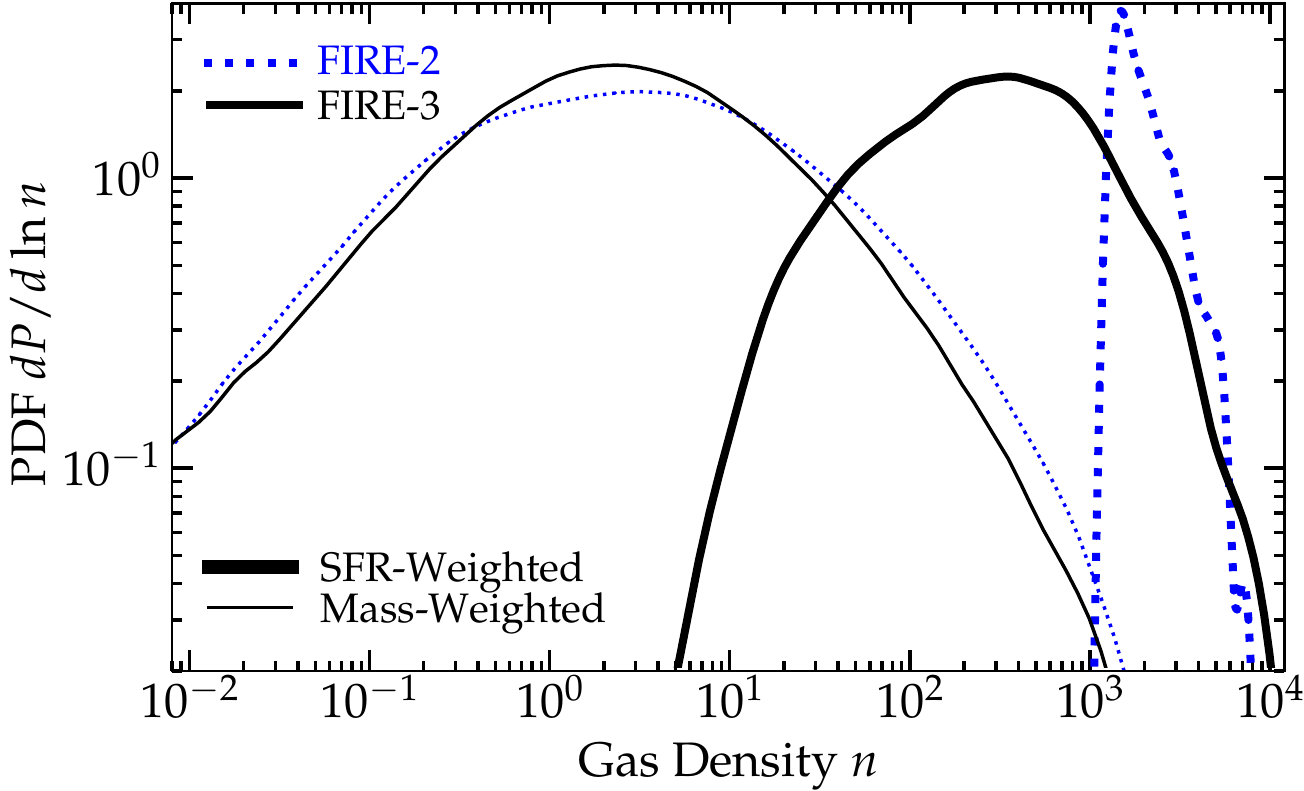}
	\vspace{-0.1cm}
	\caption{Probability distribution function (PDF) of ISM gas densities $n$, for the same controlled restart of a Milky Way-like galaxy ({\bf m12i}) as Fig.~\ref{fig:phase}, for \firetwo\ ({\em blue dotted}) and \firex\ ({\em black solid}), again for gas inside $r<10\,$kpc. We compare the PDF weighted by gas mass ({\em thin}) and by star formation rate (SFR; {\em thick}). \firetwo\ requires gas to exceed an explicit density threshold $n>n_{\rm crit}=1000\,{\rm cm^{-3}}$ to form stars, leading to the obvious ``spike'' at $\sim n_{\rm crit}$ in the SFR-weighted PDF, and forcing some lower-density gas to ``pile up'' at these densities before it can form stars (producing the slightly-elevated high-$n$ tail in the mass-weighted PDF). This sharp numerical feature could artificially impose features on e.g.\ properties of GMCs or star clusters formed. In \firex, per \S~\ref{sec:sf.criteria}, we remove the explicit density threshold, allowing gas at any densities to fragment and form stars if it is a converging flow, locally self-gravitating/bound at the resolution scale, and Jeans unstable with a thermal Jeans mass in the stellar mass range. Although this still effectively restricts SF to dense, collapsing, cold ($T\lesssim 100\,$K) gas, that gas can come from a wider range of $n$ (with a median to $n\sim 200-300\,{\rm cm^{-3}}$, at this resolution), without numerically ``forcing'' that gas to get denser before it is assumed to fragment to stellar-mass scales.
	\label{fig:sfr}}
\end{figure}

\subsection{Star Formation Criteria} 
\label{sec:sf.criteria}

In \firex, we allow star formation in gas which meets all of the following three criteria:

\begin{enumerate}

\item{\bf Self-Gravitating:} Following \citet{hopkins:virial.sf}, we calculate the virial parameter $\alpha_{i}(t_{i}) \equiv (\| \nabla \otimes {\bf v} \|_{i}^{2} + 2\,[k_{i}\,v_{f,\,i}]^{2} ) / (8\pi\,G\,\rho_{i})$ at time $t_{i}$, where $k_{i}\equiv 1/\Delta x_{i}$, is the cell wavenumber, $\Delta x_{i} \equiv (\rho_{i}/m_{i})^{1/3}$ the cell size in terms of the gas cell density and mass, $v_{f,\,i}^{2}\equiv c_{s,\,i}^{2} + v_{A,\,i}^{2}$ is the fast wavespeed in terms of the thermal sound speed $c_{s,\,i}$ and \Alf\ speed $v_{A,\,i}$, $\nabla\otimes{\bf v}$ the velocity gradient tensor and $\| \nabla \otimes {\bf v} \|^{2}  \equiv \sum_{ij} |\nabla_{i} \otimes {\bf v}_{j}|^{2}$ the Frobenius norm. The normalization is chosen so that $\alpha = {\rm KE}/{\rm PE}$, the ratio of kinetic to potential energy, exactly for a uniform density/temperature sphere with a uniform velocity gradient tensor (and this reduces to the usual Jeans criterion when kinetic energy is negligible). Note this includes kinetic, thermal, and magnetic energies explicitly: we do not include radiation or cosmic ray energies as the diffusion/streaming escape time for those is vastly faster than the free-fall time. We then require ${\alpha}_{i}<1$.

\item{\bf Jeans Unstable:} We require the thermal Jeans mass $m_{{\rm J},\,i} \approx M_{\odot}\,(c_{s,\,i}/0.2\,{\rm km\,s^{-1}})^{3}\,(\rho_{i}/3000\,m_{p})^{-1/2}$ be less than the maximum of $100\,M_{\odot}$ or the cell mass $m_{i}$. Physically, this requires that thermal fragmentation (which sets the smallest scale to which structures should fragment) should occur and can go down to {\em un-resolved scales} (so resolved collapsing structures are still captured), or to the mass scale of actual stars.

\item{\bf Converging Flow:} Since star formation occurs via un-resolved fragmentation in a turbulent medium, there is no reason it should strictly require that the flow on larger (resolved) scales be globally converging \citep{hopkins:frag.theory,guszejnov:gmc.to.protostar.semi.analytic,vazquez.semadeni:2019.global.gmc.toy.model}, and indeed in simulations which resolve ISM or GMC turbulence this tends to be an essentially ``stochastic'' random variable on cloud scales \citep{hopkins:virial.sf,murray:2017.turb.collapse}. However, to help ensure against star formation in regions being rapidly disrupted, we do not allow star formation in cells with $\nabla \cdot {\bf v} > 0$ (using our standard higher-order gradient estimator) if the expansion timescale $1/\nabla \cdot {\bf v}$ is shorter than the longer of the star formation or self-gravitating free-fall time (this is also large enough to ensure spurious stochastic fluctuations do not dominate). Conversely, to help ensure against artificially delaying star formation in rapidly collapsing regions, we set the star formation time to the minimum of the free-fall time or the compression time $-1/\nabla \cdot {\bf v}$ in regions with $\nabla \cdot {\bf v} < 0$. We show in Appendix~\ref{sec:alternatives:sf} that this criterion is largely redundant with the existing virial+Jeans criteria and has little effect on our results.

\end{enumerate}

The effects of the different \firex\ and \firetwo\ star-formation criteria on the density distribution of star-forming gas are illustrated in Fig.~\ref{fig:sfr}. In \firetwo, we adopted similar (slightly simpler) versions of the self-gravity and Jeans criteria (see \citealt{hopkins:fire2.methods}, Appendix~C). The main difference in those particular criteria is the slightly stricter (including magnetic energy and more accurately accounting for thermal energy) version of the self-gravity criterion in \firex. However in \firetwo\ we {\em also} required the gas (1) exceed a density threshold $n>1000\,{\rm cm^{-3}}$, and (2) was molecular (scaling the SFR with the molecular mass fraction $2\,f_{\rm mol}$). However, regarding (1) as discussed in \citet{hopkins:virial.sf,hopkins:dense.gas.tracers} and a number of other studies, any constant-density SF threshold introduces an arbitrary scale and resolution-dependence to the simulation, which has no particular physical motivation, and as shown in \citet{grudic:max.surface.density,grudic:cluster.properties} this can imprint artificial features on e.g.\ the density distribution and sizes of star clusters and GMCs. Moreover it cannot deal well with situations like merging galaxy nuclei where all gas exceeds the threshold, nor un-resolved fragmentation in clouds where the mean density is below-threshold but local peaks would arise if one had sufficient resolution. In practice, the combination of a self-gravity and Jeans threshold sets a strong effective density threshold; in Milky Way-like galaxies, using the \firex\ model, most of the star formation arises from gas with $n_{\rm H} \sim 10^{2}-10^{3}\,{\rm cm^{-3}}$ at our standard resolution of $\sim 7000\,M_{\odot}$ in Fig.~\ref{fig:sfr}, so SF is still strongly clustered and peaked in high-density gas, without an arbitrary constant threshold applied. Regarding (2), numerous studies \citep[see e.g.][]{glover:2011.molecules.not.needed.for.sf} have consistently shown that molecular gas does not necessarily play a {\em causal} role in star formation: molecules simply serve as an effective {\em tracer} of gas that is dense (self-gravitating) and able to self-shield and cool effectively to low temperatures (hence Jeans unstable). At best, then, including a molecular SF criterion is redundant with our self-gravity and Jeans criteria; at worst, scaling SF with $f_{\rm mol}$ can artificially suppress star formation in low-metallicity/high-redshift or irradiated/starburst environments where molecule formation is inefficient but atomic and/or dust cooling are still effective.

Note that we have experimented with various other criteria, to be discussed in future work. In particular we have also explored various ``multi-free-fall'' models from e.g.\ \citet{padoan:2012.sfr.local.virparam} (as implemented in \citealt{semenov:local.vs.global.sfe}) or e.g.\ \citet{federrath:2012.sfr.vs.model.turb.boxes} and \citet{hopkins:excursion.ism} (as implemented in \citealt{2020arXiv200406008N}), in which the SF efficiency per freefall time is a continuous function of the virial parameter $\alpha$ (and potentially Mach number $\mathcal{M}$) rather than a step function. However we find these models result in some pathological behaviors applied outside of where they were originally calibrated: the \citet{padoan:2012.sfr.local.virparam} model, for example, predicts fully ``normal'' star formation (global galaxy-averaged efficiency $\epsilon>0.01$) in laminar hot halo or thick-disk hot gas with negligible cooling and Toomre $Q \gg 10$ if the scale-lengths are marginally resolved, while the \citet{federrath:2012.sfr.vs.model.turb.boxes} or \citet{hopkins:excursion.ism} models predict extremely high efficiencies ($\epsilon \gg 1$) in un-bound gas which is part of a SNe shell or wind undergoing a free expansion phase (with $\alpha \gg 1$ but $\mathcal{M} \gg 1$). Without re-calibrating models for the range of galactic cases, we adopt the more conservative step-function in $\alpha$.

\begin{figure}
	\includegraphics[width=0.98\columnwidth]{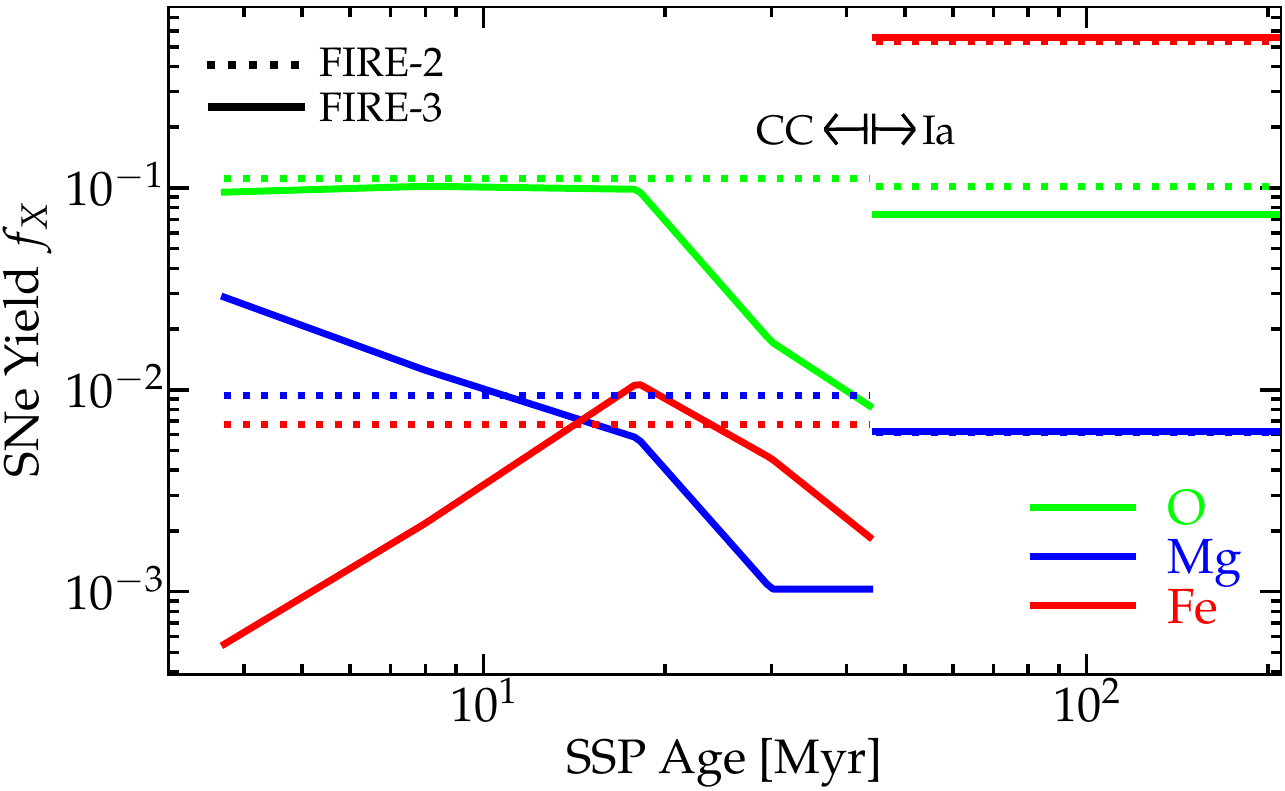}
	\includegraphics[width=0.98\columnwidth]{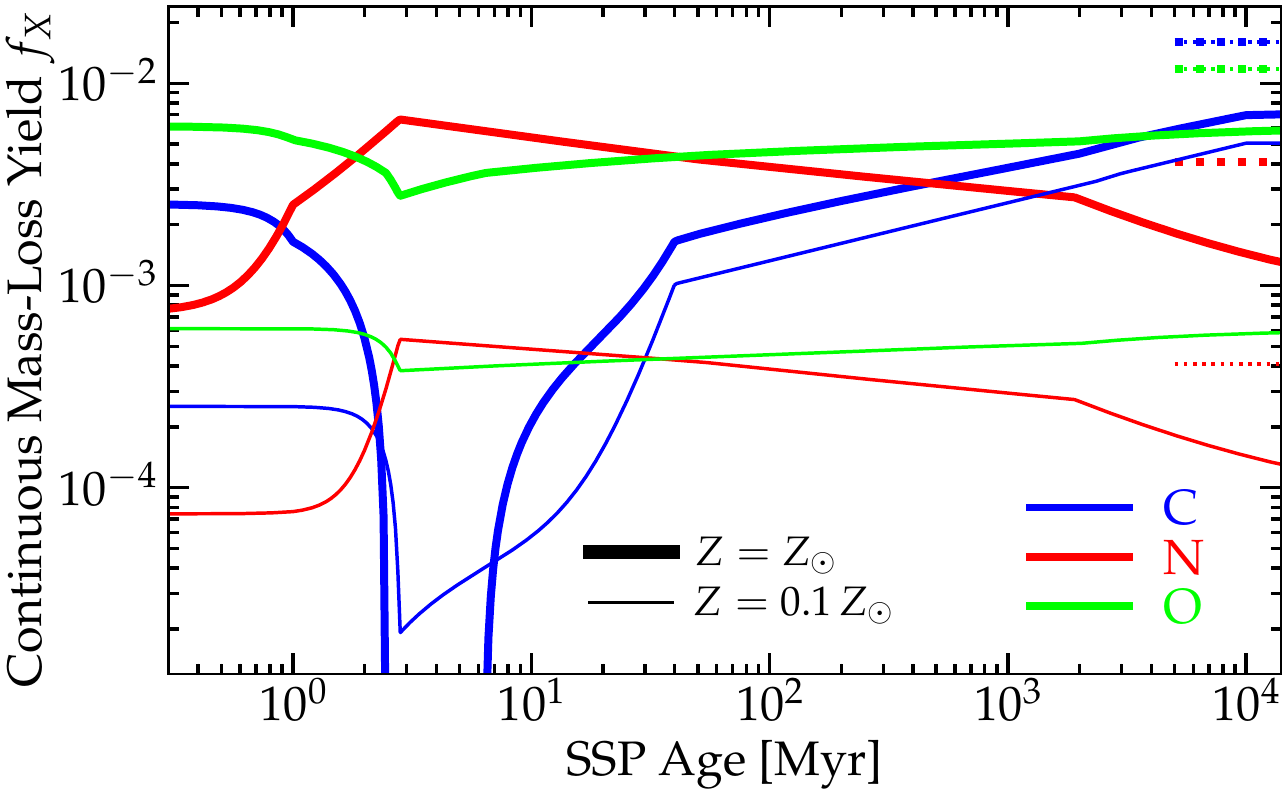}
	\vspace{-0.1cm}
	\caption{A few example yields (\S~\ref{sec:yields}) for \firetwo\ ({\em dotted}) and \firex\ ({\em solid}) vs.\ SSP age for SNe ({\em top}, core-collapse (CC) at $t<44\,$Myr and Ia at later times, as Fig.~\ref{fig:tracks}) and continuous mass-loss ({\em bottom}; primarily O/B at $t\lesssim 30\,$Myr and AGB later, as Fig.~\ref{fig:winds}). Yields are defined as fraction $f_{X}$ of ejecta mass in species $X$. The SNe yields do not depend significantly on progenitor metallicity (differences between models are larger than the mean trend); the AGB yields for CNO do, so we show both Solar ({\em thick}) and $0.1\,Z_{\odot}$ ({\em thin}) progenitor abundances. Ia yields are time-independent as we assume all Ia's are identical; core-collapse \&\ mass-loss yields were time/IMF-averaged in \firetwo\ but depend on progenitor mass and/or time in \firex. For core-collapse the time-dependence is important for detailed abundance-ratio predictions but the IMF-integrated yields are similar. For mass-loss, the progenitor metallicity-dependence of O was not captured in \firetwo, although this is a sub-dominant source of O compared to SNe.
	\label{fig:yields}}
\end{figure}

\subsection{Yields}
\label{sec:yields}

{\bf SNe Ia:} Although evidence has recently emerged for variation in SNe Ia yields \citep{2019ApJ...881...45K,2020ApJ...891...85D}, the models remain highly uncertain and the relevant species where differences emerge can be modeled with our tracer-element approach in post-processing (they do not have a large dynamical effect on cooling), so we treat all Ia's as identical explosions in-code, with ejecta mass $1.4\,M_{\odot}$. In \firetwo\ we take the mass fraction in (Z, H, He, C, N, O, Ne, Mg, Si, S, Ca, Fe) = (1, 0, 0, 3.50e-2, 8.57e-7, 0.102, 3.21e-3, 6.14e-3, 0.111, 6.21e-2, 8.57e-3, 0.531) from \citet{iwamoto:1999.sneIa.yields} (their ``W7'' model). 

For consistency we update this to the newer results from the same group using updated nuclear cross sections and reaction rates, specifically averaging the W7 and WDD2 models from \citet{leung.nomoto:2018.Ia.yield.model.update} as advocated and updated in \citet{mori:Ia.nucleo.constraints} (who argue direct SNe constraints are best matched by an equal mix of these models), giving (Z, H, He, C, N, O, Ne, Mg, Si, S, Ca, Fe) = (1, 0, 0, 1.76e-2, 2.10e-6, 7.36e-2, 2.02e-3, 6.21e-3, 0.146, 7.62e-2, 1.29e-2, 0.558). For all species which we follow where Ia's have any significant contribution to galactic abundances, this is extremely similar to yields from other groups such as the favored N100 model in \citet{seitenzahl:2013.Ia.yields}. The effect of changing the Ia yield is minimal (see Fig.~\ref{fig:yields}).

{\bf Core-collapse SNe:} In \firetwo\ (per \citet{hopkins:fire2.methods}, Appendix~A) we adopted IMF-averaged yields for core-collapse SNe (CCSNe) from \citet{nomoto2006:sne.yields}, treating each CCSNe identically, with total mass $M=10.5\,M_{\odot}$ and mass fractions (He, C, N, O, Ne, Mg, Si, S, Ca, Fe) = (3.87, 0.133, 0.0479\,{MAX}[$Z/Z_{\odot},\,1.65$], 1.17, 0.30, 0.0987, 0.0933, 0.0397, 0.00458, 0.0741). Here we update our fits to account for the fact that CCSNe at different SSP ages correspond to different progenitor masses and therefore different yields. Because each star particle represents a well-sampled IMF, it is convenient to parameterize the yields as a function of SSP age $t$ and progenitor metallicity.

We compare the time-dependent SNe yields from four independent sources: the NuGrid compilation \citep{nugrid:yields}, S16 \citep{sukhbold:yields.and.explosion.props.dense.grid}, N13 \citep{nomoto:2013.yield.update.review}, and LC18 \citep[][using the interpolation in rotation rates as a function of mass and metallicity from \citealt{prantzos:limongi.yields.interpolation}]{limongi.chieffi:2018.rotating.star.yields}. Of these, the compilation of S16 samples much more densely in progenitor mass, and importantly is the only model set that includes the entire progenitor mass range of interest from $\sim 8-120\,M_{\sun}$, so we use this as our baseline to define the time dependence at solar metallicity. Beginning from the S16 tables for stellar mass-loss and ejecta yields at different stellar masses, we add late-stage, post-He burning wind production of intermediate and heavy elements (after subtracting the progenitor surface abundance mass loss) to SNe ejecta, as these come in an extremely short period of time pre-explosion and are not included in our OB mass-loss yields, and sum different stable isotopes. We then re-bin the model grid weighting by our adopted IMF and the model sampling, with narrow bins sampling broadly similar explosion times and equal numbers of models, since the models can have sharp discontinuous behaviors with $\Delta M$ as small as $0.1\,M_{\odot}$ which our galaxy-scale simulations cannot resolve. We then use a mass-lifetime relation consistent with the adopted \citet{2014ApJS..212...14L} isochrones used for SNe rates to map to time of explosion, and re-bin the time intervals to sample uniformly in time, and re-normalize ejecta masses of H \&\ He to match those assumed in the \citet{2014ApJS..212...14L} models (as these are sensitive to the mass-loss models) conserving heavy-element masses, then fit each to a five-component piecewise power-law. At solar abundances, most of the models agree at least qualitatively once this is done. 

However, there are a couple of species, most notably Mg, where the models from S16 (and related models from the same group, including NuGrid and \citealt{woosley.weaver.1995:yields}) produce systematically low IMF-integrated yields (such that the IMF-averaged core-collapse-only yield of Mg from S16 gives [Mg/Fe]\,$\lesssim 0$, and thus it is essentially impossible to produce $\alpha$-enhanced populations as observed; see discussion in \citealt{ma:2015.fire.mass.metallicity} and \citealt{muley:2020.fire.nugrid.yield.tests}). This arises in part because of the assumption in these models that most more massive stars ($\gtrsim 20\,M_{\odot}$) do not explode, giving rise to low ``effective'' yields. We therefore more carefully interpolate these yields only between the models that do explode and extrapolate this to stars that we assume to explode in our SNe rate models (but not in S16). This increases the IMF-integrated yields for Mg, Ne, and Si to be closer to those from models like N13 which adopt a somewhat higher mass threshold for implosion.

Finally, we follow a similar procedure to \firetwo\ to evaluate whether any predicted dependence on progenitor metallicity is robust: specifically, we compare the three model sets which sample a range of progenitor $Z$ at fixed progenitor mass $M$ (NuGrid, N13, and LC18). We fit each $Z$ dependence to a linear or quadric in log-space and ask whether all three models are consistent with the same {\em sign} of the dependence (e.g.\ whether all are statistically consistent with a yield that increases, or decreases, with progenitor $Z$). We find that almost no species passes this test: all three models predict a weak enhancement in O yield ($\propto Z^{-(0-0.2)}$) from intermediate-mass ($15-25\,M_{\odot}$) stars at low progenitor $Z$, but otherwise the models either disagree on the qualitative sign of the $Z$ dependence or predict no trend. We therefore default to assuming no progenitor-$Z$ dependence for metal yields. These models and the STARBURST99 isochrones used for the CC rates (\S~\ref{sec:stellar.evol.tracks}) do consistently predict a very weak dependence of ejecta energy on progenitor $Z$ with the median best-fit $E= \epsilon_{z}\,10^{51}\,{\rm erg}$, where $\epsilon_{z}\approx{\rm max}[(\tilde{z}+10^{-4})^{-0.12},\,1]$, though any small effects of this are almost entirely offset by the slightly weaker dependence of SNe terminal momentum on metallicity adopted in \firex\ (\S~\ref{sec:sne}).

The total ejecta mass per event is well-fit by $M_{\rm ej} = 10\,M_{\odot}\,(t/6.5\,{\rm Myr})^{-\psi_{\rm CCM}}$ with $\psi_{\rm CCM}=2.22$ for $t\le 6.5$\,Myr and $\psi_{\rm CCM}=0.267$ for $t>6.5\,$Myr.  The metal yields in dimensionless mass-fraction-of-ejecta units $y_{{\rm cc},\,j}$ for species $j$ are fit by continuous piecewise power-laws of the form: 
\begin{align}
\label{eqn:cc.yields} y_{{\rm cc}, j} = & 
\begin{cases}
0 \hfill & \ (t \le t_{{\rm cc}, j,\,1}) \\ 
a_{{\rm cc}, j,1}\,(t/t_{{\rm cc}, j,1})^{\psi_{{\rm cc}, j,1}} \hfill & \ (t_{{\rm cc}, j,1} < t \le t_{{\rm cc}, j,2}) \\
... & ... \\
a_{{\rm cc}, j,n}\,(t/t_{{\rm cc}, j,n})^{\psi_{{\rm cc}, j,n}} \hfill & \ (t_{{\rm cc}, j,n} < t \le t_{{\rm cc}, j,n+1} ) \\ 
0 \hfill & \ (t \ge t_{{\rm cc}, j,n+1}) \\ 
\end{cases} 
\end{align}
with $\psi_{{\rm cc}, j,n}\equiv \ln{(a_{{\rm cc}, j,n+1}/a_{{\rm cc}, j,n})}/\ln{{(t_{{\rm cc}, j,n+1}/t_{{\rm cc}, j,n})}}$, and $t_{{\rm cc},\,j} = $(3.7, 8, 18, 30, 44)\,Myr. The coefficients
$a_{{\rm cc},j,n}$ for species tabulated are given in Table~\ref{tbl:cc.yields}.

\begin{footnotesize}
\ctable[caption={{\normalsize Fitting-Function Coefficients for Core-Collapse Yields in Eq.~\ref{eqn:cc.yields}}\label{tbl:cc.yields}},center,
]{lccccc}{
}{
\hline\hline
Species & $a_{{\rm cc},\,j,\,1}$ & $a_{{\rm cc},\,j,\,2}$  & $a_{{\rm cc},\,j,\,3}$ & $a_{{\rm cc},\,j,\,4}$ & $a_{{\rm cc},\,j,\,5}$ \\
\hline
He & 0.461 & 0.330 & 0.358 & 0.365 & 0.359  \\
C & 0.237 & 8.57e-3 & 1.69e-2 & 9.33e-3 & 4.47e-3  \\
N & 1.07e-2 & 3.48e-3 & 3.44e-4 & 3.72e-3 & 3.50e-3  \\
O & 9.53e-2 & 0.102 & 9.85e-2 & 1.73e-2 & 8.20e-3   \\
Ne & 2.60e-2 & 2.20e-2 & 1.93e-2 & 2.70e-3 & 2.75e-3  \\
Mg & 2.89e-2 & 1.25e-2 & 5.77e-3 & 1.03e-3 & 1.03e-3   \\
Si & 4.12e-4 & 7.69e-3 & 8.73e-3 & 2.23e-3 & 1.18e-3  \\
S & 3.63e-4 & 5.61e-3 & 5.49e-3 & 1.26e-3 & 5.75e-4  \\
Ca & 4.28e-5 & 3.21e-4 & 6.00e-4 & 1.84e-4 & 9.64e-5  \\
Fe & 5.46e-4 & 2.18e-3 & 1.08e-2 & 4.57e-3 & 1.83e-3 \\
\hline
\multicolumn{6}{c}{Time Boundaries $t_{{\rm cc},\,j,\,n}$ [Myr]} \\
\hline
-- & $t_{{\rm cc},\,j,\,1}$ & $t_{{\rm cc},\,j,\,2}$  & $t_{{\rm cc},\,j,\,3}$ & $t_{{\rm cc},\,j,\,4}$ & $t_{{\rm cc},\,j,\,5}$ \\
-- & 3.7 & 8 & 18 & 30 & 44 \\
\hline\hline
}
\end{footnotesize}

\begin{footnotesize}
\ctable[caption={{\normalsize Fitting-Function Coefficients for Stellar Mass-Loss Yields in Eq.~\ref{eqn:massloss.yields.y}}\label{tbl:massloss.yields}},center,star
]{lcccccc}{
}{
\hline\hline
Term & $x_{j,1}$ & $x_{j,2}$ & $x_{j,3}$ & $x_{j,4}$ & $x_{j,5}$ & $x_{j,6}$ \\
\hline\hline
\multicolumn{7}{l}{HHe ($y_{\rm HHe}$; $\psi_{{\rm HHe},0}=3$)} \\
\hline
$t_{{\rm HHe},\,n}$ [Gyr] & 0.0028 & 0.01 & 2.3 & 3.0 & 100 & -- \\ 
$a_{{\rm HHe},\,n}$ & 0.4\,{MIN}[$(z_{\rm CNO}+0.001)^{0.6},\,2$] & 0.08 & 0.07 & 0.042 & 0.042 & -- \\
\hline
\multicolumn{7}{l}{CNO ($y_{\rm CNO}$; $\psi_{{\rm CNO},0}=3.5$)} \\
\hline
$t_{{\rm CNO},\,n}$ [Gyr] & 0.001 &  0.0028 & 0.05 & 1.9 & 14 & 100 \\ 
$a_{{\rm CNO},\,n}$ & 0.2\,{MIN}[$z_{\rm CNO}^{2}+10^{-4},\,0.9$] & 0.68\,{MIN}[$(z_{\rm CNO}+0.001)^{0.1},\,0.9$] & 0.4 & 0.23 & 0.065 & 0.065 \\
\hline
\multicolumn{7}{l}{HC ($y_{\rm HC}$; $\psi_{{\rm HC},0}=3$)} \\
\hline
$t_{{\rm HC},\,n}$ [Gyr] & 0.005 & 0.04 & 10 & 100 & -- & -- \\ 
$a_{{\rm HC},\,n}$ & $10^{-6}$ &  0.001 & 0.005 & 0.005 & -- & -- \\
\hline\hline
}
\end{footnotesize}

The results for a couple species of interest are shown in Fig.~\ref{fig:yields}. While the IMF-integrated yields are broadly similar to \firetwo, the progenitor mass-dependence can lead to a order-of-magnitude different ratios of e.g. [Mg/O] or [Mg/Fe] for SNe which explode at different SSP ages, which is important for predicting detailed internal abundance patterns and spreads of galaxies \citep{muley:2020.fire.nugrid.yield.tests}.

{\bf Stellar Mass-Loss:} Continuous mass-loss yields (from e.g. OB and AGB outflows) are treated as follows. Given some total mass-loss rate $\dot{M}$, the mass fraction in each species $X$ is given by $f_{\rm X}$. For heavy elements, models predict negligible enrichment in dredged-up outflows, so $f_{\rm X} = f_{{\rm X},\,0}$, the initial surface abundance. Of the species tracked, however, we  follow more complicated returns for He, C, N, O. In \firetwo\ we adopted the age-independent stellar mass (IMF) and Hubble-time-integrated values given in \citet{hopkins:fire2.methods} Appendix A: ($f_{\rm He}$, $f_{\rm C}$, $f_{\rm N}$, $f_{\rm O}$)=(0.36, 0.016, 0.0041, 0.0118\,{MAX}[$Z/Z_{\odot},\,1.65$]). This notably ignores any distinction between OB and AGB outflows, let alone more detailed progenitor mass (hence time, in a population sense) dependence. In \firex, we update this to follow the full time-dependence following the combination of the same models used for the mass-loss rate tabulation from \citet{2014ApJS..212...14L} and a comparison yield set given by the compilation of AGB+OB winds from \citet{2015ApJS..219...40C} and \citet{limongi.chieffi:2018.rotating.star.yields} as sampled and interpolated after calibration to observational constraints on rotation and other properties in \citet{prantzos:limongi.yields.interpolation}. We follow a similar procedure as for CCSNe to rebin the latter models in time, and to vet the metallicity dependence, including only dependences where both models are consistent with the same sign of the progenitor metallicity dependence (and in those cases using the mean of the logarithmic dependence of yield on progenitor $Z$).

While we could use a pure look-up table for the yields, we prefer to parameterize these in the following manner, as it aids in physical interpretation and also guarantees against un-physical extrapolation of yields beyond pre-computed values in the references above. For He, $f_{\rm He} = f_{{\rm He},\,0}\,(1-y_{\rm HeC}) + y_{\rm HHe}\,f_{{\rm H},\,0}$, i.e.\ some He can be lost to heavier species ($y_{\rm HeC}$) and some produced by H burning ($y_{\rm HHe}$). For N, $f_{\rm N} = f_{{\rm N},\,0} + y_{\rm CN}\,f_{{\rm C},\,0} + y_{\rm ON}\,f_{{\rm O},\,0}$, i.e.\ we model secondary production of N from both initial C and O. For C and O, $f_{\rm C} = f_{{\rm C},\,0}\,(1-y_{\rm CN}) + y_{\rm HeC}\, f_{{\rm He},\,0} + y_{\rm HC}\,f_{{\rm H},\,0}\,(1-y_{\rm HHe})$, $f_{\rm O} = f_{{\rm O},\,0}\,(1-y_{\rm ON})$, i.e.\ the amount going into secondary N is lost (subtracted) while some primary C is produced. In the above, $y_{\rm HeC}=y_{\rm HC}$, $y_{\rm CN} = {\rm MIN}[ 1,\ 0.5\,y_{\rm CNO}\,(1+x_{\rm OC}) ]$, $y_{\rm ON} = y_{\rm CNO} + (y_{\rm CNO} - y_{\rm CN})\,x_{\rm OC}^{-1}$, $x_{\rm OC} = f_{{\rm O},0}/f_{{\rm C},0}$. The yields are then entirely determined by the initial abundances and $y_{\rm HHe}$, $y_{\rm CNO}$, $y_{\rm HC}$ which can be fit as a function of population age and metallicity to continuous piecewise power-laws similar to those used for the wind mass-loss rates, 
\begin{align}
\label{eqn:massloss.yields.y} y_{j=({\rm HHe},\,{\rm CNO},\,{\rm HC})} = & 
\begin{cases}
a_{j,1}\,(t/t_{j,1})^{\psi_{j,0}} \hfill & \ (t \le t_{j,\,1}) \\ 
a_{j,1}\,(t/t_{j,1})^{\psi_{j,1}} \hfill & \ (t_{j,1} < t \le t_{j,2}) \\
... & ... \\
a_{j,n}\,(t/t_{j,n})^{\psi_{j,n}} \hfill & \ (t_{j,n} < t \le t_{j,n+1} )
\end{cases} 
\end{align}
with $\psi_{j,n}\equiv \ln{(a_{j,\,n+1}/a_{j,\,n})}/\ln{{(t_{j,\,n+1}/t_{j,\,n})}}$ required by continuity for $n>0$. 
For $y_{\rm HHe}$: 
$\psi_{{\rm HHe},0}=3$, 
$t_{{\rm HHe},\,n}=$(0.0028, 0.01, 2.3, 3.0, 100)\,Gyr, 
$a_{{\rm HHe},\,n}=$(0.4\,{MIN}[$(z_{\rm CNO}+0.001)^{0.6},\,2$] , 0.08, 0.07, 0.042, 0.042). 
For $y_{\rm CNO}$: 
$\psi_{{\rm CNO},\,0}=3.5$, 
$t_{{\rm CNO},\,n}=$(0.001, 0.0028, 0.05, 1.9, 14, 100)\,Gyr, 
$a_{{\rm CNO},\,n}=$(0.2\,{MIN}[$z_{\rm CNO}^{2}+10^{-4},\,0.9$], 0.68\,{ MIN}[$(z_{\rm CNO}+0.001)^{0.1},\,0.9$] , 0.4, 0.23, 0.065, 0.065). 
For $y_{\rm HC}$: 
$\psi_{{\rm HC},0}=3$, 
$t_{{\rm HC},\,n}=$(0.005, 0.04, 10, 100)\,Gyr, 
$a_{{\rm HC},\,n}=$($10^{-6}$, 0.001, 0.005, 0.005). These are summarized in Table~\ref{tbl:massloss.yields}.
Here the metallicity of primary interest is the CNO metallicity, as this drives the dependence of the light element fusion, so we take 
$z_{\rm CNO} = (Z_{\rm C,0} + Z_{\rm N,0} + Z_{\rm O,0}) / (Z_{\rm C} + Z_{\rm N} + Z_{\rm O})_{\odot}$. 

The resulting fits for CNO are shown in Fig.~\ref{fig:yields}. The dynamical effect of the new yields is minimal given the smaller mass-loss rates relative to \firetwo, but they capture a number of phenomena in detail (with both time and progenitor-metallicity dependence) that were not captured in \firetwo, and the new fits are more robust to prevent extrapolation to un-physical regimes (e.g.\ excessive N return at high progenitor $Z$ if C and O are relatively depleted).

\subsection{Optional Physics: Cosmic Rays}
\label{sec:crs}

As noted above, some \firex\ simulations explicitly evolve the cosmic ray (CR) population and dynamics. While our focus  here is on the ``default'' \firex\ methods, we summarize the default CR implementation for completeness. In \firetwo\ simulations with explicit CR dynamics 
\citep{su:2018.stellar.fb.fails.to.solve.cooling.flow,su:turb.crs.quench,su:2021.agn.jet.params.vs.quenching,chan:2018.cosmicray.fire.gammaray,chan:2021.cosmic.ray.vertical.balance,hopkins:cr.mhd.fire2,hopkins:2020.cr.outflows.to.mpc.scales,hopkins:2020.cr.transport.model.fx.galform,hopkins:cr.transport.constraints.from.galaxies,ji:fire.cr.cgm,ji:20.virial.shocks.suppressed.cr.dominated.halos,trapp:2021.radial.transport.fire.sims}, we treated CRs with a simplified approximation developed in \citet{chan:2018.cosmicray.fire.gammaray}: we evolved just the total CR energy density as a relativistic fluid obeying a simple two-moment equation with a somewhat ad-hoc ``streaming plus diffusion'' approximation. In \firex\ simulations including CR dynamics, our treatment of CRs is updated to match that presented in \citet{hopkins:cr.multibin.mw.comparison}: we evolve the full CR distribution function/spectrum for multiple species with a recently-derived rigorous two-moment formulation including all terms up to leading order in $\mathcal{O}(u/c)$ (where $u$ is the MHD fluid velocity). Specifically, we explicitly evolve the distribution function of a gyrotropic population of CRs $f=f({\bf x},\,{\bf p},\,t,\,s,\,...)$ as a function of time $t$, position ${\bf x}$, CR momentum/rigidity/energy ${\bf p}$, and species $s$, using the two-moment closure scheme derived formally from the full CR equations of motion in \citet{hopkins:m1.cr.closure}, which accurately captures both the strong-scattering and free-streaming regimes, with the appropriate (anisotropic) forces on the gas from Lorentz force and parallel scattering terms and thermalized collisional/radiative losses as described in \citet{hopkins:cr.multibin.mw.comparison}. We follow the full CR spectrum from $\sim$\,MeV-TeV energies, including all relevant adiabatic, diffusive re-acceleration, streaming loss, catastrophic, spallation, pionic, annihilation, radioactive decay, ionization, Coulomb, secondary production, and radiative loss (inverse Compton, Bremstrahhlung, synchrotron) processes, using the local plasma properties (e.g.\ ${\bf B}$, $n_{\rm ion}$, $u_{\rm rad}$) determined self-consistently in-code. First-order Fermi acceleration is treated as injection with a single power-law spectrum $\propto R^{-4.2}$ in rigidity, with $10\%$ of the initial ejecta kinetic energy from SNe and fast stellar winds injected at reverse shock formation (and $2\%$ of that in leptons). Following standard practice in the Galactic CR literature, by default in \firex\ we assume the effective (pitch-angle averaged) scattering rate can be approximated with a simple power-law as $\bar{\nu}=10^{-9}\,{\rm s^{-1}}\,\beta\,(R/{\rm GV})^{-0.6}$ (though variations with spatially-variable diffusion coefficients, such as those in \citealt{hopkins:2021.sc.et.models.incompatible.obs}, can be studied as well). This is calibrated in \citet{hopkins:cr.multibin.mw.comparison} to simultaneously reproduce, for MW-like galaxies, observations of the CR spectra, $\gamma$-ray emissivities, ionization rates, and secondary-to-primary ratios for a wide range of species (electrons, positrons, protons/H, anti-protons,  $^{7}$Be, $^{9}$Be, $^{10}$Be, B, C, N, O) at energies $\sim$\,MeV-TeV. In \firex, if we are not trying to model specific MW observations, we only explicitly simulate the proton and electron spectra (sufficient for predicting observables such as synchrotron, $\gamma$-ray emission, as well as CR effects on gas).

\subsection{Optional Physics: Black Holes}
\label{sec:bhs}

A number of \firex\ simulations will also include supermassive BHs/AGN growth and feedback processes, as compared to the purely stellar feedback processes described above. All of the numerical methods used for our BH simulations have been presented and extensively tested in a series of papers  \citep{hopkins:m31.disk,hopkins:zoom.sims,hopkins:inflow.analytics,hopkins:qso.stellar.fb.together,hopkins:2021.bhs.bulges.from.sigma.sfr,daa:BHs.on.FIRE,daa:20.hyperrefinement.bh.growth,su:turb.crs.quench,su:2021.agn.jet.params.vs.quenching,torrey:2020.agn.wind.bal.gal.fx.fire,grudic:starforge.methods,guszejnov:2020.starforge.jets,ma:2021.seed.sink.inefficient.fire,wellons:prep}. However many of these studies only considered one aspect of BH growth or feedback or dynamics, and others considered or compared a variety of methods. Therefore for clarity and convenient reference we summarize the complete default numerical implementation of BHs adopted for \firex\ simulations with ``live BHs'' here. This is largely motivated by the theoretically and empirically preferred models identified in the extensive study of $\sim1000$ \firetwo\ simulations in \citet{wellons:prep} and \citet{su:2021.agn.jet.params.vs.quenching}.

\subsubsection{Seeding, Dynamics, \&\ Mergers}
\label{sec:methods:bhs:seeding}

\begin{enumerate} 

\item{\em Seeding:} When a gas cell meets all criteria to create a star particle of mass $\Delta m$, it can instead be converted into a BH seed with a probability strongly weighted to high acceleration/surface density scales and low metallicities, 
$p_{\rm BH} \equiv [ 1 - \exp{\{-(\Delta m/M_{0})\,f_{s,\,Z}(Z/Z_{s,\,0})\,f_{s,\,a}(\bar{a}_{\rm grav}/\bar{a}_{s,\,0}) \}} ]\,\Theta$. 
Here $\Theta =0$ or $=1$ is a step function which requires (1) the cell meets all of the star formation criteria (the gas flow is locally-converging, the cell is self-gravitating and bound including thermal+kinetic+magnetic support, and Jeans unstable); (2) no other BH already exists within a distance $r = 10\,R_{s,\,0}$; and (3) the local acceleration scale $\bar{a}_{\rm grav} \equiv |\langle {\bf a}(<R_{s,\,0})_{\rm grav} \rangle | \equiv {G\,M_{\rm enc,\,tot}(r < R_{s,\,0})}/{R_{s,\,0}^{2}}$ (where $M_{\rm enc,\,tot}(r < R_{s,\,0})$ is the total mass of {\em all} species inside $R_{s,\,0}$) exceeds $0.01\,\bar{a}_{s,\,0}$. The functions $f_{s,\,Z}(x)=(1+x+x^{2}/2)^{-1}$ and $ f_{s,\,a}(x) \equiv 1-\exp{(-x^{2})}$ are arbitrary but chosen to give a steep (but not step-function like) cutoff at $Z>Z_{s,\,0}$ and $\bar{a}_{\rm grav} < \bar{a}_{s,\,0}$, where we choose $Z_{s,\,0} = 0.001$ and $\bar{a}_{s,\,0}  = 2\,\langle \dot{p}_{\ast}/m_{\ast}\rangle \sim 2\times10^{-7}\,{\rm cm\,s^{-2}}$ (equivalent to an ``effective surface density'' $\Sigma_{\rm eff} \equiv M_{\rm enc,\,tot}(r < R_{s,\,0}) / \pi\,R_{s,\,0}^{2} \sim 5000\,{\rm M_{\odot}\,pc^{-2}}$, with $\langle \dot{p}_{\ast}/m_{\ast}\rangle$ defined below). We take $R_{s,\,0}\equiv {\rm MAX}[100\,{\rm pc},\,H_{s}]$ (where $H_{s}$ is the interaction kernel\footnote{A cubic spline with $H_{s}=4\,\Delta x_{\rm gas}$ where $\Delta x_{\rm gas}$ is the mean inter-gas-cell spacing of the cells inside the kernel.} radius of compact support) and $M_{0} \approx 10^{6}\,M_{\odot}$ (a constant which determines the average number of seeds formed per unit stellar mass). This is motivated by models where BHs form preferentially in very high-acceleration (or high-surface density/``gravitational pressure'') regions where dense, bound/long-lived star clusters form \citep{grudic:sfe.cluster.form.surface.density,grudic:max.surface.density,2019MNRAS.487..364L,shi:2020.imbh.budget.star.clusters,ma:2021.seed.sink.inefficient.fire}, ensuring they do not simply ``wander'' about the galaxy as random stars. All seeds are formed with an initial BH mass $M_{\rm BH} =100\,M_{\odot}$, zero ``reservoir'' mass, and total dynamical mass of the spawning cell.

\item{\em Dynamics:} BHs follow the usual gravity equations with force softening set to $1/2$ the minimum star particle softening.  As shown explicitly in \citet{ma:2021.seed.sink.inefficient.fire}, realistic IMBHs cannot efficiently ``sink'' under the influence of dynamical friction if they are ``free wandering'' through the galaxy in a timescale shorter than the Hubble time unless they are already super-massive, with masses $\gtrsim 10^{7}\,M_{\odot}$ (where our numerical resolution can easily capture the relevant dynamical friction processes). But IMBHs as low mass as $\sim 100\,M_{\odot}$ can remain firmly ``anchored'' (modulo effects like gravitational recoil from mergers) at the center of dense, bound star clusters where (by construction) we attempt to model their formation. However, at FIRE resolution, we only marginally resolve dense star clusters and certainly cannot resolve processes important to this ``anchoring'' like mass segregation and core collapse within the cluster (which depend on individual stellar N-body processes in the cluster center), so we employ the method from \citet{wellons:prep} where the BHs are artificially ``drifted'' at a fraction of the local velocity dispersion towards the local stellar+DM+BH binding-energy extremum, in order to prevent them from being artificially ejected from under-resolved clusters. 

Within the interaction kernel of the BH ``$a$,'' it identifies all collisionless (star, DM, BH) neighbors ``$b$'' (we exclude gas because the gas can be in outflows up to $\sim 10^{5}\,{\rm km\,s^{-1}}$, which easily leads to spurious ``forces''), and identifies the neighbor with the minimum {\em total} energy (not potential) in the frame of the BH, weighted by a spatial kernel, i.e.\ the minimum of $\Psi_{b} \equiv \tilde{w}_{ab}^{-1}\,(\Phi_{b} + (1/2)\,|{\bf v}_{b}-{\bf v}_{a}|^{2})$ (where $\tilde{w}_{ab} \equiv 1+ |{\bf x}_{b}-{\bf x}_{a}|^{2}\,(h_{a}^{-2} + [4\,\epsilon_{a}]^{-2})$, with $h_{a} \sim H_{a}/3$ the mean inter-particle separation in the kernel and $\epsilon_{a}$ the BH force softening above). This prevents the BH being ``pulled'' to neighbors moving with very large velocities (which can easily occur in e.g.\ mergers) or very large separations. The BH then does not ``jump'' to position ${\bf x}_{b}$ but moves smoothly, with ${\bf x}_{a} \rightarrow {\bf x}_{a} + ({\bf x}_{b}-{\bf x}_{a})\,(1 - \exp{[-v_{s}\,\Delta t_{a}/|{\bf x}_{b}-{\bf x}_{a}|]}) \approx {\bf x}_{a} + \hat{\bf x}_{ba}\,v_{s}\,\Delta t_{a}$ where $\Delta t_{a}$ is the timestep, and $v_{s} = {\rm MAX}(v_{ff},\,c_{0})$ with $c_{0} = 10\,{\rm km\,s^{-1}}$ (a typical sound speed, set as a minimum) and $v_{ff}^{2} = -2\,\Psi_{b}$ (the free-fall velocity into the local potential minimum). This effective velocity is artificial, and strong kicks can still produce ``over-shoot'' of BHs, so it decays as ${\bf v}_{a} \rightarrow {\bf v}_{a} + \langle \delta{\bf v} \rangle\,(1 - \exp{[-a_{s}\,\Delta t_{a}/|\delta {\bf v}|]})$, where $\langle \delta {\bf v} \rangle$ is the mean kernel-weighted relative velocity of the particles $b$, and $a_{s} = {\rm MIN}[ |\delta {\bf v}|/\tau ,\,{\rm MAX}[ G\,M_{a}^{k}/H_{a}^{2},\, -2\,\Psi_{b} / H_{a} ]]$ (where $M_{a}^{k}$ is the total mass in the kernel, so the latter two terms represent two estimates of the local free-fall acceleration on the scale of the kernel $H_{a}$, and the decay time of the velocity is limited by the $|\delta {\bf v}|/\tau$ term to not be faster than $\tau \equiv 1\,{\rm Myr}$). 

\item{\em Mergers:} Similarly, because we cannot resolve their small-scale dynamics, two BH particles are merged in a timestep (conserving all relevant quantities) if they reside within the same interaction kernel, have overlapping force softenings, and are gravitationally bound to one another.

\item{\em Timesteps:} In addition to all the usual dynamical timestep criteria (for gravity, etc.), we impose additional timestep restrictions on the BH particles to ensure (1) their accretion rate and/or wind outflow rates do not exceed the equivalent mass flux of $<0.1$ gas element per timestep, that (2) the BH and/or disk cannot grow by more than $\sim 1\%$ per timestep, and (3) that the BH cannot take a longer timestep than the shortest timestep of an interacting gas element.

\end{enumerate}

\subsubsection{Accretion}
\label{sec:methods:bhs:accretion}

BH accretion follows the methods described in \citet{wellons:prep}. Each BH particle evolves an accretion ``reservoir'' (e.g.\ an un-resolved \citealt{shakurasunyaev73}-type disk) of mass $M_{\rm acc}$, separate BH mass $M_{\rm BH}$, and total bound dynamical mass $M_{\rm dyn}=m_{i}$. 

{\em Accretion from the ISM:} The reservoir accretes continuously from the surrounding gas at a rate 
\begin{align}
\dot{M}_{\rm acc} &\equiv \eta_{\rm acc}\,\left(1-f_{\rm w,\,\ast} \right)\,M_{\rm gas}\,\Omega \\
\eta_{\rm acc} &\equiv \frac{0.01\,[(M_{\rm BH}+M_{\rm acc})/M_{\rm d}]^{1/6}}{1 + 3\,(M_{\rm gas}/M_{\rm d})\,(M_{\rm d}/10^{9}\,M_{\odot})^{1/3}} \\
f_{\rm w,\,\ast} &\equiv \left[1 + \frac{\bar{a}_{\rm grav}}{\langle \dot{p}_{\ast} / m_{\ast} \rangle} \right]^{-1}
\end{align}
where $M_{\rm gas}$, $M_{\rm d}$, $\Omega=\sqrt{G\,M_{\rm enc,\,tot}/R^{3}}$ refer to the gas and disk mass and dynamical frequency evaluated within a radius $R$.\footnote{We measure properties such as $M_{\rm gas}$, $M_{\rm d}$, $\Omega$ inside the BH interaction kernel ($R=H_{s}$). We estimate the ``disk'' mass $M_{\rm d}$ following \citet{hopkins:inflow.analytics}, assuming cold gas forms a disk and using the total angular momentum ${\bf J}_{\rm tot,\,\ast}$ of collisionless stars (and dark matter) in the kernel to decompose into an isotropic+thin disk combination with $M_{\rm d,\,\ast} = {\rm min}[M_{\rm tot,\,\ast},\,7\,|{\bf J}_{\rm tot,\,\ast}| / 4\,(G\,M_{\rm tot,\,\ast}\,R)^{1/2}]$. Alternatively we can use the kinematic decomposition in \citet{angles.alcazar:grav.torque.accretion.cosmo.sim.implications,daa:BHs.on.FIRE}, with similar results, but some extra computational cost.} The efficiency parameter $\eta_{\rm acc}$ is calibrated from orders-of-magnitude higher-resolution simulations in \citet{lodato:2006.local.grav.transport.to.smbh.seed,levine:sim.mdot.pwrspectrum,hopkins:zoom.sims,hopkins:inflow.analytics,hopkins:m31.disk,hopkins:qso.stellar.fb.together,hopkins:2021.bhs.bulges.from.sigma.sfr,daa:20.hyperrefinement.bh.growth} that directly resolve gas inflows on scales between the ``true'' accretion disk and resolved scales here ($\sim 0.01-100\,$pc), and physically represents the effects of dominant ``gravitational torques'' driving angular momentum loss in gas as the key accretion process \citep{daa:20.hyperrefinement.bh.growth}. The term $f_{\rm w,\,\ast}$ represents the fraction of gas lost by ejection from stellar feedback on un-resolved scales before reaching the accretion disk, calibrated again to high-resolution simulations (\citealt{grudic:sfe.cluster.form.surface.density,wada:torus.mol.gas.hydro.sims,hopkins:qso.stellar.fb.together}; for a review see \citealt{hopkins:2021.bhs.bulges.from.sigma.sfr}): here $\bar{a}_{\rm grav} / \langle \dot{p}/m_{\ast} \rangle$ is just the dimensionless ratio of the gravitational acceleration inwards $\bar{a}_{\rm grav}$ to the momentum flux per unit mass $\langle \dot{p}_{\ast} / m_{\ast} \rangle \sim 10^{-7}\,{\rm cm\,s^{-2}}$ from e.g. radiation and winds from a young stellar population, which determines what fraction of the gas inside the un-resolved kernel can remain bound \citep{colin:2013.star.cluster.rhd.cloud.destruction,2017MNRAS.472.4155G,2017MNRAS.471.4844G,2018ApJ...859...68K,grudic:2019.imf.sampling.fx.on.gmc.destruction}. Finally, we account for the fact that if $M_{\rm acc} \gg M_{\rm BH}$ the newly-accreted reservoir gas would fragment and form stars \citep[e.g.][]{goodman:qso.disk.selfgrav}, so we impose an upper limit $M_{\rm acc} \le 10\,M_{\rm BH}$ and do not allow accretion into the reservoir to exceed this, but this has weak effects.

{\em Accretion onto the BH:} Motivated by a simple analytic \citet{shakurasunyaev73} $\alpha$-disk model, the BH then accretes from the reservoir at a rate 
\begin{align}
\dot{M}_{\rm BH} &\equiv \frac{M_{\rm acc}}{t_{\rm acc}} \\ 
t_{\rm acc} &\equiv 42\,{\rm Myr}\,\left[1 + \frac{M_{\rm BH}}{M_{\rm acc}} \right]^{0.4}
\end{align}
We have tested making $t_{\rm acc}$ a factor $\sim 10$ shorter, or changing the power-law index from $0.4$ to $0$ or $1$, which have small effects (although changing the normalization by large factors does somewhat regulate the ``burstiness'' of feedback, see e.g.\ \citealt{rosas-guevara:fx.of.viscous.time.cosmo.bh.sims}); the salient feature is that $\dot{M}_{\rm BH}$ becomes comparable to the Eddington limit when $M_{\rm acc} \sim M_{\rm BH}$, and smaller when $M_{\rm acc}$ is very small. Because this expression naturally asymptotes at around the Eddington limit for a massive disk, we find it makes no difference to our results if we impose a strict Eddington limit to accretion (and therefore opt not to do so). 

{\em Angular Momentum:} We track the total angular momentum ${\bf J}_{\rm S}$ of each sink (mass $M_{\rm S}=M_{\rm BH}+M_{\rm acc}$). When the sinks lose mass (to e.g.\ outflows or radiation), lacking information about detailed angular momentum transport within the sink, we simply assume the mass lost carries the same specific angular momentum, i.e.\ ${\bf j}_{\rm S} = {\bf J}_{\rm S} / M_{\rm S}$ is constant. When two sinks are merged, ${\bf J}_{\rm S}$ is added directly, along with the total angular momentum of the pair of sinks in the center-of-mass frame (exactly conserving total ${\bf J}$ following \citealt{hubber:2013.res.criteria.in.star.clusters.strong.scattering.effects}). Because the accretion rate $\dot{M}_{\rm acc}$ is calculated continuously from the integral/mean gas properties within the BH kernel, we update ${\bf J}_{\rm S} \rightarrow {\bf J}_{\rm S} + (\dot{M}_{\rm acc}\,\Delta t)\,\langle {\bf j}_{\rm kernel} \rangle$ from accretion continuously assuming the accreted gas has the same mean specific angular momentum of the gas in the kernel $\langle {\bf j}_{\rm kernel} \rangle$.

We refer to \citet{wellons:prep} for an extensive survey of accretion models, but simply note as reviewed therein that our default model is both well-motivated physically by basic physical arguments (e.g.\ angular momentum transfer physics), but it also provides a much better match to small-scale simulations compared to e.g.\ Bondi-Hoyle scalings (which can be incorrect by $\sim 4-8$ orders of magnitude in regimes where they do not apply),\footnote{It is worth noting that our favored accretion model {\em does} reduce, up to the usual order-unity-prefactor ambiguity, to the Bondi-Hoyle accretion rate in the limit where Bondi-Hoyle accretion would actually be the correct sub-grid accretion model: namely when the Bondi radius is resolved, the BH dominates the gravity (over gas+stars+dark matter) on resolved scales and gas self-gravity is negligible, the gas is homogeneous, and gas has negligible angular momentum. But our accretion rate estimator is also valid in limits where e.g.\ gas is in a cold, self-gravitating disk, a situation in which the Bondi-Hoyle estimator fails catastrophically.} as well as BH-host galaxy scaling relations.

\begin{figure}
	\includegraphics[width=0.98\columnwidth]{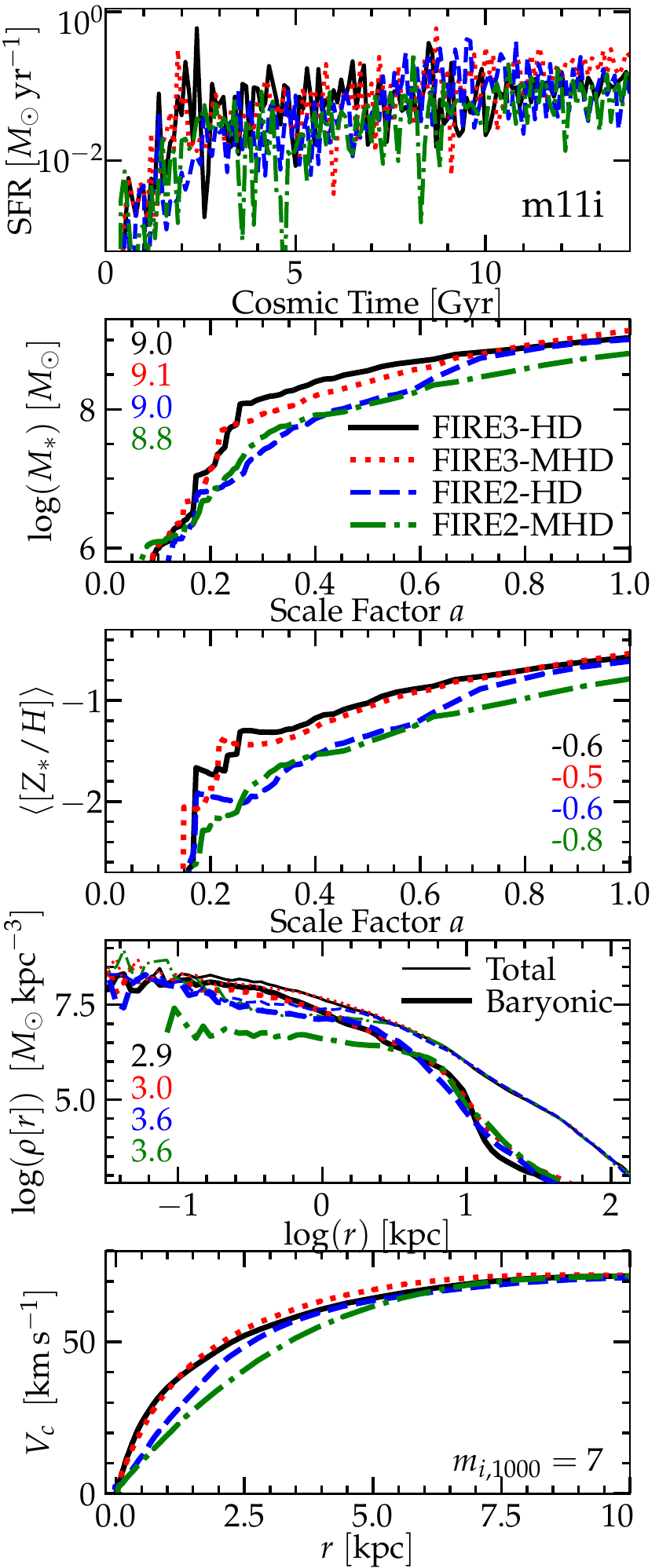}
	\vspace{-0.1cm}
	\caption{Comparison of a fully-cosmological \firetwo\ (with MHD active, or pure hydrodynamic [HD]) and \firex\ simulations of an LMC-mass dwarf galaxy ({\bf m11i}) from $z\sim100$ to $z=0$ (see \S~\ref{sec:galaxy.compare}). 
	{\em Top:} Archeological SFR vs.\ cosmic time (averaged in $\sim100\,$Myr intervals).
	{\em Second:} Stellar mass vs.\ scale factor $a$. Numbers give the value $\log{(M_{\ast}/M_{\odot})}$ at $z=0$.
	{\em Third:} Mean stellar metallicity vs.\ $a$. Numbers give $\langle [ Z_{\ast}/H ] \rangle$ at $z=0$.
	{\em Fourth:} Spherically-averaged radial density profile of all mass (including dark-matter; {\em thin}) and baryonic ({\em thick}) at $z=0$, versus galacto-centric radius $r$. Numbers give the stellar effective radius $R_{\rm eff,\,\ast}$ in kpc at $z=0$.
	{\em Bottom:} Circular velocity ($V_{\rm c}^{2} \equiv G\,M_{\rm tot}(<r)/r$) profile. 
	In \firetwo\ and \firex, MHD effects are weak. \firex\ features slightly more early (high-$z$) star formation in small dwarf progenitors, compared to \firetwo, owing to improvements in the UVB, metal-free cooling physics, star formation criteria, SNe treatment, and stellar isochrones (see \S~\ref{sec:galaxy.compare}). 
	\label{fig:galaxy.compare}}
\end{figure}

\subsubsection{Feedback}
\label{sec:methods:bhs:feedback}

{\bf Radiative Feedback:} We assume the accretion disk emits a bolometric luminosity 
\begin{align}
\dot{E}_{\rm rad}^{\rm BH} &\equiv L \equiv \epsilon^{\rm BH}_{r}\,\dot{M}_{\rm BH}\,c^{2}
\end{align}
with $\epsilon^{\rm BH}_{r}=0.1$ by default. This is injected at the BH location, then the radiation transport and its effects are numerically treated with the same RHD method as the stellar radiative feedback \citep{hopkins:radiation.methods}, assuming the template from \citet{shen:bolometric.qlf.update} for an un-obscured AGN spectrum at wavelengths from radio through $\gamma$-rays, so it self-consistently couples to dust (accounting for sublimation above dust temperatures $T_{\rm dust} > T_{\rm sub} \approx 2000\,$K), photo-electric, photo-ionization, Lyman-Werner, and Compton heating mechanisms,\footnote{In our default LEBRON RHD method, photo-ionization essentially follows a Stromgren approximation around the BH with (for our spectral template) $\dot{N}_{\rm ion} = 5.5\times10^{54}\,{\rm s^{-1}}\,(L_{\rm bol}/10^{45}\,\rm{erg\,s^{-1}})$. Photo-electric heating is evolved but only important for shifting gas from CNM to WNM (temperatures $<10^{4}\,$K) in a narrow range around the BH before the UV flux is fully-absorbed by dust. For the Compton terms we calculate an obscuration-independent Compton temperature $\approx 4\times10^{7}\,$K \citep[see also][]{sazonov04:qso.radiative.heating} with appropriate heating/cooling rates as a function of the local radiation energy density \citep{cafg:2012.egy.cons.bal.winds}. For radiation pressure, we first calculate the short-range single-scattering absorption as $L^{0}_{\rm abs} = (1-e^{-\langle\tau_{\nu} \rangle})\,L$ with $\langle \tau_{\nu} \rangle$ evaluated from the flux-mean opacity in the kernel around the emitting particle, before propagating the un-absorbed and re-radiated luminosity. Given our resolution limitations we make the conservative assumption that single-scattering occurs within this kernel and only re-radiated IR ($L_{\rm IR}=L^{0}_{\rm abs}$) is allowed to exert radiation pressure outside the kernel (acceleration ${\bf a} = \kappa\,{\bf F}/c = \kappa_{\rm IR}\,L_{\rm IR}\,\hat{\bf r} / (4\pi\,c\,r^{2})$).} as well as single and multiple-scattering radiation pressure (with the total coupled photon momentum flux usually $\sim 1\, L/c$). The fraction of the mass converted into radiation ($=[\epsilon^{\rm BH}_{r}/(1-\epsilon^{\rm BH}_{r})]\,\dot{M}_{\rm BH}$) is removed from the reservoir. 

{\bf Mechanical Feedback (Non-Relativistic):} Non-relativistic mechanical outflows from the accretion disk are treated with the numerical method described in detail in \citet{grudic:starforge.methods} (see also \citealt{su:2021.agn.jet.params.vs.quenching,wellons:prep}). We assume an outflow with mass-loading 
\begin{align}
\dot{M}_{\rm w,\,BH} = \dot{M}_{\rm BH}
\end{align} 
is launched from the disk, with an at-launch velocity 
\begin{align} 
v_{\rm w} = v_{3000}\,3000\,{\rm km\,s^{-1}},
\end{align} 
giving a kinetic energy-loading
\begin{align}
\dot{E}_{\rm w}^{\rm BH} &\equiv \epsilon_{\rm w}^{\rm BH}\,\dot{M}_{\rm BH}\,c^{2} \approx 5\times10^{-5}\,v_{3000}^{2}\,\dot{M}_{\rm BH}\,c^{2}
\end{align}
corresponding momentum-loading $\sim 0.1\,v_{3000}\,L/c$, and a much smaller specific internal thermal ($T_{\rm w} \sim 10^{4}\,$K) and magnetic (${\bf B}_{\rm w}$ set to the local ambient ISM value) energy.\footnote{Metallicity and other ``passive'' scalar fields are also taken to have ambient ISM values.} The mass $\dot{M}_{\rm w,\,BH}$ is removed from the reservoir and injected onto the domain via the creation or spawning of new hyper-resolved wind-cells which are created around the BH with the desired mass, energy, velocity, etc. These hyper-cells have an initial mass resolution $\sim1000$ times better than the typical gas cell mass $\Delta m_{0}$ in our simulations ($\Delta m_{\rm hyper}\approx 0.001\,\Delta m_{0}$), designed to ensure that phenomena such as diffuse jet/wind cavities and reverse shocks can be resolved; they are de-refined when they mix fully with ambient gas. The outflows are perfectly-collimated at injection, with positions and velocities following along the bipolar axis ${\bf j}_{\rm S}$ of the net specific angular momentum of the BH, spawned symmetrically (in pairs with at most one pair per timestep) to ensure manifest linear momentum conservation.

{\bf Jets (Relativistic Mechanical Feedback):} In simulations which explicitly evolve CR dynamics, relativistic jets are treated as in \citet{su:2021.agn.jet.params.vs.quenching} by injecting CRs in an infinitely-collimated jet, attached to the spawned mechanical feedback hyper-cells. The injection properties (e.g.\ spectral shapes) are, for simplicity, taken to be identical to acceleration in SNe shocks (\S~\ref{sec:crs}), except the CR acceleration efficiency is given by 
\begin{align}
\dot{E}_{\rm cr}^{\rm BH} &\equiv \epsilon^{\rm BH}_{\rm cr}\,\dot{M}_{\rm BH}\,c^{2}
\end{align}
 with (by default) $\epsilon^{\rm BH}_{\rm cr} \sim 10^{-3}$. This corresponds to $\sim 0.2-2\%$ of the kinetic energy of accreting gas at the ISCO, depending on BH spin (as compared to e.g.\ the canonical $\sim 10\%$ of SNe shock energy which goes into accelerating CRs). Again, the rest mass converted into energy is removed from the reservoir.

\begin{figure}
	\includegraphics[width=0.98\columnwidth]{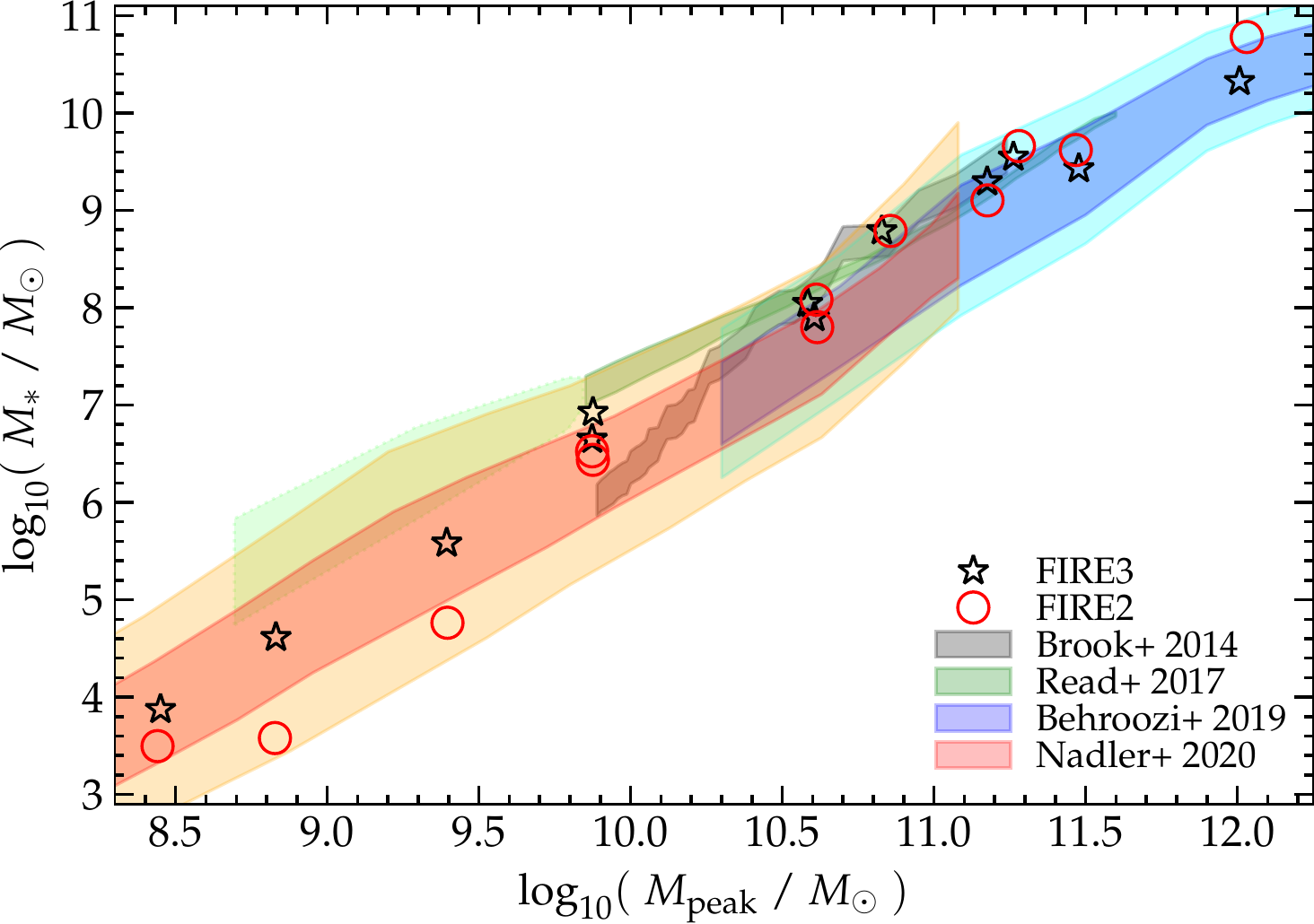}
	\vspace{-0.1cm}
	\caption{Comparison of the $z=0$ stellar masses of a subset of initial \firex\ simulations compared to their identical counterparts run with the \firetwo\ models, as a function of peak halo mass (only centrals plotted, so this is present virial mass). We compare observational constraints from abundance matching, lensing, \&\ kinematics (shaded). \citet{brook:2014.mgal.mhalo.local.group} from the local group, and \citet{read:2017.shallow.mgal.mhalo} from local field (lighter region is extrapolated) do not consider scatter in $M_{\ast}(M_{\rm halo})$, so shaded range shows the uncertainty in their mean relation. \citet{nadler:2020.abundance.matching.w.lmc.fx} and \citet{behroozi:2019.sham.update} explicitly model scatter so we show the $\pm1\,\sigma$ (darker) and $\pm2\,\sigma$ (lighter) regime. Note the absolute values here can be significantly altered by other physics (e.g.\ cosmic rays, AGN). But consistent with Fig.~\ref{fig:galaxy.compare}, we see very similar masses at intermediate/high halo masses, while for small dwarfs with [Z/H]$\lesssim -2$ we somewhat enhanced star formation in \firex.
	\label{fig:masses}}
\end{figure}

\section{Illustrative Example of The Effects on Galaxy Formation}
\label{sec:galaxy.compare}

Fig.~\ref{fig:galaxy.compare} shows a preliminary example of the combined effect of all of the updates to the ``default'' \firex\ model synthesized in this paper on a fully-cosmological simulation. Specifically, we simulate the halo {\bf m11i} from \citet{hopkins:fire2.methods,hopkins:cr.mhd.fire2,elbadry:fire.morph.momentum}, which was well-studied in \firetwo, with the default or ``base'' version of the \firex\ code (e.g.\ not including explicit CR dynamics or BHs), from its cosmological initial condition at $z\sim 100$ to $z=0$ (with mass resolution $\sim 7000\,\msun$). This is a dwarf galaxy with a final stellar mass broadly similar to e.g.\ the LMC, and a convenient test owing to (a) relatively low computational cost and (b) being intermediate-mass (low-enough mass that e.g.\ metallicity-dependence and UVB effects can be important, but high-enough mass that it has sustained star formation to $z=0$ and non-trivial gas phase structure). 

We stress that it is not our intention in this paper to present a detailed survey of the effects of the updated \firex\ physics on all galaxy formation properties, halo masses, etc., nor is it to present a new ``suite'' of simulations -- rather it is to clearly document the changes in the code from \firetwo, and provide useful fitting functions for all of the various terms (updated cooling functions, stellar evolution tabulations, yields, etc.) for use by the community. Nonetheless it is useful to see at least one example of how the collective effect of these changes influences the formation of a galaxy. 

We compare both the MHD and hydrodynamic (HD) versions of the same galaxy. We can see that in \firex, like \firetwo, MHD makes little difference to global galaxy properties \citep{su:2016.weak.mhd.cond.visc.turbdiff.fx,hopkins:cr.mhd.fire2}. The clear systematic difference is that \firex\ robustly produces somewhat stronger star formation at early times (high redshifts) in small, low-metallicity dwarf galaxy progenitors. This owes to a combination of effects: later reionization, more accurate accounting for metal-free atomic+molecular cooling, weaker ``early'' stellar mass-loss feedback at low metallicities, the more accurately total-energy conserving SNe treatment (which limits the coupled SNe momentum in super-bubble situations with strong outflows), and less restrictive star formation criteria, in \firex. We have tested swapping each of these ``pieces'' of \firex\ with their \firetwo\ analogues and find that no single one explains the entire difference, but each contributes comparably, with the improved SNe treatment playing the largest individual role (owing to the scaling of SNe terminal momentum with density and metallicity, the effects of the improved algorithm are strongest in extremely low-mass, low-metallicity systems), followed by the more accurate treatment of cooling in the extremely metal-poor ($\Hmol$-dominated) limit. The enhanced earlier star formation in \firex\ leads to slightly larger stellar masses by $z=0$, and slightly higher central galaxy densities and circular velocities, but the effect by $z=0$ is modest (a factor $\sim 1.5$). The SFRs are actually quite similar between \firex\ and \firetwo\ after  $z\lesssim 2-3$, when the \firetwo\ galaxy has reached a metallicity $Z \sim 0.01\,Z_{\odot}$, which is well after reionization, and high enough metallicity that metal-line cooling dominates over molecular, minimizing the differences between the cooling physics and also making it much easier for both runs to cool to $\sim 10\,$K (minimizing the differences between the SF criteria, as the simulations can more easily meet the Jeans criteria).

In Fig.~\ref{fig:masses}, we compare the $z=0$ stellar masses of our galaxies with their peak ($z=0$ virial) halo masses, in an initial subset of a few runs. We only plot the one or two most-massive (best-resolved) central galaxies in the high-resolution region, for the volumes {\bf m09}, {\bf m10v}, {\bf m10q}, {\bf m11a},  {\bf m11d}, {\bf m11f}, {\bf m11h}, {\bf m11i}, {\bf m11q}, {\bf m12i}, from our \firetwo\ study in \citet{hopkins:fire2.methods}, using identical definitions and ICs as therein. These have a mass resolution $\approx 250\,M_{\odot}$ for all the halos with masses $\lesssim 2\times10^{10}\,M_{\odot}$ and $\approx 7000\,M_{\odot}$ at higher masses. We compare to various observational estimates of the stellar mass-halo mass relation at $z\sim0$, for reference. Our intention here is not to compare with observations in detail (which requires a larger sample where we can analyze scatter and evolution in these relations, separating satellites vs.\ centrals, etc., and can depend systematically on other physics not included in these runs but described above, such as cosmic rays and AGN), but to identify any systematic trend in \firetwo\ versus \firex\ results. Consistent with Fig.~\ref{fig:galaxy.compare}, we see that at the lowest $z\sim 0$ stellar masses (and metallicities), $\lesssim 10^{6}-10^{7}\,M_{\odot}$, the \firex\ simulations produce systematically higher stellar masses by up to a factor of a few. The differences owe to the same changes producing the higher SFRs/masses at early times in Fig.~\ref{fig:galaxy.compare} (when that galaxy is also lower-mass). Naively, the increase actually appears to bring the \firex\ galaxies into somewhat better agreement with observationally inferred stellar masses, and is still smaller than the difference between various observationally derived $M_{\ast}-M_{\rm halo}$ relations at the lowest (ultra-faint) masses (compare e.g.\ \citealt{read:2017.shallow.mgal.mhalo} and  \citealt{nadler:2020.abundance.matching.w.lmc.fx}). 

Above intermediate $z\sim0$ stellar masses $\gtrsim 10^{6}-10^{7}\,M_{\odot}$, there is little difference in the $z\sim0$ stellar masses, until we reach the most massive ($\sim10^{12}\,M_{\odot}$) halos, where the stellar masses decrease by up to a factor of $\sim 2$ (again in slightly better agreement with observations). Preliminary low-resolution simulations of other Milky Way-mass halos suggest this result is robust.\footnote{We have confirmed that the change in stellar masses at $M_{\rm halo} \gtrsim 10^{12}\,M_{\odot}$ owes primarily to the explicit accounting for the gas relative velocities around SNe in \firex\ (\S~\ref{sec:appendix:sne}). Specifically, when there is net {\em inflow} towards a region with SNe ($\beta<0$, in Eqs.~\ref{eqn:beta.E} \&\ \ref{eqn:beta.C}) and the inflow speed becomes comparable to the shell velocity at which a SNe remnant in a stationary medium would become radiative ($\sim 200\,{\rm km\,s^{-1}}$, which can occur in the central regions of massive galaxies), then the SNe shock energy and temperature are enhanced by this relative velocity and the onset of the cooling/snowplow phase is delayed, leading to somewhat more efficient stellar feedback \citep[see e.g.][]{ostrikermckee:blastwaves}.} At masses $\gg 10^{12}\,M_{\odot}$ we have not yet run any high-resolution \firex\ simulations to $z\sim0$ but preliminary low-resolution tests confirm the expected behavior from \firetwo\ \citep[e.g.][]{wellons20_rotation,parsotan:2021.mock.fire.profiles.z2}, namely that although the stellar masses are somewhat reduced in \firex\ versus \firetwo, simulations without AGN feedback such as those here invariably produce excessively large central stellar densities, velocity dispersions/circular velocities, and stellar masses at halo masses $\gg 10^{12}\,M_{\odot}$. 

In summary, the net effect of this is that (absent AGN and CRs) \firex\ produces a slightly ``flatter'' $M_{\ast}-M_{\rm halo}$ relation below $\sim 10^{12}\,M_{\odot}$, with $M_{\ast} \propto M_{\rm halo}^{2.1}$ in \firetwo\ and $M_{\ast} \propto M_{\rm halo}^{1.8}$ in \firex\ (with the same normalization around $M_{\rm halo} \sim 10^{10.5}-10^{11}\,M_{\odot}$). The offset is small, but appreciable over a sufficiently large dynamic range in $M_{\rm halo}$. We have also validated the arguments above by running a limited series of low-resolution experiments where we adopt \firex\ physics but arbitrarily multiply the SNe energies by a factor $\sim (0.1\,Z_{\odot}/Z)^{0.3}$ (i.e.\ making feedback slightly stronger at low metallicities, and slightly weaker at high metallicities), which we find eliminates most of the differences in both stellar masses and star formation histories between \firex\ and \firetwo.

\section{Conclusions}
\label{sec:conclusions}

We have presented an extensive update of the treatment of stellar evolution physics and nucleosynthetic yields, as well as low-temperature/high-density ISM cooling and chemistry, which together constitute the \firex\ version of the FIRE project. We present all of the changes from \firetwo\ to \firex, and insofar as possible, provide detailed and accurate fitting functions for all quantities in a form which can be directly implemented in other galaxy formation models, and synthesize our survey of state-of-the-art updates to e.g.\ stellar evolution tracks.

This paper has two purposes. First, to serve as a reference for future \firex\ studies, which will use these updated models to study a wide variety of problems in galaxy formation, as with \firetwo. Future work will study in detail how the updated stellar models alter galaxy formation (if at all), although as we noted above, most of the qualitative conclusions of \firetwo\ are unlikely to change, as the updates here for many of the ``most important'' quantities are minor. Integrated SNe rates and energetics, stellar total luminosities, and integrated total metal yields change by at most tens of percent, while quantities like the detailed cooling and chemistry of dense, cold ($<10^{4}\,$K) ISM gas are largely irrelevant to ISM dynamics in metal-rich galaxies (like the Milky Way) so long as the cooling times are shorter than dynamical times \citep[see e.g.][]{hopkins:fire2.methods}, although they can be important at early times or in dwarf galaxies at low metallicities. But the updated models here are still important both (a) because they represent more accurately our best attempt to represent the state-of-the-art inputs from stellar evolution and nucleosynthetic yields, so we can ``anchor'' the galaxy formation models as best as possible in what we believe we know about stars, and (b) because they enable improved predictive power for observational tracers and diagnostics (e.g.\ CO or CII emission, or detailed stellar abundance patterns within galaxies). 

Our second purpose is to make public these updated inputs in a convenient form which can be used easily and directly in other codes, to improve the accuracy with which these physics are treated in other non-FIRE simulations as well. It is worth noting that almost all of the fitting functions here -- especially those for stellar evolution and yields -- are new and have not appeared previously in the literature. Most involve considerable effort to synthesize multiple sources and identify the most robust/stable trends, vetting to remove erroneous or extraneous extrapolations (in order to cover the entire dynamic range of interest accurately), and additional modeling to translate them into a convenient form for galaxy simulations (see e.g.\ \S~\ref{sec:yields}). 

We stress that in no way do we intend to imply that the physics here are ``complete'' or ``final.'' In future work, we will continue to explore the role of other physics not part of the ``default'' or ``base'' \firex\ model, e.g.\ cosmic rays (\S~\ref{sec:crs}) and supermassive black holes (\S~\ref{sec:bhs}). Meanwhile, stellar astrophysics remains an incredibly active and rapidly-evolving field, with an exponentially-growing amount of new data from massive spectroscopic and astrometric and time-domain surveys as well as new probes such as gravitational waves. It is therefore almost certain that the input stellar evolution and nucleosynthetic models here will continue to evolve: our goal here is update our galactic models to ``catch up'' with advances in our understanding of stellar evolution from the last two decades, with the intention of future updates reflecting the state-of-the-art in the field.

\acknowledgments{
We thank Alessandro Lupi for helpful suggestions, and the many collaborators who have contributed to the development of the FIRE project. Support for PFH was provided by NSF Research Grants 1911233 \&\ 20009234, NSF CAREER grant 1455342, NASA grants 80NSSC18K0562, HST-AR-15800.001-A. Numerical calculations were run on the Caltech compute cluster ``Wheeler,'' allocations AST20016, AST21010 \&\ TG-AST140023 supported by the NSF and TACC, and NASA HEC SMD-16-7592.
AW received support from:NSF grants CAREER 2045928 and 2107772; NASA ATP grants 80NSSC18K1097 and 80NSSC20K0513; HST grants GO-14734, AR-15057, AR-15809, and GO-15902 from the Space Telescope Science Institute (STScI), which is operated by the Association of Universities for Research in Astronomy, Inc., for NASA, under contract NAS5-26555; a Scialog Award from the Heising-Simons Foundation; and a Hellman Fellowship.
Support for MYG was provided by NASA through the NASA Hubble Fellowship grant \#HST-HF2-51479 awarded  by  STScI.
MBK acknowledges support from NSF CAREER award AST-1752913, NSF grants AST-1910346 and AST-2108962, NASA grant NNX17AG29G, and HST-AR-15006, HST-AR-15809, HST-GO-15658, HST-GO-15901, HST-GO-15902, HST-AR-16159, and HST-GO-16226 from STScI.
CAFG was supported by NSF through grants AST-1715216 and CAREER award AST-1652522; by NASA through grant 17-ATP17-0067; by STScI through grant HST-AR-16124.001-A; and by the Research Corporation for Science Advancement through a Cottrell Scholar Award and a Scialog Award.
We ran simulations using: the Extreme Science and Engineering Discovery Environment (XSEDE), supported by NSF grant ACI-1548562; Frontera allocations AST21010 and AST20016, supported by the NSF and TACC; Pleiades, via the NASA HEC program through the NAS Division at Ames Research Center.
}

\datastatement{The data supporting the plots within this article are available on reasonable request to the corresponding author. A public version of the GIZMO code is available at \gizmourl. FIRE data releases are publicly available at \FIREpublicurl.} 

\bibliography{ms_extracted}

\clearpage

\begin{appendix}

\section{Comparison of SF Criteria and Sub-Grid Cooling Physics}
\label{sec:alternatives}

\subsection{Effects of Different SF Criteria}
\label{sec:alternatives:sf}

\begin{figure}
	\includegraphics[width=0.98\columnwidth]{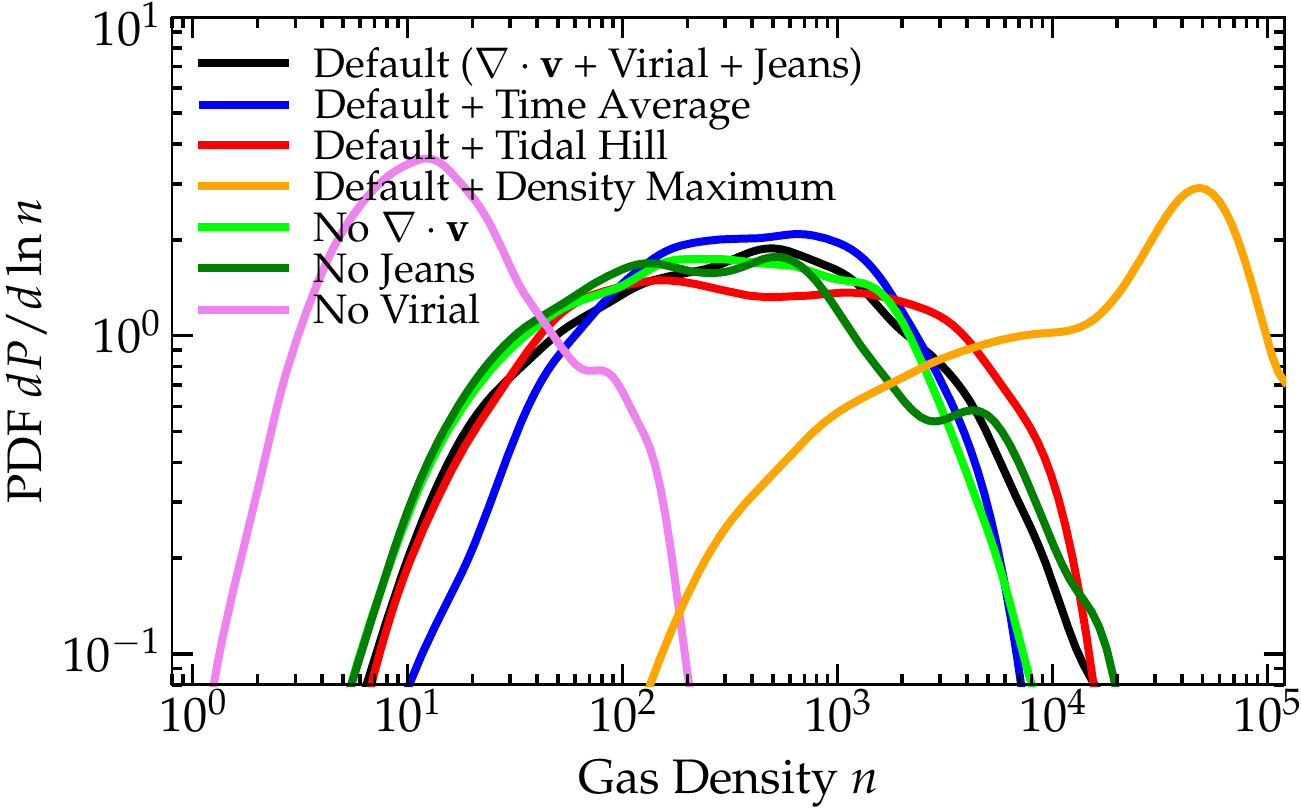}
	\vspace{-0.1cm}
	\caption{PDF of star-forming gas (weighted by SFR), as Fig.~\ref{fig:sfr}, in otherwise identical \firex\ simulations with different SF criteria. We compare our default criteria, which include a converging flow ($\nabla\cdot {\bf v}$), locally self-gravitating (virial), and thermally-unstable/fragmenting to stellar scales (Jeans) criteria (\S~\ref{sec:sf.criteria}), to runs which additionally time-average the virial criterion, add a tidal Hill  (gravitational forces converge along all axes) or density maximum criteria, or runs which remove each criteria (\S~\ref{sec:alternatives:sf}). Tidal hill, converging flow, and Jeans criteria are secondary. The virial criteria dominates of our default. The density maximum produces artificial collapse of entire GMCs around their density peak, rather than allowing fragmentation: the densities run away until we force them to stop (in the run here, we simply turn everything at $ \gtrsim 10^{5}\,{\rm cm^{-3}}$ into stars immediately, to prevent the timesteps from becoming too small).  
	\label{fig:sf.crit}}
\end{figure}

Briefly, we test how different SF criteria influence our \firex\ results.
We first note that our virial criterion, based on the local velocity field, can change rapidly (on of order a few timesteps), so we consider a model following \citet{grudic:sfe.gmcs.vs.obs,guszejnov:2020.mhd.turb.isothermal.imf.cannot.solve} where we replace the ``instantaneous'' virial parameter $\alpha_{i}$ with a rolling time-average $\bar{\alpha}_{i}(t)$ defined by $d\bar{\alpha}_{i}/dt = (\alpha_{i}-\bar{\alpha}_{i})\,[(1+\bar{\alpha}_{i})/(1+\alpha_{i})]\,\tau^{-1}$, where $\tau\equiv t_{\rm ff,\,i}/8$ is shorter than the local free-fall time $t_{\rm ff,\,i}$ (so that the indicator $\bar{\alpha}_{i}$ can adjust before collapse to infinite density occurs). This has only a small effect on the distribution of star-forming densities, slightly reducing star formation in low-density gas owing to the ``time lag'' of a few $\tau$ before $\bar{\alpha}_{i}$ can decay to match $\alpha_{i}$ when e.g.\ gas cools rapidly. In some very rare circumstances, however, we find this lag can become comparable to the free-fall time, artificially suppressing star formation until collapse proceeds to extremely high densities. 

We next consider two additional SF criteria. First, a ``tidal Hill'' criterion as implemented in \citet{grudic:starforge.methods}: specifically requiring that the tidal tensor $\mathbb{T}({\bf x}) \equiv \nabla \otimes {\bf g}({\bf x})$ (where ${\bf g}({\bf x}) \equiv \nabla \Phi_{\rm grav}({\bf x})$ is the gravitational acceleration field at position ${\bf x}$) has three negative eigenvalues. This physically means that the gravitational tidal field is compressive along all three directions, so we can think of it as a stricter version of our convergent flow ($\nabla \cdot {\bf v}$) criterion. But we see in Fig.~\ref{fig:sf.crit} this has little effect: this is because essentially all the gas that meets our virial criterion must also meet this tidal Hill criterion, and the virial criterion is more strict because it allows for non-gravitational forces resisting collapse. Thus although this can catch some special edge cases important for single-star formation simulations like those in \citet{grudic:starforge.methods}, we find it is redundant in \firex, and do not include it by default. 

Second, we consider a ``density maximum'' criterion -- requiring that a cell be the highest-density of all its interacting neighbors (all cells which could interact hydrodynamically with it). This can lead to unphysical behaviors shown in Fig.~\ref{fig:sf.crit}. What occurs is the following: in a marginally-resolved GMC, this criterion essentially means that only the global density maximum of the entire GMC is eligible to form stars. But recall, the SF timescale we impose is the free-fall time, which of course is (by definition) also the same timescale that an isothermal self-gravitating, collapsing cloud (required by our virial and $\nabla \cdot {\bf v}$ criteria) requires to collapse to formally infinite density. So the marginally-resolved clouds, instead of fragmenting, collapse to essentially arbitrarily high densities -- the only reason the star-forming gas density PDF cuts off in Fig.~\ref{fig:sf.crit} at $\sim 10^{5}\,{\rm cm^{-3}}$ is because we impose a purely-numerical upper limit at which the gas is immediately turned into stars, to avoid infinitely-small timesteps. Thus, this criteria, while sensible for simulations which resolve {\em individual} proto-stars in a sink-particle-like manner (requiring  that one resolve the thermal Jeans mass of cold molecular gas, i.e.\ gas cell masses of $\sim 10^{-3}\,M_{\odot}$), does {\em not} make physical sense for galaxy-scale simulations where a ``star particle'' represents either more than one star or indeed any kind of sub-fragmentation. We find similar results for other common sink-particle criteria used in single-star formation simulations (e.g.\ requiring that no other sinks be within the interacting neighbor set, or requiring the cell be the potential minimum instead of density maximum). 

Next, we compare turning off each of our default SF criteria in turn. Per \S~\ref{sec:sf.criteria}, the converging flow criterion has little effect. And at least for a $z=0$, Milky Way metallicity galaxy, the Jeans criterion is also secondary, though this can be more important for metal-poor galaxies where cooling to very low temperatures is less efficient (see \S~\ref{sec:galaxy.compare} and discussion below). The virial criterion is clearly the most important of our adopted SF criteria: removing it, the peak of SF moves to $\sim 10\,{\rm cm^{-3}}$ gas, which is the minimum density that can meet our Jeans criterion at the $\sim 100\,$K CNM temperatures, but this is primarily non-self-gravitating diffuse HI (especially in the presence of magnetic fields, where it has a plasma $\beta \ll 1$). Of course, if we remove both the Jeans and virial criteria (leaving only converging-flow), then we see un-physical star formation even in hot gas at very low densities. 

\begin{figure}
	\includegraphics[width=0.99\columnwidth]{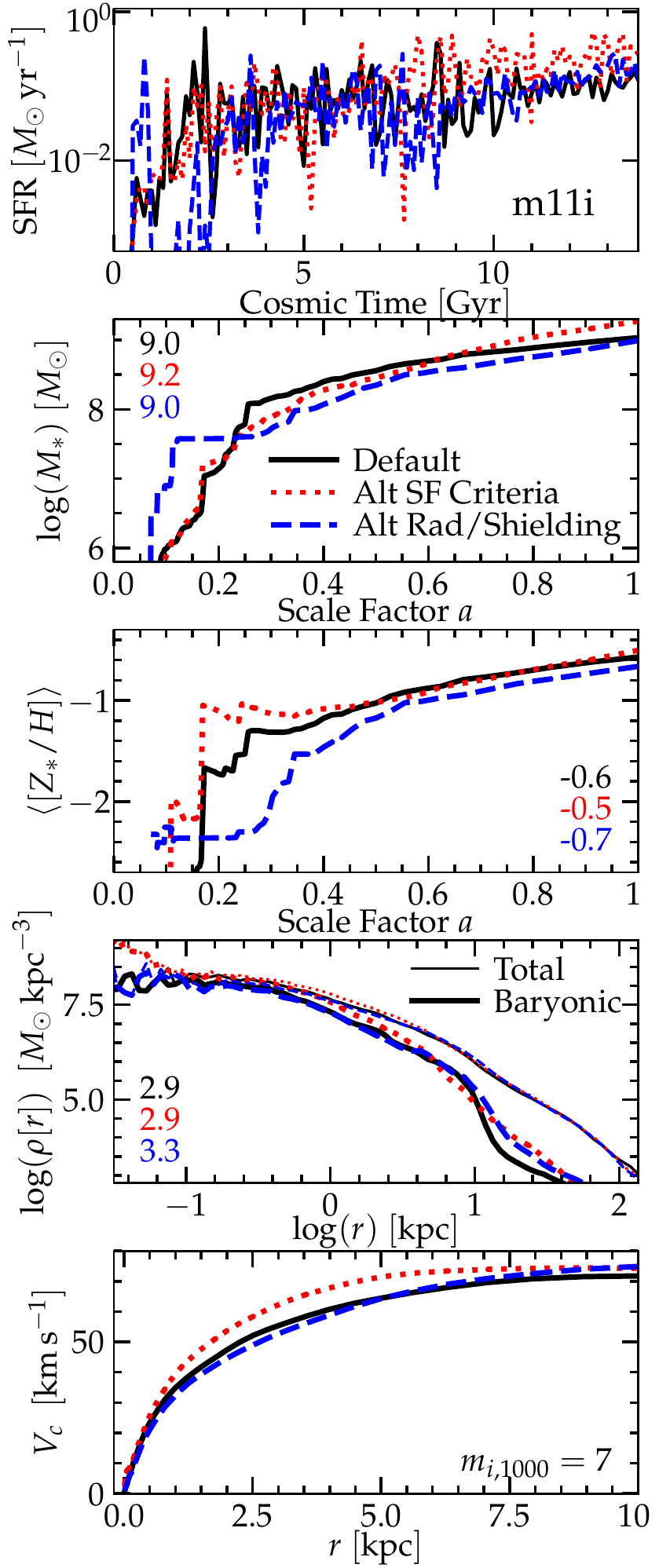}
	\vspace{-0.2cm}
	\caption{Comparison of the fully-cosmological {\bf m11i} simulation as Fig.~\ref{fig:galaxy.compare} (\S~\ref{sec:galaxy.compare}), but testing two variant \firex\ models. The ``Alt SF Criteria'' model adds the tidal Hill and density maximum criterion as Fig.~\ref{fig:sf.crit}, with a maximum density enforced: despite this artificially forcing SF into much more dense clumps, the global galactic SF history is similar, as it is feedback-regulated. The ``Alt Rad/Shielding'' criterion both takes $C_{m}=1$ (as Fig.~\ref{fig:phase.clumping}) and assumes the incident Lyman-Werner background (before line self-shielding) is always the Habing/Milky Way value (instead of using the in-code dynamically-estimated radiation field). This makes cooling less efficient at low metallicity, leading to somewhat more ``bursty'' SF as gas must clump further to cool.
	\label{fig:galaxy.compare.altsfcool}}
\end{figure}

Fig.~\ref{fig:galaxy.compare.altsfcool} compares the effect of these additional SF criteria on a cosmological simulation as Fig.~\ref{fig:galaxy.compare}, specifically the ``density maximum'' criterion which produces artificial clumping to very high densities. Even in this case, the effect is modest, consistent with our \firetwo\ studies showing that {\em global} galactic SF properties were very insensitive to e.g.\ the density threshold or SF efficiency per dynamical time (see \citealt{hopkins:fire2.methods}). What does change is the resolution-scale structure of SF, in particular the density of e.g.\ star clusters which form, which are artificially over-dense if we enforce a criterion like the density maximum, as shown in \firetwo\ simulations in e.g.\ \citet{ma:2020.globular.form.highz.sims}.

Before moving on, it is important to note that while the density PDF in Fig.~\ref{fig:sf.crit} is insensitive to many variations in the SF criteria at fixed numerical resolution, it {\em does} vary with resolution, as it should. Up to an order-unity constant, our Jeans+virial criteria are equivalent to the statement that SF occurs when the ``effective'' Jeans length ($\lambda_{\rm J,\,eff} \sim \delta v_{\rm eff}/\sqrt{G\,\rho}$, where $\delta v_{\rm eff}$ is the total signal speed including e.g.\ thermal, turbulent, and magnetic/\Alf\ speeds) or Jeans mass ($M_{\rm J,\,eff} \sim (4\pi/3)\,(\lambda_{\rm J}/2)^{3}\,\rho$) becomes unresolved ($M_{\rm J,\,eff} \lesssim \Delta m_{i}$ or $\lambda_{\rm J,\,eff} \lesssim \Delta x_{i}$, where $\Delta m_{i}$ and $\Delta x_{i}$ are the cell mass/size), which occurs above some critical density $\rho_{\rm crit}(\Delta m_{i},\,\Delta x_{i},\,\delta v_{\rm eff},\,...)$. In other words, we convert gas cells to star particles when the fragmentation cascade can no longer be explicitly resolved (so the PDF of star-forming gas in Fig.~\ref{fig:sf.crit} shifts to higher densities at higher resolution). Assuming turbulent motions dominate $\delta v_{\rm eff}$ in a standard super-sonic cascade or observed ISM linewidth-size relation (expected on scales $\Delta x \gtrsim 0.1\,$pc; \citealt{bolatto:2008.gmc.properties}), then for a Lagrangian code like ours (fixed $\Delta m_{i}$ but variable $\Delta x_{i}$), the critical density scales as $\rho_{\rm crit} \propto \Delta m_{i}^{-1/2} \propto N_{\rm cells}^{-3/2}$, and this scaling is explicitly confirmed in FIRE resolution tests in Table~3 of \citealt{hopkins:fire2.methods}. If thermal or magnetic support dominates $\delta v_{\rm eff}$ (expected on smaller scales), then $\rho_{\rm crit} \propto \Delta m_{i}^{-2} \propto N_{\rm cells}^{-6}$ scales more rapidly (see \citealt{guszejnov:2020.mhd.turb.isothermal.imf.cannot.solve}). In either case, by not including a fixed density threshold, our SF criteria can automatically be appropriately adaptive to resolution.

\begin{figure}
	\includegraphics[width=0.99\columnwidth]{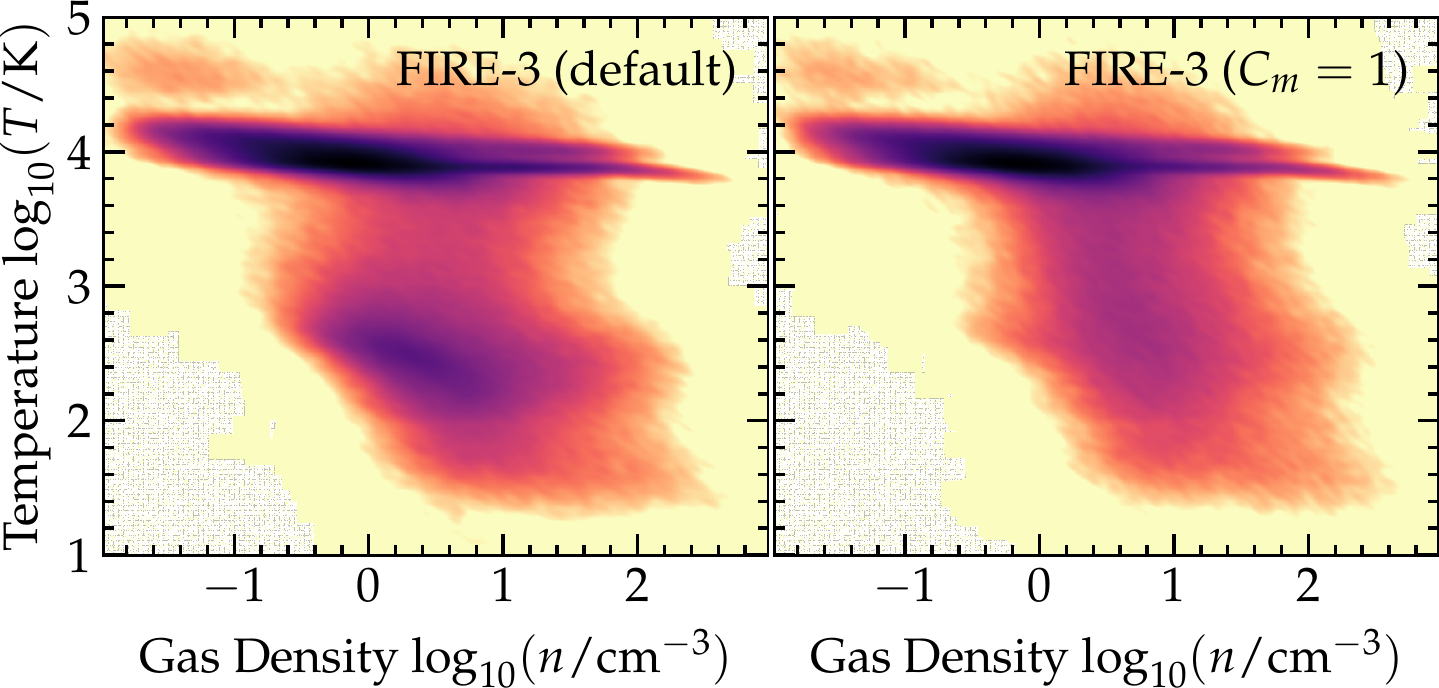}
	\vspace{-0.1cm}
	\caption{Phase diagram of ISM gas, as Fig.~\ref{fig:phase}, in otherwise identical \firex\ simulations with different treatments of the sub-resolution-scale clumping factor $C$ used for the molecular line self-shielding calculation; we compare our default ({\em left}; \S~\ref{sec:cooling}) to a run with no clumping ($C_{m}=1$ always). In the latter, cold gas does not appear until slightly higher resolved densities $n \gtrsim$\, a few ${\rm cm^{-3}}$, and the phase separation between CNM and WNM is ``smeared out.'' 
	\label{fig:phase.clumping}}
\end{figure}

\subsection{Molecular Cooling/Formation Physics} 
\label{sec:alternatives:mol}

Figs.~\ref{fig:phase.clumping} \&\ \ref{fig:galaxy.compare.altsfcool} consider the effect of variations to the cooling physics module, particularly in the shielding treatment, which (unlike e.g.\ the cooling rate functions) has non-trivial uncertainties. In Fig.~\ref{fig:phase.clumping}, we specifically compare our default treatment to a simulation where we assume that a clumping factor $C_{m}=1$ for all $m$, as defined in \S~\ref{sec:cooling} -- i.e.\ no clumping factor at all, compared to the standard clumping factor estimated for a lognormal density distribution in supersonic turbulence, which we use as our default. This is used exclusively in calculating the molecular formation/destruction rates, where larger $C_{m}$ promotes molecule formation  via collisional processes, and suppresses molecule destruction by enhancing the line self-shielding. Removing this clumping, we see as expected that cold gas cannot form until somewhat higher densities, when self-shielding without a clumping factor is more efficient -- but this is just a factor of a few shift in the density $n$ where $T \sim 100\,$K gas appears, corresponding to the expectation for $C_{m} \sim$\,a few. What is slightly more surprising is that this also suppresses the separation between CNM and WNM phases, making the results more similar to \firetwo. That occurs because, without accounting for clumping, line self-shielding being less efficient leads to a less-rapid phase transition in thermally-unstable neutral gas. In Fig.~\ref{fig:galaxy.compare.altsfcool}, we consider a variant model where we both remove clumping and assume the incident UV background is always the MW-like (Habing) value, ignoring the actual RHD in the code. Although this can significantly change the thermal state of the cold gas, it has surprisingly small effects on the bulk galaxy properties. The star formation is slightly more ``bursty,'' which suppresses the total mass and central density via producing more concentrated episodes of SNe feedback, as the assumption that the gas always sees an external Habing-level UVB means cold gas (required to meet our Jeans criterion) cannot form until very large resolved column densities build up, which would otherwise fragment earlier (leading, fore example, to the SFR actually ``overshooting'' the model with self-consistent radiation treatment in the initial starburst at early times). This type of behavior is qualitatively consistent with the results in recent RHD studies \citep{kannan:photoion.feedback.sims,rosdahl:2015.galaxies.shine.rad.hydro,kimm:lyman.alpha.rad.pressure,emerick:rad.fb.important.stromgren.ok,hopkins:radiation.methods}, arguing that accounting for local radiation tends to have a ``smoothing'' effect on star formation, reducing burstiness.

\section{Updated Mechanical Feedback Algorithm} 
\label{sec:appendix:sne}

\subsection{Expressions for Individual Coupling Events}
\label{sec:appendix:sne:one}

As detailed in the text, our numerical algorithm for coupling mechanical feedback (e.g.\ SNe) largely follows that derived in detail and extensively tested in \citet{hopkins:sne.methods}, Appendix~E,  with minor updates. We summarize the complete algorithm including these updates here. Consider an ``event'' where a star particle ``$a$'' (with coordinate position ${\bf x}_{a}$, velocity ${\bf v}_{a}$, total particle/cell mass $m_{a}$) injects, in a given timestep $\Delta t$ (taking the code from initial time $t^{(0)}  \rightarrow t^{(1)} = t^{(0)} + \Delta t$), some ejecta mass $m_{{\rm ej},\,a}$ (which can be spread across different explicitly-followed species ``$s$,'' as $m_{{\rm ej},\,a} \equiv \sum_{s} m_{{\rm ej},\,a,\,s}$). In the rest-frame of the star particle, assume the ejecta is isotropic with total kinetic energy ${\rm KE}_{{\rm ej},\,a}$ (and some arbitrary set of scalar energies ``$j$'' including e.g.\ thermal and cosmic ray energies $U_{{\rm ej},\,a,\,j}$). The ejecta is  coupled to some set of gas cells ``$b$,'' which consist of all cells for which there is a non-vanishing oriented hydrodynamic face area ${\bf A}_{ba}$ that can be constructed between them and the star particle $a$ (this is defined rigorously in \citet{hopkins:sne.methods}, but essentially includes all cells $b$ which are either ``neighbors of $a$'' or for which ``$a$ is a neighbor of $b$''). If we only have one event site $a$ (we will generalize below), then the update to the conserved quantities (e.g.\ $m$, $U$, and momentum ${\bf p}$) of cell $b$ is given by:
\begin{align}
\label{eqn:flux.m.coupling} m_{b,\,s}^{(1)} &= m_{b,\,s}^{(0)} + |\bar{\bf w}_{ba}|\,m_{{\rm ej},\,a,\,s}  \\ 
\label{eqn:flux.u.coupling} U_{b,\,j}^{(1)} &= U_{b,\,j}^{(0)} + |\bar{\bf w}_{ba}|\,U_{{\rm ej},\,ab,\,j}  \\ 
\label{eqn:flux.p.coupling} {\bf p}_{b}^{(1)} &= {\bf p}_{b}^{(0)} + \Delta m_{ba}\,{\bf v}_{a} + \Delta {\bf p}_{ba} \\ 
\label{eqn:delta.p.defn}\Delta {\bf p}_{ba} &\equiv \psi_{a}\,\Delta m_{ba}\,\left(1 + \frac{m_{b}}{\Delta m_{ba}} \right)^{1/2}\,\left(\frac{2\,\epsilon_{a}}{m_{{\rm ej},\,a}} \right)^{1/2} \hat{\bf{w}}_{ba} \\
\label{eqn:epsilon} \epsilon_{a}  &\equiv {\rm KE}_{{\rm ej},\,a} + \frac{1}{2}\sum_{b} m_{{\rm ej},\,a}\,w^{\prime}_{ba}\,|{\bf v}_{ba}|^{2}  
\end{align}
where $\Delta m_{ba} \equiv |\bar{\bf w}_{ab}|\,m_{{\rm ej},\,a}$, ${\bf v}_{ba} \equiv {\bf v}_{b} - {\bf v}_{a}$, and $\bar{\bf w}_{ba} = |\bar{\bf w}_{ba}|\,\hat{\bf w}_{ba}$ is a normalized vector weight function defined by
\begin{align}
\label{eqn:w.prime} w^{\prime}_{ba} &\equiv \frac{|\bar{\bf w}_{ba}|}{1+\Delta m_{ba}/m_{b}} \\
\label{eqn:vector.weight.normalized} \bar{\bf w}_{ba} &\equiv \frac{{\bf w}_{ba}}{\sum_{c}\,|{\bf w}_{ca}|} \\ 
\label{eqn:vector.weight.normalized.sub1} {\bf w}_{ba} &\equiv \omega_{ba}\, \sum_{+,\,-}\,\sum_{\alpha}\,(\hat{\bf x}_{ba}^{\pm})^{\alpha}\,\left( f_{\pm}^{\alpha} \right)_{a} \\ 
\label{eqn:vectornorm} \left( f_{\pm}^{\alpha} \right)_{a} &\equiv \left\{ \frac{1}{2}\,\left[1 +  \left( \frac{\sum_{c}\,\omega_{ca}\,|\hat{\bf x}_{ca}^{\mp}|^{\alpha}}{\sum_{c}\,\omega_{ca}\,|\hat{{\bf x}}_{ca}^{\pm}|^{\alpha}} \right)^{2}\right]\right\}^{1/2} \\
\label{eqn:vector.weight.def.sub1} (\hat{\bf x}^{+}_{ba})^{\alpha} &\equiv {|{\bf x}_{ba}|^{-1}}\,{\rm MAX}({\bf x}_{ba}^{\alpha},\,0)\,{\Bigr|}_{\alpha=x,\,y,\,z} \\
\label{eqn:vector.weight.def.sub2} (\hat{\bf x}^{-}_{ba})^{\alpha} &\equiv {|{\bf x}_{ba}|^{-1}}\,{\rm MIN}({\bf x}_{ba}^{\alpha},\,0)\,{\Bigr|}_{\alpha=x,\,y,\,z} 
\end{align}
Here ${\bf x}_{ba} \equiv {\bf x}_{b}-{\bf x}_{a}$ and $\omega_{ba}$ is the scalar weight function:
\begin{align}
\label{eqn:solidangle}\omega_{ba} & \equiv \frac{1}{2}\,\left(1-\frac{1}{\sqrt{1+({\bf A}_{ba}\cdot \hat{\bf x}_{ba})/(\pi\,|{\bf x}_{ba}|^{2})}}\right) 
\end{align}
with ${\bf A}_{ba}$ again the oriented hydrodynamic area of intersection of faces between $b$ and $a$ (defined in \GIZMO\ identically to \citealt{hopkins:sne.methods}). The function $\psi_{a}$ is introduced to ensure manifest energy conservation even when the surrounding velocities ${\bf v}_{ba}$ take on arbitrary values, and to account for the effects of a finite cooling radius, and is defined as:
\begin{align}
\label{eqn:psi} \psi_{a} =& 
\begin{cases}
\tilde{\psi}_{a}^{\rm E} \hfill & \ (m_{b} < \tilde{m}_{t,\,ab}) \\ 
{\rm MIN} \left[ \tilde{\psi}_{a}^{\rm C} ,\ \tilde{\psi}_{a}^{\rm E} \right] \hfill & \ (m_{b} \ge \tilde{m}_{t,\,ab}) \\ 
\end{cases}
\end{align}
where 
$\tilde{\psi}_{a}^{\rm E} \equiv {\rm MIN}[\psi_{a}^{\rm E}(\beta_{\psi,\,a}^{0}),\,\psi_{a}^{\rm E}(\beta_{\psi,\,a})]$, 
$\tilde{\psi}_{a}^{\rm C} \equiv \psi_{a}^{\rm C}\,\sqrt{(\Delta m_{ba} + p_{t,\,ba}^{2}/2\,\epsilon_{a})/m_{b}}$, 
$\tilde{m}_{t,\,ab} \equiv p_{t,\,ba} / (v_{t,\,ba} + {\rm MAX}[0,\, \psi_{a}^{\rm C}\,{\bf v}_{ba}\cdot \hat{\bf x}_{ba}])$, 
$\beta_{\psi,\,a}^{0} \equiv \xi_{\rm tk}^{0} / 2\,(1+\xi_{\rm tk}^{0})^{1/2}$, 
and $v_{t,\,ba} \equiv 2\,\epsilon_{a}/p_{t,\,ba}$, and $\psi_{a}^{\rm E}$, $\psi_{a}^{\rm C}$ are given by:
\begin{align}
\label{eqn:psi.E} \psi_{a}^{\rm E}(\beta_{\psi,\,a}) &\equiv \sqrt{1+\beta_{\psi,\,a}^{2}} - \beta_{\psi,\,a} \\ 
\label{eqn:beta.E} \beta_{\psi,\,a} &\equiv \sum_{b}\,{\bf v}_{ba}\cdot \hat{\bf w}_{ba}\,\left({\frac{w^{\prime}_{ba}\,m_{b}}{2\,\epsilon_{a}}}\right)^{1/2} \\
\label{eqn:psi.C} \psi_{a}^{\rm C} &\equiv \frac{1}{1+\Delta m_{ba}/m_{b}}\,{\rm MIN}\left[ \psi_{a}^{\rm E}(\beta_{\psi,\,a}^{0}),\ \frac{1}{2\,\beta_{\phi,\,a}} \right] \\ 
\label{eqn:beta.C} \beta_{\phi} &\equiv \sum_{b}\,{\bf v}_{ba}\cdot \hat{\bf w}_{ba}\,\frac{w_{ba}^{\prime}}{v_{t}} 
\end{align}
Meanwhile, the coupled thermal energy is given by:
\begin{align}
\label{eqn:u.thermal} U_{{\rm ej},\,ab}^{\rm th} =&
\begin{cases}
 U_{{\rm ej},\,a}^{{\rm th},\,0}   +   g_{U,\,a}(\beta_{\psi,\,a}) \,\epsilon_{a}  \hfill & (m_{b} < \tilde{m}_{c,\,ab}) \\
 0 \hfill & (m_{b} \ge \tilde{m}_{c,\,ab}) \\
\end{cases} \\
 g_{U,\,a}(\beta_{\psi,\,a}) &\equiv {\rm MAX}\left[ 0,\ \frac{\xi_{\rm tk}^{0} - 2\,\beta_{\psi,\,a}\,(1+\xi_{\rm tk}^{0})}{1+\xi_{\rm tk}^{0} } \right] 
\end{align}
with $\tilde{m}_{c,\,ab} \equiv p_{t,\,ba}/{\rm MAX}[0,\, v_{t,\,ba} + \psi_{a}^{\rm C}\,{\bf v}_{ba}\cdot \hat{\bf x}_{ba}]$. Here $\xi_{\rm tk}^{0}$ is a dimensionless constant which defines the thermal-to-kinetic energy ratio at the end of the self-similar energy-conserving phase of an idealized remnant in a static medium (by default we take the exact homogeneous-background \citealt{sedov:book} value $\xi_{\rm tk}^{0}=2.54$). In \GIZMO, the thermal heating term is added to the ``hydrodynamic work'' term which is spread over timestep of cell $b$ so that it can be included self-consistently in the fully-implicit thermochemistry routines (i.e.\ this is not operator-split from the cooling operation, since operator splitting this term in dense gas tends to introduce spurious noise in the gas temperature).
The function $p_{t,\,ba} = p_{t}(\epsilon_{a},\,\rho_{b},\,Z_{b},\,T_{b},\,...)$ is the ``terminal momentum'' for a blastwave with the given kinetic energy and ambient gas properties, defined in \firex\ as given in the text (\S~\ref{sec:sne}).\footnote{For our adopted equations and $p_{t,\,ba}(...)$, the asymptotic outward radial momentum coupled in the momentum-conserving limit for an explosion with ejecta energy ${\rm KE}_{{\rm ej},\,a}=10^{51}\,{\rm erg}$, in a stationary (${\bf v}=\boldsymbol{0}$), homogeneous, isotropic background medium with Solar metallicity and ISM density $n=1\,{\rm cm^{-3}}$ becomes $\int d{\bf p} \cdot \hat{\bf x} \approx \sum_{b} \Delta {\bf p}_{ba} \cdot \hat{\bf x}_{ba} \approx p_{t,\,ba}(\epsilon_{a}=10^{51}\,{\rm erg},\,n=1\,{\rm cm^{-3}},\,Z=Z_{\odot},\,...)/(1+\xi_{\rm tk}^{0})^{1/2} \approx 2.5\times10^{5}\,{\rm M_{\odot}\,km\,s^{-1}}$.}

\subsection{Limiting Behaviors and Differences from \firetwo}

Although the expressions above are rather opaque written in compact form, they have simple interpretations given alongside their rigorous derivation in \citet{hopkins:sne.methods}, and the equations above are nearly identical to those therein  with only a couple of subtle differences. In the energy-conserving limit ($m_{b} < \tilde{m}_{t/c,\,ab}$, $\psi_{a} \rightarrow \tilde{\psi}_{a}^{\rm E}$), these expressions guarantee that the  coupled momentum $\Delta {\bf p}_{ba}$ (1) manifestly conserves total linear momentum in the rest frame ($\sum_{b} \Delta {\bf p}_{ba} = \boldsymbol{0}$), (2) is distributed numerically and statistically isotropically in the explosion rest frame (meaning that no artificial or numerical grid directions or bias can be imprinted on the ejecta momentum), (3) has a flux into all neighbor cells which scales proportionally the solid angle subtended by that cell  as seen by the star in its rest frame ($\Delta \Omega_{ab}$), and (4) produces a change in total kinetic+thermal energy across all  cells which {\em exactly} matches the total energy injected, i.e.\ the net increase in the total energy of all gas  cells after the event is coupled is identically the desired and conserved injected energy  (e.g.\ $\sim 10^{51}\,{\rm erg}$, for a SNe). This last property requires the non-trivial values of $\tilde{\psi}_{a}^{\rm E}$, which account for the relative motion of gas around a star, ${\bf v}_{ba} \ne \boldsymbol{0}$, and the fact that kinetic energy is a non-linear function of momentum. For the case with negligible motion (${\bf v}_{ba} \approx \boldsymbol{0}$), $\psi\rightarrow 1$ and the algorithm becomes identical to that in \firetwo; however these terms account for the fact that if, for example, gas is receding/in outflow around a star ($\langle {\bf v}_{ba} \cdot \hat{\bf w}_{ba} \rangle > 0$, so $\beta_{\psi,\,a} > 0$ and $\psi_{a} < 1$), adding a given momentum (directed radially-outwards from the star) to the gas involves a higher energy cost (it must accelerate gas which is already receding rapidly). 

In this (energy-conserving) limit, the only difference between the expressions above and those derived in \citet{hopkins:sne.methods} is that we more carefully specify the behavior of the thermal versus kinetic energy coupling, which was left completely general in that derivation. Importantly, consider the limit where the weighted gas motion (representing some very massive column of gas) around the star is approaching/inflow ($\langle {\bf v}_{ba} \cdot \hat{\bf w}_{ba} \rangle < 0$, so $\beta_{\psi,\,a} < 0$ and $\psi_{a}^{\rm E} > 1$). Then adding some small ejecta momentum $\Delta {\bf p}_{ba} \propto \hat{\bf w}_{ba}$ to the gas directed radially outwards from the star will {\em decelerate} the gas cells (decreasing their kinetic energy). If one coupled {\em only} kinetic energy (or enforced a fixed ratio of kinetic-to-thermal-energy coupled), the only total-energy-conserving solution would be to ``reverse'' the direction of the arbitrarily-massive external gas column -- i.e.\ enhancing the coupled momentum by some enormous ${\psi}_{a}^{\rm E} \gg 1$. This is clearly unphysical. Instead in this limit we take $\psi_{a} \rightarrow \tilde{\psi}_{a}^{\rm E} \equiv  {\rm MIN}[\psi_{a}^{\rm E}(\beta_{\psi,\,a}^{0}),\,\psi_{a}^{\rm E}(\beta_{\psi,\,a})]$ in Eq.~\ref{eqn:psi}, i.e.\ assign the ``normal'' momentum one would obtain, with the residual kinetic energy (representing the deceleration of  the ambient material, or equivalently the enhanced shock-front velocity) converted into thermal energy (the $g_{U,\,a}\,(\beta_{\psi,\,a})\,\epsilon_{a}$ term in Eq.~\ref{eqn:u.thermal}). It is straightforward to verify this ensures all desired conservation properties.

In the momentum-conserving limit ($m_{b} > \tilde{m}_{t/c,\,ab}$, $\psi_{a} \rightarrow \tilde{\psi}_{a}^{\rm C}$), the scale at which the feedback is coupled is larger (in both enclosed mass, and, implicitly, radius) than the scale at which the blastwave reaches its asymptotic/terminal momentum $p_{t}$ and becomes radiative, continuing in the momentum-conserving phase. In this case the above expressions give a scalar radial outward momentum coupled ($=\sum_{b} |\Delta {\bf p}_{ba}|$) equal to $p_{t}/(1+\xi_{\rm tk}^{0})^{1/2}$, with negligible thermal energy ($U_{{\rm ej},\,ab}^{\rm th}\rightarrow 0$), as in \firetwo. This satisfies all same desired conservation properties above, except the energy condition is now an inequality (because the blastwave is radiative, the total energy increase must less or equal to the total injected energy). The only differences between the above and the expressions derived in \citet{hopkins:sne.methods} are that (1) the MIN condition in Eq.~\ref{eqn:psi} explicitly ensures the coupled kinetic energy in this  limit cannot be larger than the energy-conserving case, and (2) the threshold values $\tilde{m}_{t/c,\,ab}$ for this case account in slightly more detailed fashion for the local relative velocity.\footnote{As noted in \citet{hopkins:sne.methods}, $p_{t}/v_{t}$ is the swept-up mass at which the shell would reach the terminal momentum if ${\bf v}_{ba} = \boldsymbol{0}$. The $\psi_{a}^{\rm C}\,v_{ba}\cdot\hat{\bf x}_{ba}$ term accounts for the fact that if the gas is, for example, receding from the star faster than some speed $v_{0} > v_{t}$, the velocity at which the terminal mass is reached and at which the shell will cool must be larger accordingly. The presence of non-zero relative velocity also breaks the strict degeneracy between cooling and terminal momentum criteria.}

\subsection{Accounting for Multiple Events in the Same Timestep}

The expressions in \S~\ref{sec:appendix:sne:one} above ensure energy conservation as desired, for a single feedback ``event.'' But an additional complication arises if there are multiple events from different  source star particles $a$, which all influence the same gas cell $b$ within a single numerical timestep. This can arise fairly easily if, for example, a dense cluster of young stars forms and blows away most of its surrounding gas, so many star particles ``see'' the same surrounding set of gas cells. In this limit, one can simply linearly add coupled scalar quantities such as mass or thermal energy independently from each event, i.e.\ $m_{b}^{(1)} = m_{b}^{(0)} + \Delta m_{b}^{\rm tot}$ where $\Delta m_{b}^{\rm tot} \equiv \sum_{a} \Delta m_{ba}$, retaining manifest conservation. But now define the linear momentum  change from Eq.~\ref{eqn:flux.p.coupling} for a single event $\Delta {\bf p}_{ba}^{\prime} \equiv \Delta m_{ba}\,{\bf v}_{a} + \Delta {\bf p}_{ba}$, and note that the corresponding kinetic energy change is, by careful construction  of the $\psi$ terms above, given by
\begin{align}
\Delta {\rm KE}_{ba} &\equiv {\rm KE}_{b}^{(1)} - {\rm KE}_{b}^{(0)} = \frac{|{\bf p}_{b}^{(0)} + \Delta {\bf p}_{ba}^{\prime}|^{2}}{2\,(m_{b}^{(0)} +  \Delta m_{ba})}   - \frac{|{\bf p}_{b}^{(0)}|^{2}}{2\,m_{b}^{(0)}} 
\end{align}
Next note that if one naively then independently added the total linear momentum change from all events, ${\bf p}_{b}^{(1)} = {\bf p}_{b}^{(0)} + \Delta {\bf p}_{b}^{\rm tot}$ with $\Delta {\bf p}_{b}^{\rm tot} \equiv \sum_{a} \Delta {\bf p}_{ba}^{\prime}$, one would obtain a total naive kinetic energy change: 
\begin{align}
\Delta {\rm KE}_{b}^{\rm naive} \equiv  \frac{|{\bf p}_{b}^{(0)} + \Delta {\bf p}_{b}^{\rm tot}|^{2}}{2\,(m_{b}^{(0)} +  \Delta m_{b}^{\rm tot})}   - \frac{|{\bf p}_{b}^{(0)}|^{2}}{2\,m_{b}^{(0)}} \ne \Delta {\rm KE}_{b}^{\rm tot} \equiv \sum_{a} \Delta {\rm KE}_{ba}
\end{align}
unless there is only one event $a$. The problem is that kinetic energy is a non-linear function of momentum and mass, so there are non-linear cross-terms between all independent events $a$ (e.g.\ $|\Delta {\bf p}_{b}^{\rm tot}|^{2} = (\Delta {\bf p}^{\prime}_{ba_{1}} + \Delta {\bf p}^{\prime}_{ba_{2}} + ...) \cdot (\Delta {\bf p}^{\prime}_{ba_{1}} + \Delta {\bf p}^{\prime}_{ba_{2}} + ...)$, so all possible $\Delta {\bf p}^{\prime}_{ba_{n}} \cdot \Delta {\bf p}^{\prime}_{ba_{m}}$ appear). If, say, many events accelerate a gas cell away from a group of stars in the same timestep, this could allow the actual total energy to increase by much more than the total injected mechanical energy from feedback. 

The conceptually simplest solution to this is to ensure that all stellar feedback events are processed in serial: i.e.\ the total change from state $(0)\rightarrow(1)$ for all gas cells $b$ is processed for one event $a_{1}$ before evaluating the ``initial'' properties (now those from state $(1)$) for event $a_{2}$. While this can often be true in many regions of a simulation given our relatively small  timesteps, it is computationally prohibitive to ensure this is always and everywhere the case for all cells in a massive-parallel galaxy-scale simulation: forcing the entire simulation to process all stellar feedback operations strictly in serial would bottleneck the run. It is also possible in principle to solve simultaneously for all $\psi_{ba}$ terms for all cells and events, but this would require an impractically enormous implicit solver (far larger than $\mathcal{O}(N^{2})$). 

Instead, we retain the desired manifest energy conservation by adopting the approximate ``corrected'' momentum update, ${\bf p}_{b}^{(1)} \rightarrow {\bf p}_{b}^{(0)} + f_{{\bf p},\,b}^{\rm corr}\,\Delta{\bf p}_{b}^{\rm tot}$, with
\begin{align}
f_{{\bf p},\,b}^{\rm corr} 
&\equiv \frac{-{\bf p}_{b}^{(0)}\cdot  \Delta{\bf p}_{b}^{\rm tot}+\sqrt{|{\bf p}_{b}^{(0)}\cdot  \Delta{\bf p}_{b}^{\rm tot}|^{2}-\tilde{p}^{2}_{b}\,|\Delta{\bf p}_{b}^{\rm tot}|^{2}}}{|\Delta{\bf p}_{b}^{\rm tot}|^{2}} \\
\tilde{p}^{2}_{b} &\equiv |{\bf p}_{b}^{(0)}|^{2}\,\frac{\Delta m_{b}^{\rm tot}}{m_{b}^{(0)}} + 2\,(m_{b}^{(0)} + \Delta m_{b}^{\rm tot})\,\Delta {\rm KE}_{b}^{\rm tot}
\end{align}
which exactly guarantees the desired energy change
\begin{align}
\Delta {\rm KE}_{b} &\equiv  \frac{|{\bf p}_{b}^{(0)} + f_{{\bf p},\,b}^{\rm corr}\,\Delta {\bf p}_{b}^{\rm tot}|^{2}}{2\,(m_{b}^{(0)} +  \Delta m_{b}^{\rm tot})}   - \frac{|{\bf p}_{b}^{(0)}|^{2}}{2\,m_{b}^{(0)}} = \Delta {\rm KE}_{b}^{\rm tot}
\end{align}
while minimizing the change to the momentum update, and reduces exactly to the normal solution ($f_{{\bf p},\,b}^{\rm corr} \rightarrow 1$) if there is only one event $a$ influencing a given cell $b$ or if the non-linear cross-terms are small.

\end{appendix}

\end{document}